\documentclass[
 reprint,
 amsmath,amssymb,
 aps, prb,
]{revtex4}

\usepackage{amsfonts}
\usepackage{latexsym}
\usepackage{graphicx}
\usepackage[usenames]{color}
\usepackage[utf8]{inputenc}
\usepackage[english,bulgarian]{babel}
\usepackage{hyperref}

\expandafter\def\expandafter\UrlBreaks\expandafter{\UrlBreaks
  \do\a\do\b\do\c\do\d\do\e\do\f\do\g\do\h\do\i\do\j%
  \do\k\do\l\do\m\do\n\do\o\do\p\do\q\do\r\do\s\do\t%
  \do\u\do\v\do\w\do\x\do\y\do\z\do\A\do\B\do\C\do\D%
  \do\E\do\F\do\G\do\H\do\I\do\J\do\K\do\L\do\M\do\N%
  \do\O\do\P\do\Q\do\R\do\S\do\T\do\U\do\V\do\W\do\X%
  \do\Y\do\Z}

\addto\captionsbulgarian{%
}

\begin{document}
\title{Measuring Plank constant with colour LEDs and compact disk}

\author{Vassil~N.~Gourev}
\email[E-mail: ]{gourev@phys.uni-sofia.bg}
\affiliation{Department of Atomic Physics, Faculty of Physics,\\
 St.~Clement of Ohrid University at Sofia,\\
5 James Bourchier Blvd., BG-1164 Sofia, Bulgaria}

\author{Stojan~G.~Manolev}
\email[E-mail: ]{manolest@yahoo.com}
\affiliation{Middle School ``Goce Delchev'',\\
Purvomaiska str. 3, MKD-2460 Valandovo}

\author{Vasil~G.~Yordanov}
\email[E-mail: ]{vasil.yordanov@gmail.com}
\affiliation{Faculty of Physics,\\
 St.~Clement of Ohrid University at Sofia,\\
5 James Bourchier Blvd., BG-1164 Sofia, Bulgaria}

\author{Todor~M.~Mishonov}
\email[E-mail: ]{mishonov@gmail.com}
\affiliation{Department of Theoretical Physics, Faculty of Physics,\\
St.~Clement of Ohrid University at Sofia,\\
5 James Bourchier Blvd., BG-1164 Sofia, Bulgaria}



\begin{abstract}
This problem was given on the Open Experimental Physics Olympiad (OEPO) ``The day of the photon'' on 25 April 2015.
The Olympiad was a part of the celebration of the International Year of Light (IYL) \url{http://www.light2015.org/Home/Event-Programme/2015/Competition/Bulgaria-Second-Experimental-Physics-Olympiad--25-April-2015-in-Sofia.-The-Day-of-the-Photon-in-the-International-Year-of-the-Light.html} and was organized by the Sofia Branch of the Union of Physicist in Bulgaria and the Regional Society of Physicists of Strumica, Macedonia.
\end{abstract}

\maketitle

\section{Problem}
\begin{figure}[ht]
\includegraphics[scale=0.88]{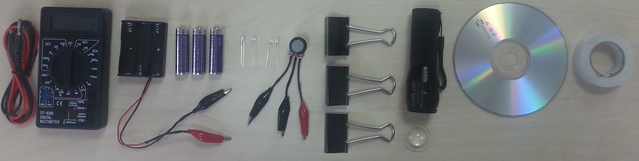}
\caption{Details from the experimental set for measuring the Plank constant $h$: 
Wires ending with plug and crocodile clip,
multimeter,
battery holder and 3 batteries of 1.5 V,
4 LEDs,
potentiometer with wires ending with crocodile clips,
3 binder clips,
tourch and its lens,
Compact Disc (CD)
double stick tape,
photo-resistor,
non transparent tube,
plasticine,
paper clip
}
\label{Fig:Set_all} 
\end{figure}

\begin{figure}[ht]
\includegraphics[scale=0.87]{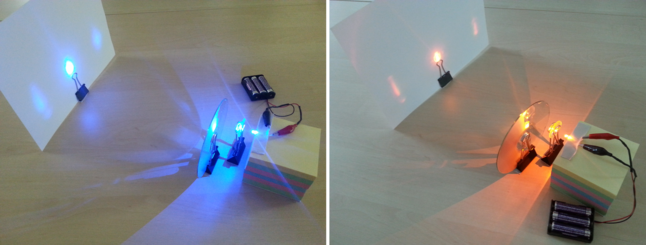}
\caption{Experimental setup for measuring the wavelength of LED light: 
The LED is connected with wires to the power supply.
The light is collimated with lens then it pass the tracks of the transparent compact disk
and one can see a bright image of the LED on the screen.
Together with the central maximum two pale diffraction maximums can be observed on the screen.
Diffraction angles are different for the different colour of the used LEDs, yellow and blue on the pictures.
The wavelength of the light can be determined by measuring the diffraction angles.
}
\label{Fig:experimental_setup_blue_amber} 
\end{figure}

\begin{figure}[ht]
\includegraphics[scale=0.15]{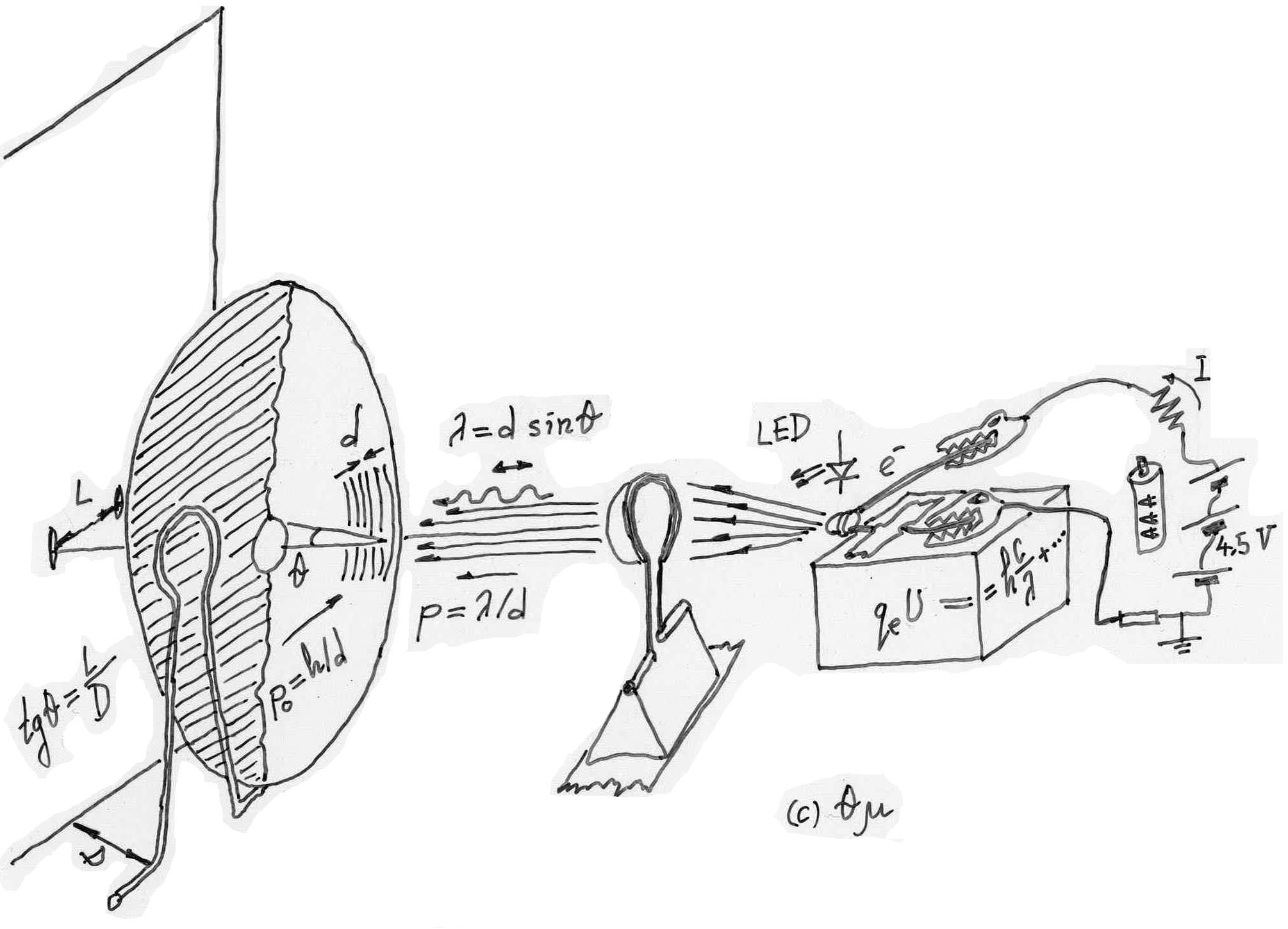}
\caption{Scheme of the setup shown on Figure~\ref{Fig:experimental_setup_blue_amber}.
The LED is connected with wires to the power supply using crocodile clips.
A photon from the collimated light has momentum $p=h/\lambda$, which is perpendicular to the CD.
The photon obtains momentum $p_0=h/d$ in direction perpendicular to the direction of the tracks of the CD, 
and for the diffraction angle $\theta$ we have $\tg\theta=p_0/p$.
$D$ is the distance between the screen and the CD, and $L$ is the distance between the central and the first diffraction maximum; $\tg\theta=L/D.$ 
The binder clips, which hold the lens and the CD, are drawn schematically on the picture.
}
\label{Fig:setup_scheme} 
\end{figure}

\begin{figure}[h]
\includegraphics[scale=0.2]{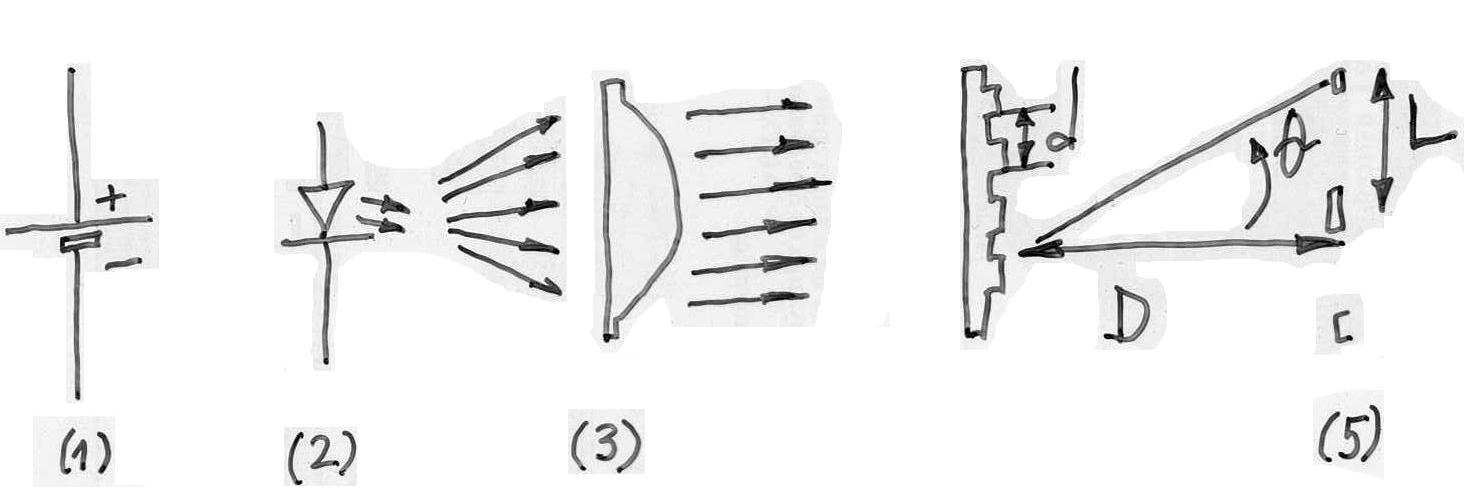}
\caption{The energy path: 1) Chemical energy in the battery, 2) The energy of the light, 3) The energy focus, 4) Diffraction, 5)  Absorption and heat.}
\label{Fig:optical_scheme1} 
\end{figure}

\begin{figure}[ht]
\includegraphics[scale=0.92]{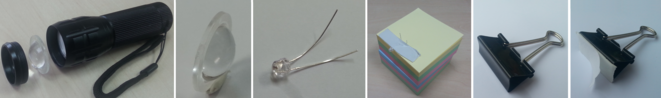}
\caption{Ring shaped screw, flashlight lens, flashlight,
LED, LED sticked with stick tape on a paper notes cube, binder clip sticked with stick tape.}
\label{Fig:set_diode_lens_disk} 
\end{figure}

Measure the Plank constant $h$, using the experimental set and the experimental setup shown on Figure~\ref{Fig:Set_all},  Figure~\ref{Fig:experimental_setup_blue_amber},
Figure~\ref{Fig:setup_scheme}, Figure~\ref{Fig:optical_scheme1}, and Figure~\ref{Fig:set_diode_lens_disk}, and the known values for the speed of light $c=299792458$~m/s 
and the charge of the electron $q_e=1.602176565 \times$$10^{-19}$~C, 
as well as the distance between the tracks of the compact disk $d=1.50\, \mu$m.

\begin{figure}[ht]
\includegraphics[scale=0.92]{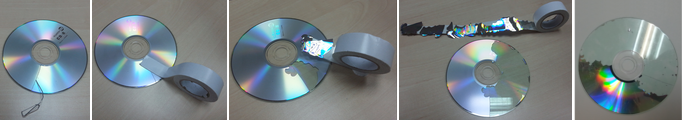}
\caption{The technology of peeling the foil from the compact disk:
1) The foil is scratched radial with the paper clip.
2) A stick tape is stuck on the scratch.
3) The tape is pulled and the foil stay sticked on it.
4) Sticking and pulling of the stick tape is repeated until the half part of the disc is peeled from the foil.}
\label{Fig:set_disk_foil_detachment} 
\end{figure}

\section{Photoeffect. Introduction}

The photo-effect is phenomenon of ejecting electron from the metal surface by an incident photon.
The energy conservation gives the relation between the photon energy  
$h\nu$, the kinetic energy of electron and the work function parametrized with voltage.
In the LED there is an inverse process of internal photo-effect:
an electron and a hole with small energies recombine and emit a photon.
The relation between the threshold voltage $U_c$ at which the LED starts to emit light and the energy of the photon is
\begin{equation} 
 q_eU_c\approx h\nu+\mbox{const},
\label{photoeffect}
\end{equation}
where the $q_e$ is the charge of the electron.
The constant in the equation above has small dependence on the material from which the LED is made.
The kinetic energies of the recombining electrons are negligible,
i.e. much smaller than the energy of the emitted photon.
Approximately we can assume that different diodes in the experimental set are made of materials with similar material constant. 

\section{Electrical part}

\subsection{Measuring the voltage drop $U_\mathrm{c}$ at which the LED starts to emit light}
\label{U_c_measurements}
\begin{enumerate}
\item
\textit{Redraw the electrical schematics on Figure~\ref{Fig:Potentiometer}a omitting the ampere meter (replacing it with a wire). 
Wire the available components in this circuit.}

Put the batteries in the battery holders.
Connect the diodes according to the schematics on Figure~\ref{Fig:MaxLight}.
Connect the power supply to the diode using the wires that ends with crocodile clips.
Before each of the crocodile clips there is a resistor, which limits the current through the LED and it can not dazzle you.
Even though the light is not so bright do not look directly in the LED. 

In this subtask you may not connect the ammeter and replace it with a wire.
Connect some of the LED with the potentiometer as it is shown on the circuit on 
Figure~\ref{Fig:Potentiometer} or more precisely with the schematics that you have redrawn without ammeter.
Connect the two lateral electric terminals of the potentiometer with the batteries using the wires and the crocodile clips from the battery holder.
You may use also the additional wires that terminates with crocodile clips.
Connect the voltmeter with the LED leads.
Rotate the shaft of the potentiometer until the LED start to emit light.
If the LED is dark or the intensity of the light does not change when you rotate the shaft of the potentiometer then you have probably connected the LED in opposite direction or have a mistake in wiring the schematics.

\begin{figure}[ht]
\includegraphics[scale=0.65]{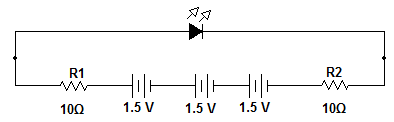}
\caption{
Electric circuit for powering the LED used in the diffraction experiment.
This circuit can also be used for testing on which direction the LED lights.
The current from the 3 sequentially connected batteries flows through the LED.
There are two resistors $R_1$ and $R_2$, which are hidden into black isolating tubes.
These resistors limit the current through the LED and protect it from damage.
The point on the vertical lines on the circuit are symbolic notation for the connection with crocodile clips.
}
\label{Fig:MaxLight} 
\end{figure}

\begin{figure}[ht]
\includegraphics[scale=0.8]{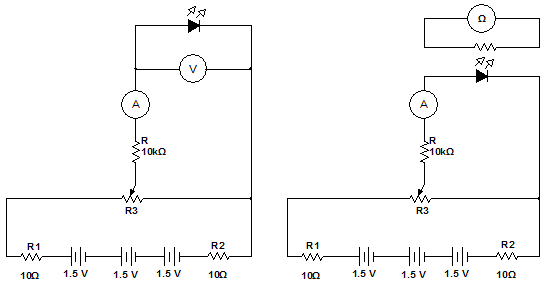}
\caption{
Adjustable voltage regulator with potentiometer for power suppling the LEDs and for investigation of $I(U).$
The voltage from the battery holder is applied to the potentiometer.
The LED is connected through a resistor $R$ to the middle terminal of the potentiometer and to one of the other terminals.
The big value of the resistor $R$ allows to investigate this part of the V-I characteristic at which the LED starts or stops to emit light.
The ammeter is connected in series with the LED, and the voltmeter is connected in parallel.
If the LED does not light in one of the end positions of the potentiometer then you can change the polarity of the batteries.
There are two resistors hidden in an insulated tubes before the crocodile clips of the battery holder leads.}
\label{Fig:Potentiometer} 
\end{figure}

\item 
\label{Item:Uc}
\textit{Measure the voltage drop $U_c$ at which each of the LED starts to emit light.} 

Use the circuit from the previous task and check if the LED light on or off when you rotate the shaft of the potentiometer.
It is required to work in a dark room and watch weak light in order to find $U_c$.
Make several measurements for each of the LEDs and find the average value of $U_c$ for each of them.
Calculate the maximum deviation and the mean value, which gives a quantity for precision of the measurement.
Put the result in a table by ordering the values of $U_c$, see Table~\ref{table:U_c_table}.
Write in the table for each of the LEDs the: colour, mean value of $U_c$, max deviation from the mean and the number of measurements.

\begin{table}[ht]
\caption{Experimental results for the voltage drop at which each LED starts to emit light.} 
\begin{tabular}{| c| l | c |}
\tableline 
\No & Colour & $U_c\mathrm{[V]}$  \\
\tableline \tableline
1& red& \\
2& yellow   & \\
3& green  & \\
4& blue     & \\
\tableline 
\end{tabular}
\label{table:U_c_table}
\end{table}

\subsection{Measuring the threshold voltage drop $\tilde{U}_\mathrm{c}$ by investigation of volt-ampere characteristic of LED}
\item 
\label{CurrentVoltageCharacteristics} \textit{Measure the Volt-Ampere (V-I) characteristic of all 4 LED}

Turn on the multimeter, which you carry with you, as an ammeter and connect it in the circuit according Figure~\ref{Fig:Potentiometer}a so that it can measure the current that flows through the LED.

Investigate the dependence between the current and the voltage $I(U)$ -- Volt-Ampere (V-I) characteristic for the available LEDs.
Fill a separate table with 25 rows for each of the LEDs.
The table has to have the following columns:
the sequence number of the measurement, the current $I$ and the voltage $U$. 
On a separate white paper prepare a table similar to the one presented on Table~\ref{Table:I(U)}.

First, check if in one of the end position of the potentiometer the LED emits light; 
change the LED direction if it does not light at all.
At the maximum brightness of the LED write the values of the current and the voltage -- this is your first measurement.
Rotate the shaft of the potentiometer and decrease the current with 20~$\mu$A;
write the values of the current and the voltage into the table until the current reach value lower than 10~$\mu$A.
Watch the voltmeter, if needed change its measuring scale.
Our advise it to rotate the shaft of the potentiometer and to write the values of the voltage on each 200~mV, until the voltage becomes zero.
Fill the columns of the table for the different colour LEDs.

\begin{table}[ht]
\caption{Volt-Ampere characteristics of the LEDs.
Sequentially the columns are for red, yellow, green and blue LED. 
You need at least 20 measurements for each diode to present well its V-I characteristic.
}
\begin{tabular}{| r | c | c | r | c | c | r | c | c | r | c |c |}
\tableline
\multicolumn{3}{|c|}{red}
& \multicolumn{3}{|c|}{yellow}
& \multicolumn{3}{|c|}{green}
& \multicolumn{3}{|c|}{blue}\\
\tableline 
\No & $I\;[\mu\mathrm{A}$] & $U$ [V] & 
\No & $I\;[\mu\mathrm{A}$] & $U$ [V] & 
\No & $I\;[\mu\mathrm{A}$] & $U$ [V] & 
\No & $I\;[\mu\mathrm{A}$] & $U$ [V] \\
\tableline \tableline
 1& & & 1& & & 1& & & 1& & \\
 2& & & 2& & & 2& & & 2& & \\
 ...& & & ...& & & ...& & & ...& & \\

\tableline 
\end{tabular}
\label{Table:I(U)}
\end{table}

\item 
\label{CurrentVoltageCharacteristics} \textit{Draw the graphics of the Voltage-Ampere characteristic of each of the 4 LEDs, using the results from the Table~\ref{Table:I(U)}.} 

Present graphically the measurements from the previous task, using one and the same coordinate system for the different colour LEDs. 
Before to start to put the points on the graphics think about the coordinate scales.
It is better if your data fits well the entire area of the squared paper.
Use different colour for the experimental points obtained from the different LEDs, so that you can distinguish the V-I characteristic of the LEDs.
It is better to use the same pencil colour as the colour of the LED.
On the $x$-axis write the voltage $U[\mathrm{V}]$.
On the $y$-axis write the current $I[\mu\mathrm{A}]$.

\item 
\label{CurrentVoltageCharacteristics_U_c_tilda} \textit{Determine the threshold voltage drop $\tilde{U}_\mathrm{c}$ using the Voltage-Ampere characteristic of each LED.}

On the V-I characteristic you can see almost linear segment with high slope at higher currents.
Draw straight line, which touches this segment.
Write the voltage of the crossing points $\tilde{U}_\mathrm{c}$ into table as it is shown on Table~\ref{table:U_c__tilde_table}. 
Compare these threshold voltages for each LED with the voltage drop when they start to emit light $U_c$ -- the last you have measured in subtask~\ref{Item:Uc}.
What is the maximum difference $\tilde{U}_\mathrm{c}-U_\mathrm{c}$ for the different colour LEDs?

\begin{table}[ht]
\caption{A template table for writing the experimental results with the values of the threshold voltage drops determined from the investigation of V-I characteristic of the LEDs.} 
\begin{tabular}{| c| l | c |}
\tableline 
\No & Colour & $\tilde{U}_c\mathrm{[V]}$  \\
\tableline \tableline
1& red & \\
2& yellow   & \\
3& green  & \\
4& blue    & \\
\tableline 
\end{tabular}
\label{table:U_c__tilde_table}
\end{table}

\item 
\label{CurrentVoltageCharacteristics_small_current} \textit{Give a quality explanation of the low slope in the V-I characteristic at low currents.} 

Take a look at the low slope in the V-I characteristic at low currents.
Notice that the slope is almost the same for each of the LEDs.
How you can explain this fact?

\subsection{Investigation of the intensity of the light, emitted from the LED, using a photo-resistor.
An alternative method for measuring the threshold voltage drop $\tilde{U}_\mathrm{c}$ of the LEDs}

\item
\label{Siemens-Ampere}
\textit{Measure the dependence of the resistance of the photo-resistor on the current flowing though the LED.}

Wire the circuit shown on Figure~\ref{Fig:Potentiometer}б.
The ammeter measures the current through the LED. You may omit the voltmeter.
Now switch the multimeter as an ohmmeter to measure the resistance of the photo-resistor $R_\varphi.$
Put the photo-resistor close to a LED that lights on its maximum intensity and observe how its resistance changes.
Put the LED and the photo-resistor into the tube and bung a plasticine at the both ends of the tube, so that to avoid disturbance of the light from the environment.
Write in a Table~\ref{Table:R_varphi(I)} the current flowing through the LED and the resistance of the photo-resistor $R_\varphi.$
Add in this table an additional column for the reciprocal value of the resistance $1/R_\varphi.$ This value is called conductance and it is measured in units $\Omega^{-1}.$
To make this experiment faster follow the instructions:
Rotate the shaft of the potentiometer, decrease the current with 20$\;\mu$A and write the values of the current and the resistance into the table.
When you reach the zero current try to investigate the small currents up to 15$\;\mu$A.
Repeat the measurements again for each of the LEDs and write the results in different tables.
Calculate the reciprocal value of the resistance $1/R_\varphi$ and write it in the third column.

\begin{table}[ht]
\caption{Dependence of the resistance of the photo-resistor on the current flowing through the photo-diode. The table should contain at least 25 rows for each of the LED}
\begin{tabular}{| r | c | c  |c | r | c | c | c | r | c | c | c | r | c | c | c |}
\tableline 
\multicolumn{4}{|c|}{червен}
& \multicolumn{4}{|c|}{жълт}
& \multicolumn{4}{|c|}{зелен}
& \multicolumn{4}{|c|}{син}\\
\tableline 
\No & $I\;[\mu\mathrm{A}$] & $R_\varphi\, [\Omega]$ &  $1/R_\varphi\, [\Omega^{-1}]$ &
\No & $I\;[\mu\mathrm{A}$] & $R_\varphi\, [\Omega]$ &  $1/R_\varphi\, [\Omega^{-1}]$ &
\No & $I\;[\mu\mathrm{A}$] & $R_\varphi\, [\Omega]$ &  $1/R_\varphi\, [\Omega^{-1}]$ &
\No & $I\;[\mu\mathrm{A}$] & $R_\varphi\, [\Omega]$ &  $/1R_\varphi\, [\Omega^{-1}]$  \\
\tableline \tableline
 1& & & &1& & & & 1& & & & 1& & & \\
 2& & & &2& & & & 2& & & & 2& & & \\
 ...& & & &...& & & & ...& & & & ...& & &\\

\tableline 
\end{tabular}
\label{Table:R_varphi(I)}
\end{table}

\item 
\label{} \textit{For all four LEDs make a graphics with the dependence of the resistance of the photo-resistor on the current flowing through the LED, using the results from the Table~\ref{Table:R_varphi(I)}.} 

On the $x$-axis of the graphics put the current through the LED $I$, 
and on the $y$-axis put the reciprocal resistance $1/R_\varphi$ in units $\Omega^{-1}.$
Use one and the same coordinate system for each of the LEDs.
It is better to make the graphics of each of the LED with different colour.
Fit the experimental points with a smooth curve.
Analyze these curves and try to explain the experimental results.

\item 
\label{} \textit{Analyze how the photo-resistor can be used to determine the threshold voltage $\tilde{U}_\mathrm{c}$.} 

\section{Optical part}

\subsection{Measure the wavelength of the light emitted from the LEDs.}

\item 
\textit{Peel the half part of the compact disc.} 

Use the paper clip to scratch the foil of the compact disc in radial direction.
Stick the double stick tape on the scratch and pull it to peel the foil of the CD as it is shown on
Figure ~\ref{Fig:set_disk_foil_detachment}.
Continue to stick and pull the stick tape until you peel half of the CD.
Grasp the CD with the binder clip as it is shown of Figure ~\ref{Fig:experimental_setup_blue_amber}.

\item 
\textit{Setting up the optical experiment.}

Stick the double stick tape on the back of the one of the binding clips. 
You can tear the stick tape easily with hands.
Grasp the white paper screen with a binding clip.
Remove the cover of the opposite side of the double stick tape and stick the binding clip with the paper screen on the table.
The screen should be placed in vertical position.
In the same way place the transparent CD parallel to the paper screen. 
The distance between the screen and the CD should be approximately 20~cm.
Be more precise in the parallel positioning of the CD and the screen.
If you are not satisfied with the accuracy of the setup you can unstick the CD with the binding clip and stick it again.
Measure the distance $D$ between the screen and the CD with accuracy 1~mm.
Graphs the lens with a binding clip and stick it on a distance around 3--4~cm in front of the CD.
Stick one of the LED on the top of the paper notes cube using the double stick tape.
The LED should be on the same height as the center of the lens.
The power supply of the LED in this diffraction experiment is shown on Figure~\ref{Fig:MaxLight}.

Light the LED.
Move  the paper notes cube with the LED until you see a bright image of the LED focused on the screen.
This is the real image of the LED.
You can remove or add paper notes until the image on the screen is on the same height as the LED with accuracy of 1--2~cm. 

Lets track the way of the light.
The light is emitted from the LED, which is almost at the focal point of the lens as it is shown on Figure ~\ref{Fig:experimental_setup_blue_amber}.
The parallel beam passes through the transparent CD and on the screen one can see a central maximum and two diffraction maximums.

Move the CD into the binding clip until the narrow light beam pass through the horizontal radius of the disc.
In this position of the CD the central maximum and the other diffraction maximums are aligned in a horizontal line on the screen.
This alignment is important for accurate measurement of the wavelength of light.

Use a colour pen to mark the central maximum and the two diffraction maximums.
Due to the lack of yellow pen you can use the black one.
The distance between the two diffraction maximums is $2L,$ 
and the distance between the central maximum and each of the diffraction maximums is $L.$

\item
\textit{What is the focal length of the lens $f$ in unites meters and its reciprocal value $1/f?$}

\item 
\textit{Measure the diffraction angles.}

Measure the distance $D$ between the CD and the screen with accuracy 1~mm.
For each of the LEDs measure the distance $L$ between the central maximum and the first diffraction maximum.
It is recommended to draw the boundaries of the light spots with a pen and after that to determine their centers.
Fill the template Table~\ref{table:Sample} with the values and the threshold voltage drop $U_\mathrm{c}$ measured in section~\ref{U_c_measurements}.

\item 
\textit{Calculation of the sine of the diffraction angle.}

Foar each of the colours calculate the diffraction angle
$\sin(\theta)=L/(L^2+D^2)^{1/2}=\sin(\arctg(L/D)).$ 
You can find a graphical solution.

Draw a right-angled triangle with catheti $L$ and $D$ and hypotenuse 100~mm.
Measure the cathetus opposed of the angle $\theta$ and parallel to $L.$ 
The length of this cathetus in decimeter units is $\sin(\theta).$ 
In Table~\ref{table:Sample} fill the columns with the angle variables and calculate the corresponding wavelengths.
The columns with the threshold voltage drop $\tilde{U}_c$ you can fill from the electrical measurements.

\item
\textit{Calculate the Plank's constant using the experimental data.}
\label{PlanckError} 

Make a graphics using the experimental data from Table~\ref{table:Sample}.
On $y$-axis write the $q_eU_c$, and on the $x$-axis $\nu$. 
Fit the experimental points with a straight line that pass maximally close to the points. 
This straight line follows the approximate equation
\begin{equation} 
 q_eU_c\approx h\nu+\mbox{const.}
\label{photoeffect}
\end{equation}
Determine the Plank's constant $h=\Delta(q_eU_c)/\Delta(\nu)$ from tangent of the angle between the straight line and the $x$-axis.
Here $\Delta$ denotes difference for the segment of the straight line; 
difference in the $x$-axis and $y$-axis segments.
How the value of the $h=\dots\,\mathrm{J\,s}$ that you have measured is in agreement with the know value of this fundamental constant? Fill the dots!
What is the deviation $100\,(h_\mathrm{exp}-h)/h=\dots\%$?
Surround and under line the result!

\item
\textit{Determine the $h$ using the V-I characteristic.}

Go to sub-section~\ref{CurrentVoltageCharacteristics_U_c_tilda}. 
From the experiment described there determine the threshold voltage drop $\tilde{U}_c$.
Repeat the same procedure to determine the Plank's constant using the $\tilde{U}_c$,
not the subjectively determined value of the voltage drop $U_c$.
Which of the both methods is better?
What could be the possible systematic errors of this experiment?
Propose a change in the experiment to improve the accuracy of the measurements.

The constant that appear in the equation~(\ref{photoeffect}) of the photo-effect has significant difference for the green LED.
Omit this experimental point for the green LED and determine $h$ from the experimental points of the rest of the LEDs.

\begin{table}[ht]
\caption{Experimental data for diffraction and the threshold voltage drop for different colour LEDs; 
$D=\dots\,\mathrm{cm}.$} 

\begin{tabular}{| c| l | c | c | c | c |c | c | c | c | c | c |}
\tableline 
\No & Цвят &  $L$ [cm] & $\tg(\theta)$ & $\theta$ [rad]& $\sin(\theta)$ & $\lambda\,[\mathrm{nm}]$
& $\nu\mathrm{[Hz]}$ & $U_c\mathrm{[V]}$ 
& $\tilde{U}_c\mathrm{[V]}$ 
& $q_eU_c\mathrm{[J]}$ & $q_e\tilde{U}_c\mathrm{[J]}$ \\
\tableline \tableline
1& червен& & & & & & $\qquad\times 10^{14}$ & & & $\qquad\times 10^{-19}$ & $\qquad\times 10^{-19}$ \\
2& жълт   & & & & & & $\qquad\times 10^{14}$ & & & $\qquad\times 10^{-19}$ & $\qquad\times 10^{-19}$ \\
3& зелен  & & & & & & $\qquad\times 10^{14}$ & & & $\qquad\times 10^{-19}$ & $\qquad\times 10^{-19}$ \\
4& син     & & & & & & $\qquad\times 10^{14}$ & & & $\qquad\times 10^{-19}$ & $\qquad\times 10^{-19}$ \\
\tableline 
\end{tabular}
\label{table:Sample}
\end{table}

\subsection{Optical experiment with mixing colour light.}
\item 
\label{item:RGB_task}
\textit{Find the colours of the shadows.}

Three light sources are shown on Figure~\ref{Fig:RGB_shadow}.
They illuminate an object and create different shadows~\cite{Bib:RGB} on the screen.
 What are the colours of the different areas marked with numbers from 1 to 6.
You can find the colours with thinking or to use the colour LEDs that you already have in order to make the experiment.
For this experiment all 3 LEDs have to be light on as it is shown of Figure~\ref{Fig:RGB_diodes}.
The red and blue LED are brighter and their intensity has to be decreased so that when their lights are mixed with the 
green LED to obtain white light.
In order to decrease the intensity of the LED they have to be powered through a current limiting resistor with appropriate values as it is shown on Figure~\ref{Fig:RGB_schema}.
The red and the green LED have to be on a distance of 1.5~cm from the blue LED in the middle.
In another experiment you can try to put also the LEDs in the vertexes of a triangle.
Use the shadow of one cent coin grasped with a binding clips as shown on Figure~\ref{Fig:RGB_holder} 

\begin{figure}[ht]
\includegraphics[scale=0.6]{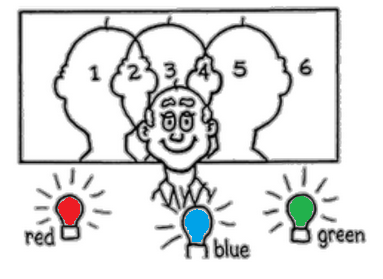}
\caption{
Shadow from 3 light sources: red, green and blue.
What is the colour in the different shadow areas?
}
\label{Fig:RGB_shadow}
\end{figure}

\begin{figure}[ht]
\includegraphics[scale=0.7]{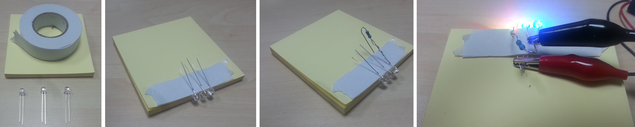}
\caption{Setup to create simultaneous light emission from red, green and blue LEDs:
1) LEDs, paper notes, double stick tape.
2) LEDs are sticked with stick tape on the paper notes. 
One of the LEDs leads is under the stick tape, the other is above it.
3) The red and the blue diode are connected with a resistor so that the light from all LEDs to be mixed into white light.
4) The crocodile clips from the battery holder wires are attached to the connected in parallel LEDs.
If some of the LED is not lighting exchange its leads. 
Place the LEDs on a distance equal to the diameter of the coin, which you will use to create the shadow on the screen.
}
\label{Fig:RGB_diodes} 
\end{figure}

\begin{figure}[ht]
\includegraphics[scale=0.55]{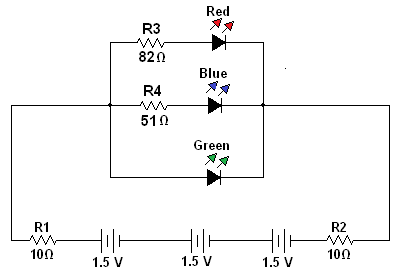}
\caption{The electric circuit of 3 LEDs connected in parallel.
The current through the red LED is limited by 82~$\Omega$ resistor,
and the current through the blue LED is limited by 51~$\Omega resistor.$
There are two $10\;\Omega$ resistors connected to the battery holder, 
which prevent from short circuit.
}
\label{Fig:RGB_schema} 
\end{figure}

\begin{figure}[h!]
\includegraphics[scale=0.6]{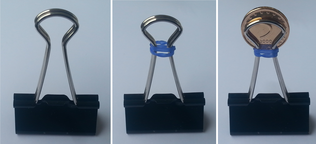}
\caption{Binder clip, ribbon band and the grasping of a coin. 
Use the coin grasped with the binder clip as an object, which you will illuminate and create colour shadows.
}
\label{Fig:RGB_holder} 
\end{figure}

\section{Theoretical part}

\begin{figure}[ht]
\includegraphics[scale=0.2]{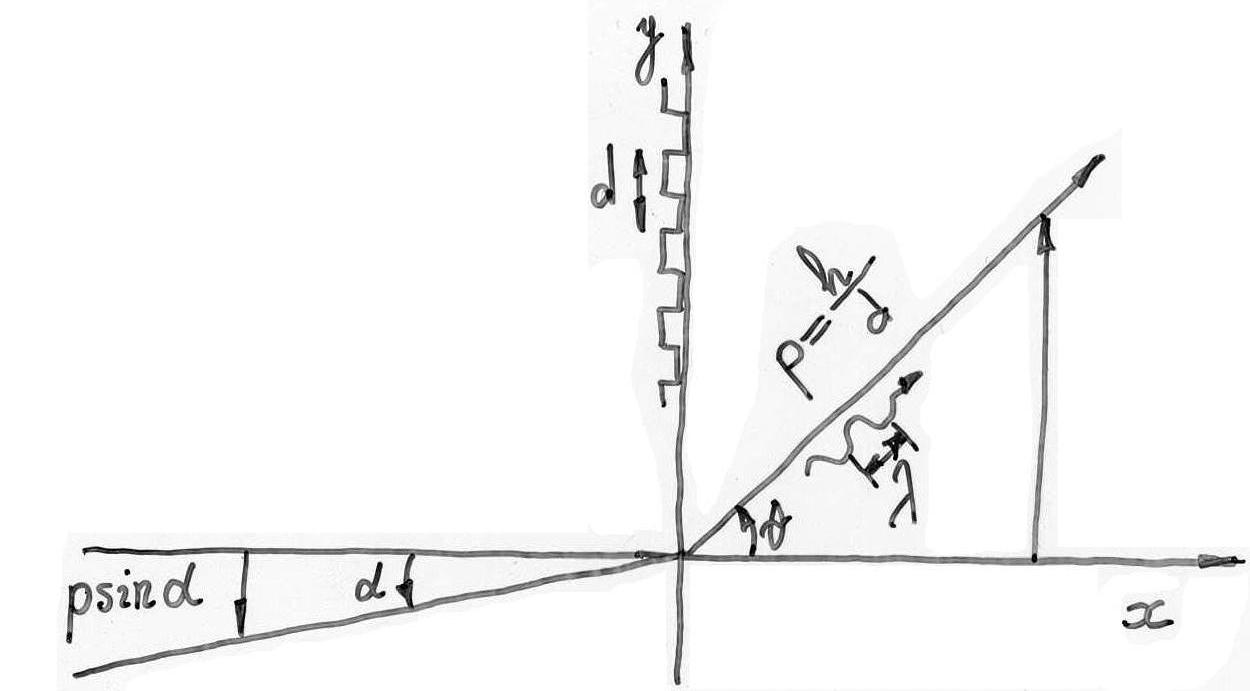}
\caption{Ridges of the diffraction grating is oriented in direction of $y$-axis.
Angles are measured starting from the $x$-axis, which is parallel of the normal of the diffraction grating.
The projection of the photon momentum after the scattering on the surface of the diffraction grating is $p\sin(\theta)$.
The projection of the photon momentum before the scattering is $p\sin(\alpha)$.
According the De Broglie formula $p=h/\lambda$.
}
\label{Fig:alpha}
\end{figure}

\item
\label{alpha}
\textit{What momentum could be hidden into a diffraction grating?}
How the formulae for the diffraction angle will changes if the light incident the diffraction grating on angle $\alpha$, 
measured from the surface normal of the diffraction grating, as it is shown on Figure~\ref{Fig:alpha}?

\item
\label{RefractionIndeks}
\textit{What is the photon momentum in water?}
How the diffraction angle will be changed if the diffraction grating is submerged in water?
The water has refractive index $n=1.33.$ 
Choose the correct answeres:\\
	A. The formulae $\lambda=d \sin(\theta)$ does not change.\\
	B. The diffraction angle increases.\\
	C. The diffraction angle decreases $n$ times.\\
	D. The sine of the angle decreases $n$ times. 

\item
\label{f<<R}
\textit{At what proportion of the focal length $f$ of the lens and the radius of the CD $R_\mathrm{CD}$ the diffraction experiment is possible?} 
Is it possible the focal length to be as long as an elbow length?

\item
\label{CombinationalFrequency}
\textit{How the frequency $\nu^\prime$ of the diffracted light will be changed if the diffraction grating is moving with speed $v$ in direction perpendicular of its ridges?}

\item
\label{Hamilton} 
\textit{Momentum of flat electromagnetic wave.}
How the photon momentum $p(E)$ depends on its energy?
To answer this question use the formulae for the speed of light $c=\lambda/T$, 
the energy $E=h/T$ and the  momentum $p=h/\lambda$ of the photon. 

\item
\label{Chain} 
\textit{Find the displacement of a vagon.}
It is very convenient to use the photon momentum when you calculate the diffraction angles.
That is why we give another task in which you have to use both the energy and momentum of the photon.
The mechanics in some sense is part of the optics.
Let's do the following thought experiment shown schematically on Figure~\ref{Fig:GedankenExperiment1Stein}. 

\begin{figure}[ht]
\includegraphics[scale=0.6]{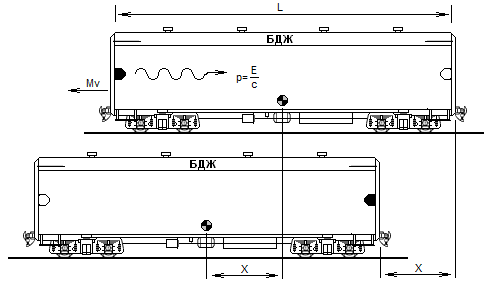}
\caption{\textit{Gedanken} experiment: Vagon with mass $M$ and light pulse with energy $E.$
The photon moves on right with momentum $p=E/c,$
and the vagon moves on left with exactly opposite momentum $MV.$
The mechanical problem is analogous to the analysis of the ``kick'' of a rifle or recoil of a gun. 
While the light pass the vagon with length $l$, the vagon displaces on the left on distance $X.$
We suppose that $X\ll l.$
The black spots on the end walls of the vagon shows where the energy is concentrated:
In the beginning the battery of the laser pointer, at the end in the heat spot where the light is absorbed.
}
\label{Fig:GedankenExperiment1Stein}
\end{figure}

Vagon with mas $M$ and length $l$ can move on rails without friction.
A light pulse from a laser pointer is flashed from left end of the vagon.
The energy of the light beam is $E$ and its momentum $p(E)$.
You have derived the formula in the previous task.
According to the momentum conservation law the vagon starts moving on the left with velocity $V=p/M.$
Let us assume that the light pulse is very short and passes the vagon for time $t=l/c.$ 
During this time the vagon moves on left on distance $X=Vt.$ 
At the end the light is absorbed in the right end of the vagon and it stops.
What happens with the light, when it is absorbed?
What is the displacement of the mass centrum of the vagon?

\item
\label{Muenhausen}
\textit{Paradox: The mass center immovable.}
Which of the following statements is correct? Where is the error?
In the previous task we get to the conclustion that the vagon is shifted on distance $X.$
This distance is small, but non zero.
However there is a exact theoreme in mechanics that for a still system without
an external forces the centrum of mass does not move .

Figuratively speaking: Baron Munchausen cannot lift himself pulling his hair.
The Baron cannot, but the vagon can -- where is the error?
As a hint let us analyze the simple example:
if there is a beech table on the left end of the vagon with mass $m$
then after the shift at the right end of the vagon at distance $l$ 
the vagon moves on left on small distance $X.$
For those quantities there is a relation $M X=m l$ 
and the mass center does not move.
Why the light can move the mass center of the vagon?

\item
\textit{How you can explain the colour of the different shadows, in the experiment described in~\ref{item:RGB_task}?}

\end{enumerate}


\newpage

\section{Original Bulgarian Text: Условие на задачата}

\section{Условие}
\begin{figure}[ht]
\includegraphics[scale=0.88]{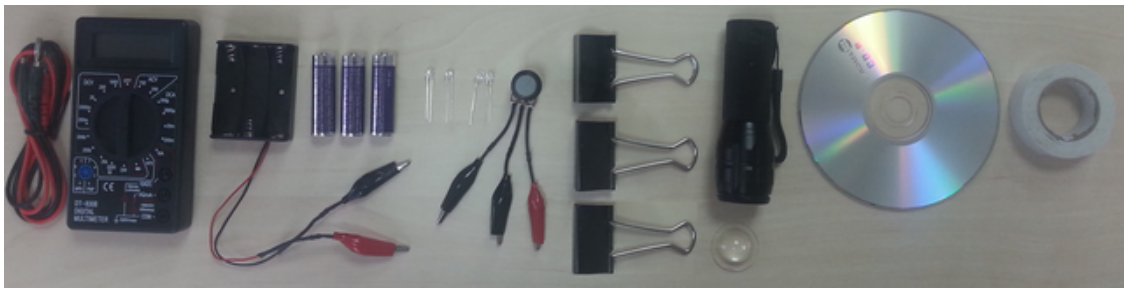}
\caption{Детайли от комплекта за измерване на константата на Планк $h$: 
Проводници със щекер и щипка,
мултиметър,
държач за батерии и 3 батерии от 1.5 V,
4 светодиода,
потенциометър с проводници и щипки,
3 скоби за хартия,
фенерче и лещата му,
компакт диск (CD)
и двойно-залепваща лента, 
фоторезистор, 
непрозрачна тръбичка,
пластелин,  
кламер. 
}
\label{Fig:BG_Set_all} 
\end{figure}

\begin{figure}[ht]
\includegraphics[scale=0.87]{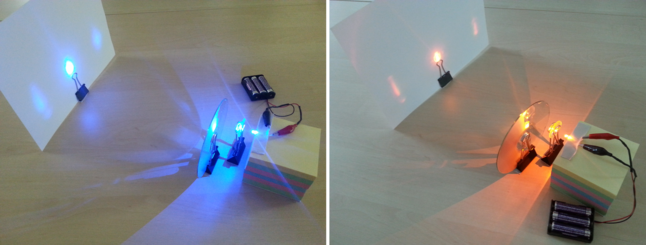}
\caption{Постановка за измерване на дължината на вълната на светлината на светодиод: 
Напрежението от батериите по жици се подава върху светодиода.
Светлината се фокусира от лещата, 
преминава през набраздената повърхност на прозрачния компактдиск
и върху екрана, като пламък от свещ, се вижда яркия образ на светодиода.
Освен централния максимум се виждат и два бледи дифракционни максимума.
Ъглите на дифракция са различни за разлините цветове, 
жълт и син на снимките. Това е метод за определяне на дължината на вълната.
}
\label{Fig:BG_experimental_setup_blue_amber} 
\end{figure}

\begin{figure}[ht]
\includegraphics[scale=0.15]{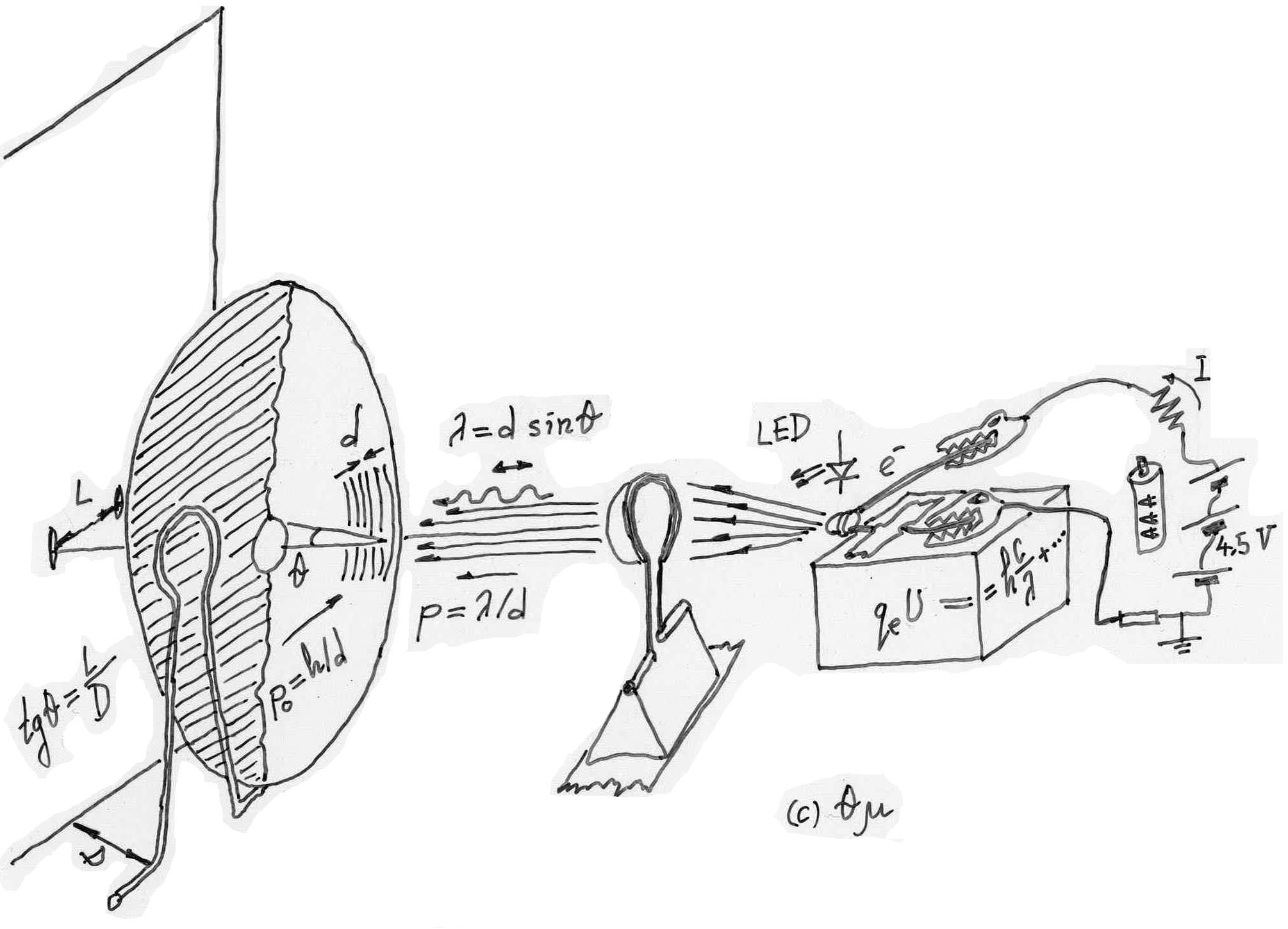}
\caption{Схема на постановките показани на Фиг.~\ref{Fig:BG_experimental_setup_blue_amber}.
Светодиода (LED) е свързан към батериите с ``крокодили''. 
Фотон от успоредния сноп от фокусираната от лещата светлина има импулс $p=h/\lambda$, който е перпендикулярен на диска.
При дифракция фотона получава импулс $p_0=h/d$ в направление препендикулярно на пътеките набраздени на компакдиска и се разсейва на ъгъл $\theta.$
$D$ e разстоянието между диска и екрана, 
а $L$ е разстоянието между централния максимум и първия дифракционен максимум; $\tg\theta=L/D.$ 
Скобите за хартия които държат лещата и диска са представени схематично.
}
\label{Fig:BG_setup_scheme} 
\end{figure}

\begin{figure}[h]
\includegraphics[scale=0.2]{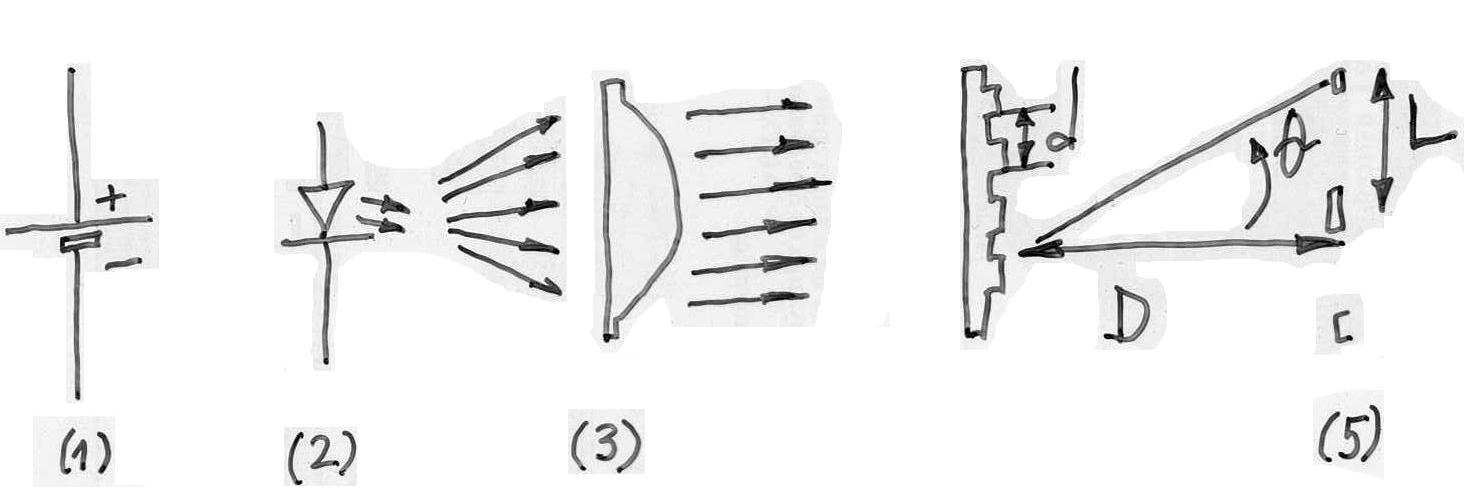}
\caption{Пътя на енергията: 1) химична в батериите 2) светлинна  3) фокусиране на светлината 4) дифракция 5) поглъщане и топлина.}
\label{Fig:BG_optical_scheme1} 
\end{figure}

\begin{figure}[ht]
\includegraphics[scale=0.92]{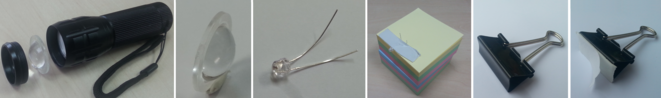}
\caption{Пръстен на винт, леща и фенерче.
Светодиод.
Лещата на фенерчето.
Компакт диск с частично отлепено фолио.
Светодиод залепен с лента върху ``постамент'' от хартиени листчета.}
\label{Fig:BG_set_diode_lens_disk} 
\end{figure}

Измерете константата на Планк $h$, 
като използвате експерименталния набор 
и експерименталната постановка
показани на 
Фиг.~\ref{Fig:BG_Set_all},  Фиг.~\ref{Fig:BG_experimental_setup_blue_amber},
Фиг.~\ref{Fig:BG_optical_scheme1},
Фиг.~\ref{Fig:BG_setup_scheme}  и
Фиг.~\ref{Fig:BG_set_diode_lens_disk}, 
а също така
известните стойности за скоростта на светлината $c=299792458$~m/s 
и заряда на електрона $q_e=1.602176565 \times$$10^{-19}$~C, 
както и разстоянието между пътеките на компакт диска $d=1.50\, \mu$m.

\section{Оценяване и организационни въпроси}

Експерименталните постановки ще останат за Вас и тяхното сглобяване те няма да се оценява.
Ще се оценяват всички предадени листове (чернови и белови) в които са дадени: таблици с експериментални резултати, графичното им представяне, правите минаващи най-близко до експерименталните точки, анализа на експерименталните резултати и отговорите на теоретичните въпроси свързани с разбирането на използваните формули за обработката на експерименталните данни. 
 
Класирането ще бъде по възрастови групи, а разпределение на подусловията по класове е: 
\begin{itemize}
\item 7--8 клас -- електрическа част: напрежения на изгасване, волтамперни характеристики на светодиодите, ом-амперни характеристики на осветяван фоторезистор 
\item 9 клас -- същите задачи като за 7--8 клас
\item 10 клас -- оптическа (измерване на дължината на вълната) и електрическа част (само напреженията на изгасване); 
\textbf{определяне на константата на Планк}
\item 11--12 клас -- всички задачи
\end{itemize}

Теоретичните задачи са свъзани с формулите използавни за обработката на експерименталните данни и имат за цел да проверят дали експериментатора разбира какво измерва. 
Те дават малко точки, 
но са са предимство при равни други условия при решени експериментални задачи. 
Теоретичните задачи ще са по-важни за класирането за най-добре решено домашно.

Съветваме учениците да решават задачите от съответната възрастова група.
Ако ученик от 7--9 клас реши всичките задачи за неговата възрастова група, може да опита да реши част от оптичните експериментални или теоретични задачи, което ще му донесе допълнителни точки.
Също така ученик от 10--12 клас може да реши задачи от електрическата част, които са за 7--9 клас, това ще му донесе допълнителни точки. 

Задачите за домашна работа може да започнете да решавате, ако сте напълно готови с всички останали задачи. 
Ще има допълнително класиране за задачите от домашно.
Ако не ви остане време, довършете задачите за домашно и ни пратете отговора днес на  емайл epo@bgphysics.eu до 24:00. 

Oбърнете се за съдействие към контролиращите учители ако имате въпроси. 
Имате за работа 4 часа и 30 мин, като през първите 2 часа не можете да напускате залата. 
При решението на задачата можете да ползвате калкулатор.  
Моля, предайте GSM-а си на контролращите учители в залата. 
Забравен у ученика GSM води до дисквалификация.

Като изключим мултиметрите,
експерименталните постановки са 
подарък от организаторите за ``кабинета по физика'' и участниците се окуражават да демонстрират решението на задачата пред съучениците си. Ако допълните постановката за иамерване на $h$ с вакуумен фотодиод заповядайте да покажете постановката си на конкурса за уреди на 6 юни 2015.

\begin{figure}[ht]
\includegraphics[scale=0.92]{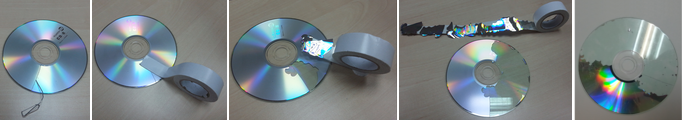}
\caption{``Технология'' за свалянето на фолиото от диска: 
1) Фолиото се издрасква (разрязва) радиално с кламера.
2) Върху разреза се залепва лента.
3) Лентата се издърпва и фолиото остава залепнало за нея.
4) Залепването и издърпването на лентата се повтаря докато половината от диска се обели.}
\label{Fig:BG_set_disk_foil_detachment} 
\end{figure}

\section{Фотоефект. Уводни думи}

При фотоефекта, фотон избива електрон от повърхността на метал 
и законът за запазване на енергията дава връзка между енергията на фотона 
$h\nu$, кинетичната енергия на електрона и отделителната работа параметризирана с напрежение. 
При светодиодите имаме обратен процес на вътрешния фотоефект: 
електрон и дупка с малки енергии рекомбинират и се излъчва фотон. 
Между праговото напрежение $U_c$, при което диода светва и енергията на фотона връзката
\begin{equation} 
 q_eU_c\approx h\nu+\mbox{const},
\label{BG_photoeffect}
\end{equation}
където $q_e$ е заряда на електрона, 
а като изключим зелените светодиоди константата 
слабо зависи от материала на светодиода.
Кинетичните енергии на рекомбиниращите електрони са пренебрежими,
т.е. много по малки от енергията на излъчения фотон.
Задачата, която ще разгледаме днес се основава на 
близките стойности на константите за различни материали,  
от които се правят светодиоди. 

\section{Електрична част}

\subsection{Измерване на праговите напрежения $U_\mathrm{c}$ на светване или угасвнае на светодиодите}
\begin{enumerate}
\item
\textit{Пречертайте схемата от Фиг.~\ref{Fig:BG_Potenciometer}а без амперметър (заменете го с проводник). Сглобете тази схема с предоставените Ви елементи.}(7--12 клас)

Поставете батериите в държача. 
И свържете диодите по схемата от Фиг.~\ref{Fig:BG_MaxLight}
Подавайте напрежението към светодиода само през края на проводниците (щипките), 
така че токът да преминава през изолираните със щлаух резистори!
Това ограничава тока през светодиода и той не може да Ви заслепи. 
Все пак не гледайте пряко светещия светодиод. 

При това подусловие можете пропуснете свързването на амперметъра, той се заменя с проводник.
Свържете сега някой от светодиодите през потенциометър,
както е показано на схемата от Фиг.~\ref{Fig:BG_Potenciometer} или по-точно от пречертаната от Вас схема без амперметър. 
Двата крайни извода на потенциометъра свържете със щипките на изводите от държача на батериите.
Можете да ползвате допълнителни проводници завършващи със щипки. 
Свържете волтметъра с жиците излизащи от светодиода. 
Въртете копчето на потенциометъра докато светодиода светне и изгасне. 
Ако светодиода е тъмен или интензивността на светлината не се променя при въртенето на потенциометъра може би сте свързали светодиода на обратно или не сте свързали правилно схемата. Проверете отново дали елемените са свързани според пречертаната от Вас схема. 
[5~т.]

\begin{figure}[ht]
\includegraphics[scale=0.65]{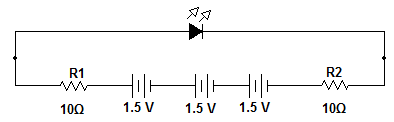}
\caption{
Захранване на светодиодите за експеримента по дифракция.
Тази схема може да се използва и за тест при каква полярност на светят светодиодите. 
Токът от 3-те последователно свързани батерии преминава през светодиода. 
Двата резистора $R_1$ и $R_2$ са скрити в черните изолиращи тръбички.
Тези резистори ограничават тока, за да може светодиода да не прегрее и да не ви заслепява.
Точките по вертикалните линии от схемата символизират свързване с щипки (``крокодилчета'').
}
\label{Fig:BG_MaxLight} 
\end{figure}

\begin{figure}[ht]
\includegraphics[scale=0.8]{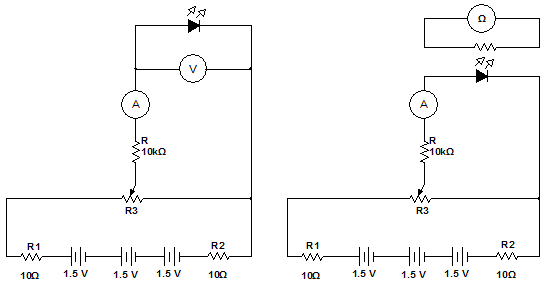}
\caption{
Потенциометрично захранване на светодиод за изследване на зависимостта $I(U).$
Напрежението от държача на батериите се подава на потенциометъра. 
Диода се свързва със средния електрод на потенциометъра и с някой от крайните изводи. 
Съпротивлението $R$ е запоено за средния извод на потенциометъра 
преди щипката (``крокодила'') и е скрито в черна изолираща тръбичка. 
Това голямо съпротивление дава възможност да се изследват тези част от ВАХ при която диода светва и изгасва. 
Амперметърът е свързан последователно на светодиода, а волтметърът успоредно.
Ако светодиода не светне в едно от крайните положения на потенциометъра,  сменете полярността на батериите.
Съпротивленията $R_1$ и $R_2$ са скрити в черни изолиращи тръбички преди крокодилите на изводите от държача на батерии.}
\label{Fig:BG_Potenciometer} 
\end{figure}

\item 
\label{Item:BG_Uc}
\textit{За 4-те светодиода измерете напрежението $U_c$, при което те изгасват или обратно започват да светят.} (7--12 клас)

При сглобената от предната точка схема проверете дали при въртене на копчето на потенциометъра светодиода светва и изгасва. 
Решението на задачата изисква да наблюдавате слаба светлина, поради което 
аудиториите са леко затъмнени. 
За всеки от светодиодите направете няколко измервания и усреднете резултата за всеки един от тях. 
Пресметнете и максималното отклонение от средната стойност,
това е оценка за точността на измерването. 
Резултатите представете таблично, като ги подредите по измереното напрежение, виж таблица~\ref{table:BG_U_c_table}
В таблицата дайте: цвят, средна стойност на напрежението, 
максимално отклонение от средната стойност, 
брой измервания за всеки светодиод. 
[10~т.]

\begin{table}[ht]
\caption{Експериментални резултати за праговите напрежения на светване или угасване на светодиодите} 
\begin{tabular}{| c| l | c |}
\tableline 
\No & Цвят & $U_c\mathrm{[V]}$  \\
\tableline \tableline
1& червен& \\
2& жълт   & \\
3& зелен  & \\
4& син     & \\
\tableline 
\end{tabular}
\label{table:BG_U_c_table}
\end{table}

\subsection{Измерване на праговите напрежения $\tilde{U}_\mathrm{c}$ чрез изследване на волтамперните характеристики на светодиода}
\item 
\label{BG_CurrentVoltageCharacteristics} \textit{За 4-те светодиода измерете тяхната волтамперна характеристика} (7--9 клас)

Включете мултиметърът, който носите със себе си, като амперметър и го свържете към схемата от Фиг.~\ref{Fig:BG_Potenciometer}а така, че да измерва тока през светодиодите. 

Изследвайте зависимостта между тока и напрежението $I(U)$ (ВолтАмперна Характеристика, ВАХ) за дадените ви светодиоди.
За всеки от светодиодите трябва да попълните отделна таблица с по 25 реда,
с номер на измерването, ток $I$ и напрежение $U$. 
Пригответе на отделен лист подобна на примерната таблица~\ref{Table:BG_I(U)}.

Първо, проверете дали в едно от крайните положения на потенциометъра диода светва; 
може да се наложи да размените изводите му.
При максимално ярък светодиод запишете напрежението и тока 
- това е първото измерване.
После въртете копчето на потенциометъра и намалявайте тока 
с около 20~$\mu$A; записвайте резултатите в таблицата, докато се получи ток по-малък от 10~$\mu$A.
След това наблюдавайте волтметъра, може да се наложи да смените обхвата.
Нашият съвет е да въртите копчето на потенциометъра 
и да записвате напрежението на светодиода през около 200~mV, 
докато достигнете до нулево напрежение.
Попълнете колоните от таблицата за различните светодиоди.
[20~т. за първия светодиод и[20~т. за трите останали светодиода]

\begin{table}[ht]
\caption{Волтамперни характеристики на светодиодите около напрежението на запалване.
Последователните колони са за червен, жълт, зелен и син светодиод. За да се опишат добре ВАХ таблицата трябва да бъде с поне 20 измервания за всеки светодиод.
}
\begin{tabular}{| r | c | c | r | c | c | r | c | c | r | c |c |}
\tableline
\multicolumn{3}{|c|}{червен}
& \multicolumn{3}{|c|}{жълт}
& \multicolumn{3}{|c|}{зелен}
& \multicolumn{3}{|c|}{син}\\
\tableline 
\No & $I\;[\mu\mathrm{A}$] & $U$ [V] & 
\No & $I\;[\mu\mathrm{A}$] & $U$ [V] & 
\No & $I\;[\mu\mathrm{A}$] & $U$ [V] & 
\No & $I\;[\mu\mathrm{A}$] & $U$ [V] \\
\tableline \tableline
 1& & & 1& & & 1& & & 1& & \\
 2& & & 2& & & 2& & & 2& & \\
 ...& & & ...& & & ...& & & ...& & \\

\tableline 
\end{tabular}
\label{Table:BG_I(U)}
\end{table}

\item 
\label{BG_CurrentVoltageCharacteristics} \textit{За 4-те светодиода представете графично тяхната волтамперна характеристика, използвайки резултатите от попълнената таблица~\ref{Table:BG_I(U)}.} (7--9 клас)

Представете графично резултатите от таблиците 
като използвате една и съща координатна система за различните светдиоди.
Преди да започнете, внимателно обмислете мащаба.  
Би било добре данните да запълват целия лист от хартията на квадратчета или милиметровата хартия. 
Добре е точките за различните светодиоди да могат да се различават. 
За тази цел използвайте моливи или химикалки с различни цветове. 
Използвайте черна химикалка за жълтия светодиод.
Върху абсцисата нанесете напреженията $U[\mathrm{V}]$,
а по ординатата - токовете $I[\mu\mathrm{A}]$.
[20~т.]

\item 
\label{BG_CurrentVoltageCharacteristics_U_c_tilda} \textit{За 4-те светодиода от графиките на ВАХ определете праговото напрежение $\tilde{U}_\mathrm{c}$.} (7--9 клас)

На ВАХ, при силни токове има почти праволинеен участък.
Начертайте права линия, която да се допира до стръмния склон на ВАХ при големи токове.
Къде тази права линия пресича абсцисата $I=0$ за различните светодиоди?
Намерете пресечните точки на правите за различните светодиоди и запишете 
съответните напрежения $\tilde{U}_\mathrm{c}$ в таблица както е показано в Таблица~\ref{table:BG_U_c__tilde_table}. 
Сравнете тези прагови напрежения с напреженията на светване и изгасване $U_c$, които измерихте в подусловие~\ref{Item:BG_Uc}, като укажете цвета на светодиода. 
Каква е максималната разлика от $\tilde{U}_\mathrm{c}-U_\mathrm{c}?$
[10~т.]

\begin{table}[ht]
\caption{Експериментални резултати за критичните напрежения получени чрез изследване на ВАХ на светодиодите.} 
\begin{tabular}{| c| l | c |}
\tableline 
\No & Цвят & $\tilde{U}_c\mathrm{[V]}$  \\
\tableline \tableline
1& червен& \\
2& жълт   & \\
3& зелен  & \\
4& син     & \\
\tableline 
\end{tabular}
\label{table:BG_U_c__tilde_table}
\end{table}

\item 
\label{BG_CurrentVoltageCharacteristics_small_current} \textit{Обяснете качествено наклона на графиките на ВАХ при малки токове.} (7--9 клас)

Сега насочете вниманието си върху полегатите склонове на ВАХ при малки токове.
Обърнете внимание, че наклона за различните светодиоди е почти еднакъв.
Как може да се интерпретира този факт?
[10~т.]

\subsection{Изследване на интензивността на светлината излъчена от светодиодите с помощта на фоторезистор. Алтернативен метод за измерване на праговите напрежения $\tilde{U}_\mathrm{c}$ на светодиода}

\item
\label{BG_Siemens-Ampere}
\textit{Измерете зависимостта на съпротивлението на фоторезистора от тока през светодиода.} (7--9 клас)

За тази цел използвайте схемата показана на Фиг.~\ref{Fig:BG_Potenciometer}б.
Амперметърът измерва тока през светодиода, но волтметъра не се поставя.
Сега включвате този мултиметър като омметър, който да измерва 
съпротивлението на фоторезистора $R_\varphi.$
Доближете фоторезистора до диод светещ с максимална интензивност, 
за да видите как се изменя съпротивлението му.
Сега трябва да премахнем влиянието на външната светлина.
Поставете светодиода и фоторезистора в тръбичката и я запушете с пластелин, 
за да не прониква светлина.
Както при измерване на ВАХ от предишната точка,
записвате тока протичащ през светодиода и съпротивлението на фоторезистора $R_\varphi.$
В таблицата~\ref{Table:BG_R_varphi(I)} оставете още една колона за реципрочната стойност на съпротивлението $1/R_\varphi,$
която също така се нарича и проводимост и се измерва в единици $\Omega^{-1}.$
Въртете копчето на потенциометъра, за да намалявате тока с около 20$\;\mu$A и пак запишете данните в таблицата.
Когато достигнете до нулев ток опитайте да се върнете обратно и да изследвате малки токове до 15$\;\mu$A.
Повторете тези измервания за всички светодиоди и запишете резултатите в различни таблици.
След приключване на измерванията пресметнете реципрочните съпротивления $1/R_\varphi$ в 3-тата колона.
 [20~т. за първия светодиод и 20~т. за трите останали светодиода]

\begin{table}[ht]
\caption{Зависимост на съпротивлението през фотодиода от тока през светодиода. Таблицата трябва да бъде с поне 25 измервания за всеки светодиод
}
\begin{tabular}{| r | c | c  |c | r | c | c | c | r | c | c | c | r | c | c | c |}
\tableline 
\multicolumn{4}{|c|}{червен}
& \multicolumn{4}{|c|}{жълт}
& \multicolumn{4}{|c|}{зелен}
& \multicolumn{4}{|c|}{син}\\
\tableline 
\No & $I\;[\mu\mathrm{A}$] & $R_\varphi\, [\Omega]$ &  $1/R_\varphi\, [\Omega^{-1}]$ &
\No & $I\;[\mu\mathrm{A}$] & $R_\varphi\, [\Omega]$ &  $1/R_\varphi\, [\Omega^{-1}]$ &
\No & $I\;[\mu\mathrm{A}$] & $R_\varphi\, [\Omega]$ &  $1/R_\varphi\, [\Omega^{-1}]$ &
\No & $I\;[\mu\mathrm{A}$] & $R_\varphi\, [\Omega]$ &  $/1R_\varphi\, [\Omega^{-1}]$  \\
\tableline \tableline
 1& & & &1& & & & 1& & & & 1& & & \\
 2& & & &2& & & & 2& & & & 2& & & \\
 ...& & & &...& & & & ...& & & & ...& & &\\

\tableline 
\end{tabular}
\label{Table:BG_R_varphi(I)}
\end{table}

\item 
\label{} \textit{За 4-те светодиода представете графично зависимостта на съпротивлението на фотодиода от тока през светодиода, използвайки резултатите от попълнената таблица~\ref{Table:BG_R_varphi(I)}.} (7--9 клас) 

Резултатите представете графично: по абсцисата - тока през светодиода $I$, 
а по ординатата - реципрочното съпротивление $1/R_\varphi$ в единици $\Omega^{-1}.$
Използвайте една координатна система за всички отделни светодиоди.
Добре е да нанасяте точките с различни цветове.
Покрай точките прекарайте водещи погледа гладки криви.
Анализирайте тези криви и се опитайте да обясните резултатите от този експеримент. 
[20~т. за първия светодиод и 20~т. за трите останали светодиода]

\item 
\label{} \textit{Анализирайте как фоторезистора може да се използва за определяне праговото напрежение $\tilde{U}_\mathrm{c}$.} (7--9 клас) [10~т.]

\section{Оптична част}

\subsection{Измерване дължината на вълната на светлината излъчвана от светодиодите. (10--12 клас)}

\item 
\textit{Обелване на половината диск.} 

Като използвате кламера разрежете фолиото от компактдиска по радиуса. 
Върху разреза залепете двойнозалепваща лента 
и като я дръпнете ще свалите фолиото под нея. 
Фиг. ~\ref{Fig:BG_set_disk_foil_detachment}.
По този начин махнете фолиото от половината диск. 
Хванете диска със скобата, както е показано на 
Фиг. ~\ref{Fig:BG_experimental_setup_blue_amber}. [0~т.]

\item 
\textit{Сглобяване на оптичната постановка.}

Сега предстои най-сложната част от задачата - запасете се с търпение и хладнокръвие:
Върху една от скобите за хартия залепете двойнозалепваща лента - тя се къса лесно с ръка. 
Поставете екрана от бял картон в скобата. 
Махнете пластмасовото фолио от лентата и фиксирайте скобата с екрана далеч пред вас. 
Екранът трябва да бъде изправен вертикално.
По същият начин фиксирайте прозрачния диск успоредно на екрана, 
примерно на педя разстояние пред него.
Постарайте се да бъдете прецизни. 
Ако не сте доволни от точността на изпълнението, 
махнете лентите и монтирайте всичко отново.
Измерете разстоянието $D$ между екрана и диска с точност до милиметър.
Пред диска поставете скобата с лещата, 
а пред лещата, примерно на два пръста разстояние, 
закрепете със залепваща лента един от светодиодите върху ``пиедестал'' от листчета. 
Светодиодът трябва да бъде на еднаква височина със центъра на лещата.
Захранването на светодиодите в експеримента по дифракция е показано на Фиг.~\ref{Fig:BG_MaxLight}.

Включете светодиода да свети. 
Движете поставката с листчета докато върху върху екрана се фокусира ярко петно. 
Това е реалният образ на светодиода през лещата. 
Добавяте и махате листчета докато образът се прожектира, 
с точност до 1-2 сантиметра, 
на същата височина, както светодиода.

Нека проследим пътя на светлината:
тя е излъчена от светодиода, 
който е почти във фокуса на лещата, 
както е показано на Фиг. ~\ref{Fig:BG_experimental_setup_blue_amber}.
Успоредният сноп преминава през прозрачния диск 
и на екрана освен централния максимум се виждат и два дифракционни максимума.

Придвижете малко диска в щипката така, 
че тесния светлинен сноп да преминава през хоризонтален радиус на диска. 
Тогава дифракционните максимуми и централният максимум се разполагат на една хоризонтала върху екрана. 
Това е важно за точното измерване на дължината на вълната, 
което е един от критериите за класиране в олимпиадата.

Използавайте цветната химикалка за да отбележите централния максимум 
и двата дифракционни максимума. 
За жълтия светодиод използвайте черния химикал.
Разстоянието между двата дифракционни максимума е $2L,$ 
а разстоянието между централия максимум и всеки от дифракционните максимуми е приблизително $L.$
[0~т.]

\item
\textit{Колко е фокусното разстояние на лещата $f$ в метри и каква е неговата реципрочна стойност $1/f?$} [15~т.]

\item 
\textit{Измерване на ъглите на дифракция.}

Запишете разстоянието $D$ между компакт-диска и екрана. Работете с точност от милиметър. 
Последователно измерете за всеки от светодиодите разстоянието $L$ от центъра на централния максимум до центъра на петното на първия дифракционен максимум. 
По-добре е да очертаете на хартиения екран светлите петна и после да определите къде според вас е центъра им. Представете редултатите таблично: цвят, разстояние $L,$ 
и като използавте праговите напрежения $U_\mathrm{c}$
попълнете таблица~\ref{table:BG_Sample}.
[20~т.]

\item 
\textit{Пресмятане на синуса от ъгъла на дифракция.}

За всеки от цветовете пресметнете синуса от ъгъла на дифракция 
$\sin(\theta)=L/(L^2+D^2)^{1/2}=\sin(\arctg(L/D)).$ 
Можете да намерите и графично решение. 
Построявате правоъгълен триъгълник с катети $L$ и $D$ и за хипотенуза 100~mm измервате срещулежащия на ъгъла $\theta$ катет успореден на $L.$ 
Дължината на този катет в дециметри дава $\sin(\theta).$ 
В таблица~\ref{table:BG_Sample} допълнете колоните с ъгловите променливи и пресметнете съответните дължини на вълните.
Колоните съдържащи критично напрежение $\tilde{U}_c$ получено чрез изследване на ВАХ остават непопълнени.
Оставете тази задача за домашно.
[15~т.]

\item
\textit{Определяне на константата на Планк по експерименталните данни.}
\label{BG_PlanckError} 

Представете графично резултатите от последната 
таблица~\ref{table:BG_Sample}.
По ординатата представете $q_eU_c$, а по абсцисата $\nu$. 
Начертайте права, която максимално добре, по ваше мнение, описва приближеното уравнение 
\begin{equation} 
 q_eU_c\approx h\nu+\mbox{const.}
\label{BG_photoeffect}
\end{equation}
От тангенса на ъгъла на наклона
$h=\Delta(q_eU_c)/\Delta(\nu)$ определете константата на Планк. 
Тук $\Delta$ означава разлика за отсечка от правата; 
разлика в абсцисата и разлика в ординатата. 
Как измерената от Вас стойност $h=\dots\,\mathrm{J\,s}$ 
се съгласува с известната стойност 
за тази фундаментална константа? Попълнете многоточието!
Колко процента е грешката $100\,(h_\mathrm{exp}-h)/h=\dots\%$?
Подчертайте резултата и го заградете в рамка!
[137~т.]

\item
\textit{Определяне на $h$ чрез ВАХ.} (Само ако Ви остане време.)

Сега изпълнете точка~\ref{BG_CurrentVoltageCharacteristics_U_c_tilda}
и извършете обработката на експерименталните данни от ВАХ.
Повторете същата процедура за определене на констатата на 
Планк като използвате не субективно определените напрежения на светване и изгасване $U_c,$ а критичните напрежения получени от анализа на ВАХ $\tilde{U}_c$.
Кой от методите е по-добър?
Посочете какви систематични грешки влияят на резултата 
и как бихте подобрили експерименталната постановка,  
за да се повиши точността на измерването.

Константата, която участва в уравнението~(\ref{BG_photoeffect}) 
за фотоефекта е значително по-различна за зеления светодиод.
Пропуснете тази точка и с помощта на останалите 3 светодиода определете $h.$
[30~т.]

\begin{table}[ht]
\caption{Експериментални данни за дифракция и прагово напрежение за фотодиоди с различни цветове; 
$D=\dots\,\mathrm{cm}.$} 

\begin{tabular}{| c| l | c | c | c | c |c | c | c | c | c | c |}
\tableline 
\No & Цвят &  $L$ [cm] & $\tg(\theta)$ & $\theta$ [rad]& $\sin(\theta)$ & $\lambda\,[\mathrm{nm}]$
& $\nu\mathrm{[Hz]}$ & $U_c\mathrm{[V]}$ 
& $\tilde{U}_c\mathrm{[V]}$ 
& $q_eU_c\mathrm{[J]}$ & $q_e\tilde{U}_c\mathrm{[J]}$ \\
\tableline \tableline
1& червен& & & & & & $\qquad\times 10^{14}$ & & & $\qquad\times 10^{-19}$ & $\qquad\times 10^{-19}$ \\
2& жълт   & & & & & & $\qquad\times 10^{14}$ & & & $\qquad\times 10^{-19}$ & $\qquad\times 10^{-19}$ \\
3& зелен  & & & & & & $\qquad\times 10^{14}$ & & & $\qquad\times 10^{-19}$ & $\qquad\times 10^{-19}$ \\
4& син     & & & & & & $\qquad\times 10^{14}$ & & & $\qquad\times 10^{-19}$ & $\qquad\times 10^{-19}$ \\
\tableline 
\end{tabular}
\label{table:BG_Sample}
\end{table}

\subsection{Оптичен експеримент със смесване на цветове.  
(7--12 клас)}
\item 
\label{item:BG_RGB_task}
\textit{Намерете цветовете на сенките.}

На Фиг.~\ref{Fig:BG_RGB_shadow} са показани схематично 3 източника на светлина 
и различни сенки~\cite{Bib:RGB} върху екрана от един осветен обект. 
Какви са цветовете в различните области номерирани от 1 до 6.
Можете да намерите цветовете с разсъждения или да използвате дадените ви светодиоди. 
За този експеримент 3-те светодиода трябва да светят едновременно,
както е показано на Фиг.~\ref{Fig:BG_RGB_diodes}.
Червеният и жълтият диод дават по-силна светлина
и за да може смесената светлина да напомня бяла те трябва да бъдат 
захранвани през компенсиращи съпротивления, виж Фиг.~\ref{Fig:BG_RGB_schema}.
Червения и синия диод трябва да са на разстояние 1.5~см от синия.
Ползвайте сянката на 1 стотинка 
закрепена както е показано на Фиг.~\ref{Fig:BG_RGB_holder} 
и ни пратете снимката да 24:00.
[20~т.]

\begin{figure}[ht]
\includegraphics[scale=0.6]{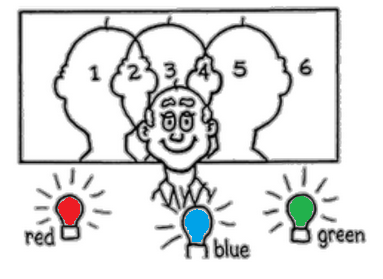}
\caption{
Сенки от 3 източника на светлина: червен (red), 
син (blue) и зелен (green) \cite{Bib:RGB}.
Посочете цвета на сенките в различните области.
}
\label{Fig:BG_RGB_shadow}
\end{figure}

\begin{figure}[ht]
\includegraphics[scale=0.7]{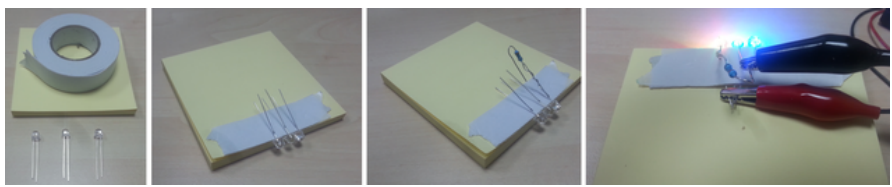}
\caption{Едновременно светене на червен, зелен и син светодиоди:
1) Диоди, листчета и двойнозалепваща лента.
2) Диодите са прикрепени с лентата върху листчетата. 
Единият електрод е под лентата, а другият е отгоре.
3) Червеният електрод и синият диод са свързани през съпротивления 
за да може светлината от 3-те диода да бъде почти бяла.
4) Крокодилите са захапали успоредно свързаните диоди.
По дългите изводи на светодиодите се свързват с отрицателните полюси на батериите.
Ако някой от диодите не свети разменете електродите му. Разположете диодите на разстояние диаметъра на монетата, която ще използвате за създаване на сянка.
}
\label{Fig:BG_RGB_diodes} 
\end{figure}

\begin{figure}[ht]
\includegraphics[scale=0.55]{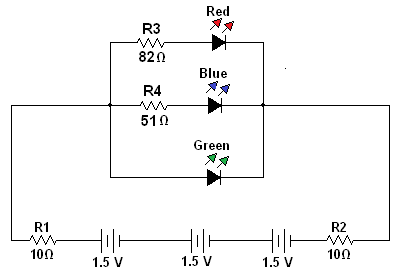}
\caption{Схема на успоредно свързване на 3 светодиода. 
Токът през червения диод се подава през 82~$\Omega$ резистор,
а синия се  свързва през съпротивление от 51~$\Omega.$
Резисторите от $10\;\Omega$, свъзани с държача на батериите,  
са скрити в шлаух. 
}
\label{Fig:BG_RGB_schema} 
\end{figure}

\begin{figure}[h!]
\includegraphics[scale=0.6]{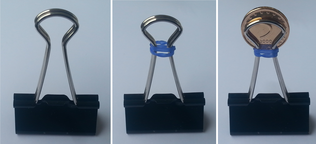}
\caption{Щипка, ластик и прикрепване на монетата. Закрепената на щипката монета използвайте за създаване на сянка при осветяване с цветните светодиоди.}
\label{Fig:BG_RGB_holder} 
\end{figure}

\section{Теоретична част. (Главно за домашно за 11 и 12 клас)}

\begin{figure}[ht]
\includegraphics[scale=0.2]{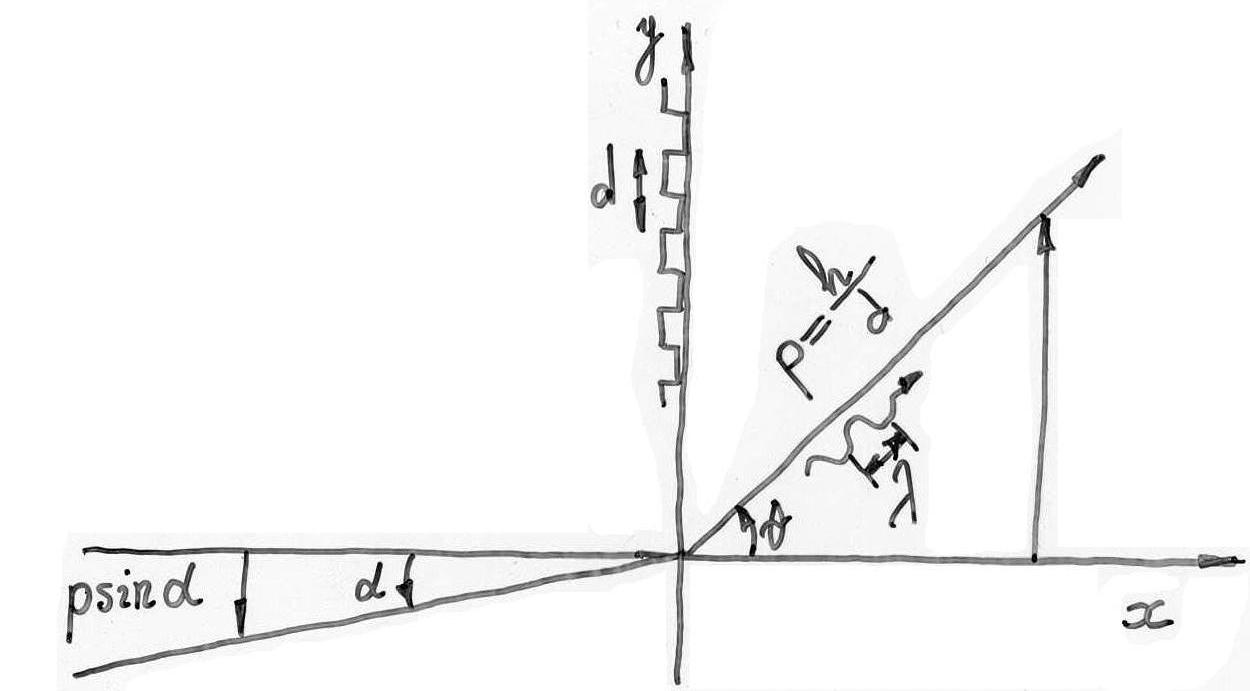}
\caption{Набраздяването на дифракционната решетка е по оста $y$.
Ъглите отчитаме от оста $x$, която е избрана по нормалата на дифракционната решетка.
Успоредната на дифракционната решетка проекция на импулса на фотона след разсейването е $p\sin(\theta)$, 
а преди разсейването $y$-компонентата на импулса е била $p\sin(\alpha)$, 
където по формулата на де Бройл $p=h/\lambda$.
}
\label{Fig:BG_alpha}
\end{figure}

\item
\label{BG_alpha}
\textit{Какъв импулс може да има в една дифракционната решетка?}
Как се променя формулата за ъгъла на дифракция, 
ако светлината пада под ъгъл $\alpha$ отчитан от нормалата на дифракционната решетка, както е показано на Фиг.~\ref{Fig:BG_alpha}? [5~т.]

\item
\label{BG_RefractionIndeks}
\textit{Какъв е импулса на фотона във водата?}
Как се изменя ъгъла на дифракция, ако работим с отразена вълна от потопен във вода диск? 
За показателя на пречупване на водата вземете приближено $n=1.33.$ 
Оградете с кръгче правилните отговори: [5~т.]\\
	А. Формулата $\lambda=d \sin(\theta)$ не се променя.\\
	Б. Ъгълът на дифракция се увеличава.\\
	В. Ъгълът на дифракция намалява $n$ пъти.\\
	Г. Синусът от ъгъла намалява $n$ пъти. 
Експериментално можете да решите тази задача за домашно,
трябва ви съд с вода и ни пратете решението до 24:00 на адрес 
epo@bgphysics.eu

\item
\label{BG_f<<R}
\textit{При какво съотношение между фокусното разстояние на лещата $f$ 
и радиуса на компактдиска $R_\mathrm{CD}$ поставения експеримент е възможен?} 
Може ли фокусното разстояние да бъде един лакът? [5~т.]

\item
\label{BG_CombinationalFrequency}
\textit{Промяна на честотата от движеща се решетка.}
Как се променя честотата на дифрактиралата светлина $\nu^\prime$, 
ако решетката се движи напречно на браздите със скорост $v$?  [5~т.]

\item
\label{BG_Hamilton} 
\textit{Импулс на плоска електромагнитна вълна.}
Използвайте формулите за скоростта на светлинната вълна $c=\lambda/T$, 
енергията $E=h/T$ и импулса на фотона $p=h/\lambda$ 
и изведете как импулса на светлинната вълна $p(E)$ зависи от енергията ѝ. [5~т.]

\item
\label{BG_Chain} 
\textit{Да се намери отместването на вагона.}
За пресмятане на ъгъла на дифракция е най-удобно да се използва импулса на фотона.
Затова добавяме още една задача, в която се разглеждат едновременно енергията и импулса на фотона.
Механиката в известен смисъл е част от оптиката.
Нека проведем мислен експеримент, който е представен схематично 
на Фиг.~\ref{Fig:BG_GedankenExperiment1Stein}. 

\begin{figure}[ht]
\includegraphics[scale=0.6]{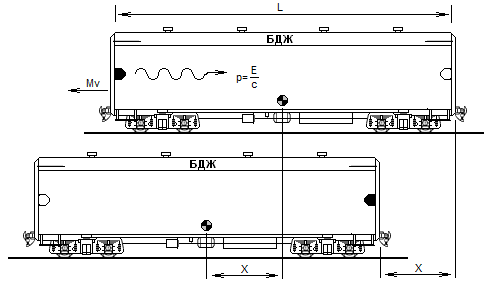}
\caption{Мислен експеримент: Вагон с маса $M$ и светлинен импулс с енергия $E.$ 
Фотонът се движи надясно с импулс $p=E/c,$
а вагона се движи наляво с точно противоположния импулс $MV.$
Механичната задача е аналогична на анализа на ``ритането'' на пушка или откат на оръдие.
Докато светлината прелита вагона с дължина $l,$ той се отмества наляво на разстояние $X.$
Предагаме, че $X\ll l.$
Точките символизират пътя на енергията: първоначално в батерията на лазерната показалка, а после в топлинното петно където светлината се е погълнала.
}
\label{Fig:BG_GedankenExperiment1Stein}
\end{figure}

Вагон с маса $M$ и дължина $l$ може да се движи без триене по релси. 
В левия край на вагона е светнато с лазерна показалка и енергията на светлинния лъч е $E.$ 
Импулса на светлината е $p(E)$. 
Формулата изведохте в предишното подусловие. 
Съгласно закона за запазване на импулса вагонът потегля наляво със скорост $V=p/M.$ 
Светлинният импулс, нека го приемем за съвсем кратък, 
прелита вагона за време $t=l/c.$ 
За това време вагонът се премества наляво на разстояние $X=Vt.$ 
Накрая, светлината се поглъща в десния край на вагона и той спира. 
Какво става със светлината, когато се погълне? 
С колко се е преместил центъра на тежестта на вагона. [5~т.]

\item
\label{BG_Muenhausen}
\textit{Парадокс: Центърът на тежестта е неподвижен. 
Кое от твърдениятя е вярно? Къде е грешката?}
В предишното подусловие стигнахме до извода, 
че вагонът се е отместил на разстояние $X.$
Това разстояние може да е малко, но не е нула. 
Съществува строга теорема от механиката, 
че едно неподвижно тяло, на което не действат външни сили 
не може да промени центъра на тежестта си. 
Образно казано: барон Мюнхаузен не може да се вдигне за косата. 
Баронът не може, но вагонът може - къде е грешката?
За подсещане, нека разгледаме прост пример: 
ако в левия край на вагона има букова маса с маса $m$, 
то при преместването ѝ в десния край на вагона на разстояние $l$, 
вагонът се премества наляво на малко разстояние $X.$ 
Тези величини са свързани със съотношението $M X=m l$ 
и центъра на тежестта не се променя. 
А защо тогава светлината може да променя центъра на тежестта на вагона? 
[5~т.]

\item
\textit{Как си обяснявате цветв на различните сенки, които са описани в точка~\ref{item:BG_RGB_task}?}
[5~т.]

\item Задача за домашно. 
Ако намерите решението на последното подусловие пишете ни днес до 24:00 на epo@bgphysics.eu,
като непременно ни напишете името на консултанта или източника на информация,
например Интернет страница или книга, който сте ползвали. 
Изобщо пратете ни домашното си със всички подусловия които решихте през следобеда.
[премия 1-камък]

\section{Задача за конкурса ``Уреди за кабинета по физика, 6 юни 2015''. (10, 11 и 12 клас)}

\item
\label{BG_ClassicalPhotoeffect}
\textit{Класически опит за фотоефект.}
Светодиодите на които днес измерихте честотата и дължината на вълната 
са добри източници на монохроматична светлина.
Изследвайте ВАХ на вакуумен светодиод (вакуумент фотоелемент или вакуумна фотоклетка),
например СЦВ-4 или RCA-930\cite{VaccumPhotoCell},  
когато катода му е осветен със светлината на различни светодиоди.
Ако не намерите подходящ елемент може да ползвате и фотоумножители като например ФЭУ-35\cite{PhotoMultiplier}, 
но тогава трябва да използвате само антимон-цезиеевия електронен емитер електрод (1) и първия ускоряващ електрод (2) без да подавате захранване.
Намерете спиращото напрежение $U_c$ за различните цветове;
този опит е описан в учебниците по физика.
Нанесете експерименталните данни на равнината енергия-честота ($E$-$\nu$).
Прекарайте права линия която максимално добре да описва експерименталните точки.
От уравнението $q_eU_c=h\nu+q_e A$,
намерете константата на Планк и отделителната работа на метала с който е покрит катода.
Оценете точността на измерването.
Този метод трябва да дава значително по-висока точност
защото за разлика от различните светодиоди 
които имаха различни ширини на забранените зони,
сега метала е един и същ с постоянна отделителна работа.
Има много съвременни ръководства, но все пак започнете търсенето в Интернет с класическата работа на Миликен\cite{Millikan:1916}.
Опитайте да изследвате и ВАХ на полупроводников фотодиод осветяван 
от различните светодиоди.
И заповядайте на конкурса \url{http://bgphysics.eu/} да демонстрирате работата на тази класическа постановка. 
Най-голям успех за олимпиадата би било реанимацията 
на този експеримент в училищата; опитайте да го направите днес за домашно. 
[137~т.]

\end{enumerate}

\newpage
\section{Решение}

Познаването на фундаменталните константи на вселената е важна част от развитието на цивилизацията.
Фундаменталните константи се измерват със скъпоструваща апаратура 
и за тяхното непрекъснато доуточняване работят първокласни експериментатори. 
За константата на Планк $6.62606957\times10^{-34}\;\mathrm{Js}$ точността се измерва в милиардни части.
От друга страна, този фундаментален квант на действие е част от света който ни заобикаля. 
Светодиодите и компактдисковете са в нашето ежедневие. 
Всеки ученик ги е виждал, учил е нещо за тях и е голям успех, 
ако ученика успее да определи тази фундаментална константа използвайки само знания от учебния материал.
Ако запомни, че в Деня на фотона е измерил константата на Планк с точност 20\% 
това може да се смята за голям успех на нашата образователна система. 
Наученото в училище успешно се е трансформирало в умение
и този млад човек ще бъде полезен за всяка квалифицирана професия.

По-долу са описани решенията за всяка от точките на условието.
В коментара често има и елементи на популярна статия, 
но те са предназначени главно за учителите.

\ref{BG_CurrentVoltageCharacteristics} и
\ref{BG_Siemens-Ampere}
В таблица~\ref{Table:BG_I(U)R_ph} е дадено едно пълно изследване на проблема, 
което малко надхвърля поставената на учениците задача,
но дава представа за цялостната физика на проблема.
За светодиодите с различни цветове са дадени данни за напрежението $U,$ тока $I$ 
и проводимостта на фоторезистора осветяван от светодиода $R^{-1}$

\begin{table}
\caption{Експериментални данни за напрежението $U,$ тока $I$ и проводимостта на фоторезистора $R^{-1}$ за червен, жълт, зелен и син светодиоди. За някои напрежения токовете са неизмеримо малки или съпротивлението на фоторезистора много голямо.}
\begin{tabular}{| r | c | c | c | c || c | c | c | c || c | c | c | c || c |c | c | c |}
\tableline  \tableline
\No & 
 $U$ [V] & $I[\mu\mathrm{A}]$ & $R_\varphi[\Omega]$ & $R^{-1}_\varphi[\Omega^{-1}]$ &
 $U$ [V] & $I[\mu\mathrm{A}]$ & $R_\varphi[\Omega]$ & $R^{-1}_\varphi[\Omega^{-1}]$ &
 $U$ [V] & $I[\mu\mathrm{A}]$ & $R_\varphi[\Omega]$ & $R^{-1}_\varphi[\Omega^{-1}]$ &
 $U$ [V] & $I[\mu\mathrm{A}]$ & $R_\varphi[\Omega]$ & $R^{-1}_\varphi[\Omega^{-1}]$ \\
\tableline
 1& 0.042 &   -   & 0    &   0    &  0.042 &   -   & 0.1 &   0    &    0.042 &   -   & 0    &   0    &      0.042 &   -   & 0 &   0     \\
 2& 0.207 &   -   & 0.1 &   0    &  0.215 &   -   & 0.1 &   0    &    0.2     &   -   & 0.1 &   0    &      0.204 &   -   & 0 &   0     \\
 3& 0.419 &   -   & 0.1 &   0    &  0.405 &   -   & 0.1 &   0    &    0.404 &   -   & 0.1 &   0    &    0.412 &   -   & 0.1 &   0     \\
 4& 0.623 &   -   & 0.1 &   0    &  0.605 &   -   & 0.1 &   0    &    0.601 &   -   & 0.1 &   0    &    0.602 &   -   & 0.1 &   0     \\
 5& 0.813 &   -   & 0.1 &   0    &  0.807 &   -   & 0.1 &   0    &    0.808 &   -   & 0.1 &   0    &    0.811 &   -   & 0.1 &   0     \\
 6& 1.006 &   -   & 0.1 &   0    &  1.028 &   -   & 0.1 &   0    &    1.011 &   -   & 0.1 &   0    &    1.022 &   -   & 0.1 &   0     \\
 7& 1.217 &   -   & 0.2 &   0    &  1.203 &   -   & 0.2 &   0    &    1.208 &   -   & 0.2 &   0    &    1.207 &   -   & 0.2 &   0     \\
 8& 1.407 &   -   & 0.3 &   0    &  1.421 &   -   & 0.2 &   0    &    1.404 &   -   & 0.2 &   0    &    1.404 &   -   & 0.2 &   0     \\
 9& 1.606 & 136.5 & 13.9 & 0.00733& 1.607 &   -   & 2.1 &   0    &    1.606 &   -   & 0.2 &   0    &    1.605 &   -   & 0.2 &   0     \\
10& 1.623 & 79.8 & 20 & 0.0125 &   1.668 & 617 & 10.5 & 0.00162 &    1.803 &   -   & 0.2 &   0    &    1.807 &   -   & 0.2 &   0     \\
11& 1.652 & 30.66 & 42 & 0.0326 &  1.696 & 202.5 & 21.5 & 0.00494 &  2.008 &   -   & 0.3 &   0    &    2.009 &   -   & 0.2 &   0     \\
12& 1.668 & 18.5 & 62 & 0.0540 &   1.721 & 76 & 41.2 & 0.0131 &     2.208 &   -   & 0.5 &   0    &    2.217 &   -   & 0.4 &   0     \\
13& 1.679 & 13.27 & 82 & 0.0753 &  1.737 & 42.6 & 61.1 & 0.0234 &   2.41 &   -   & 2.4 &   0    &     2.382 & 228 & 10.7 & 0.00439  \\
14& 1.690 & 9.87 & 106 & 0.101 &   1.75 & 27.75 & 82 & 0.0360 &     2.527 & 370.8 & 10.5 & 0.0027 &   2.402 & 127.7 & 17 & 0.00783  \\
15& 1.696 & 8.29 & 124 & 0.120 &  1.758 & 20.79 & 101 & 0.0481 &    2.593 & 71.5 & 30.6 & 0.0139 &   2.413 & 95.6 & 21.5 & 0.0104 \\
16& 1.701 & 7.07 & 143 & 0.141 &  1.767 & 15.46 & 125 & 0.0646 &   2.606 & 53.6 & 38.3 & 0.0186 &   2.446 & 43.3 & 43.1 & 0.0230  \\
17& 1.706 & 6.13 & 163 & 0.163 &  1.773 & 12.89 & 143 & 0.0775 &   2.623 & 38.43 & 50.5 & 0.0260 &  2.467 & 28.82 & 63.5 & 0.0347  \\
18& 1.711 & 5.46 & 182 & 0.183 &  1.778 & 11.15 & 160 & 0.0896 &   2.643 & 26.29 & 70.2 & 0.0380 &  2.482 & 22.35 & 81.6 & 0.0447  \\
19& 1.716 & 4.84 & 205 & 0.206 &  1.783 & 9.58 & 180 & 0.104 &    2.659 & 19.99 & 90.3 & 0.0500 &  2.499 & 17.17 & 107.1 & 0.0582  \\
20& 1.718 & 4.48 & 221 & 0.223 &  1.789 & 7.87 & 210 & 0.127 &    2.672 & 16.17 & 110.6 & 0.0618 & 2.508 & 15.16 & 122 & 0.0659  \\
22& 1.724 & 3.85 & 256 & 0.259 &  1.797 & 6.34 & 248 & 0.157 &    2.69 & 12.01 & 148.4 & 0.0832 &  2.529 & 11.64 & 162.8 & 0.0859  \\
21& 1.722 & 4.09 & 242 & 0.244 &   1.793 & 6.93 & 232 & 0.144 &     2.682 & 13.7 & 130.1 & 0.0729 &  2.519 & 13.15 & 142.1 & 0.0760  \\
\tableline  \tableline
\end{tabular}
\label{Table:BG_I(U)R_ph}
\end{table}

\begin{figure}[ht]
\includegraphics[scale=0.6]{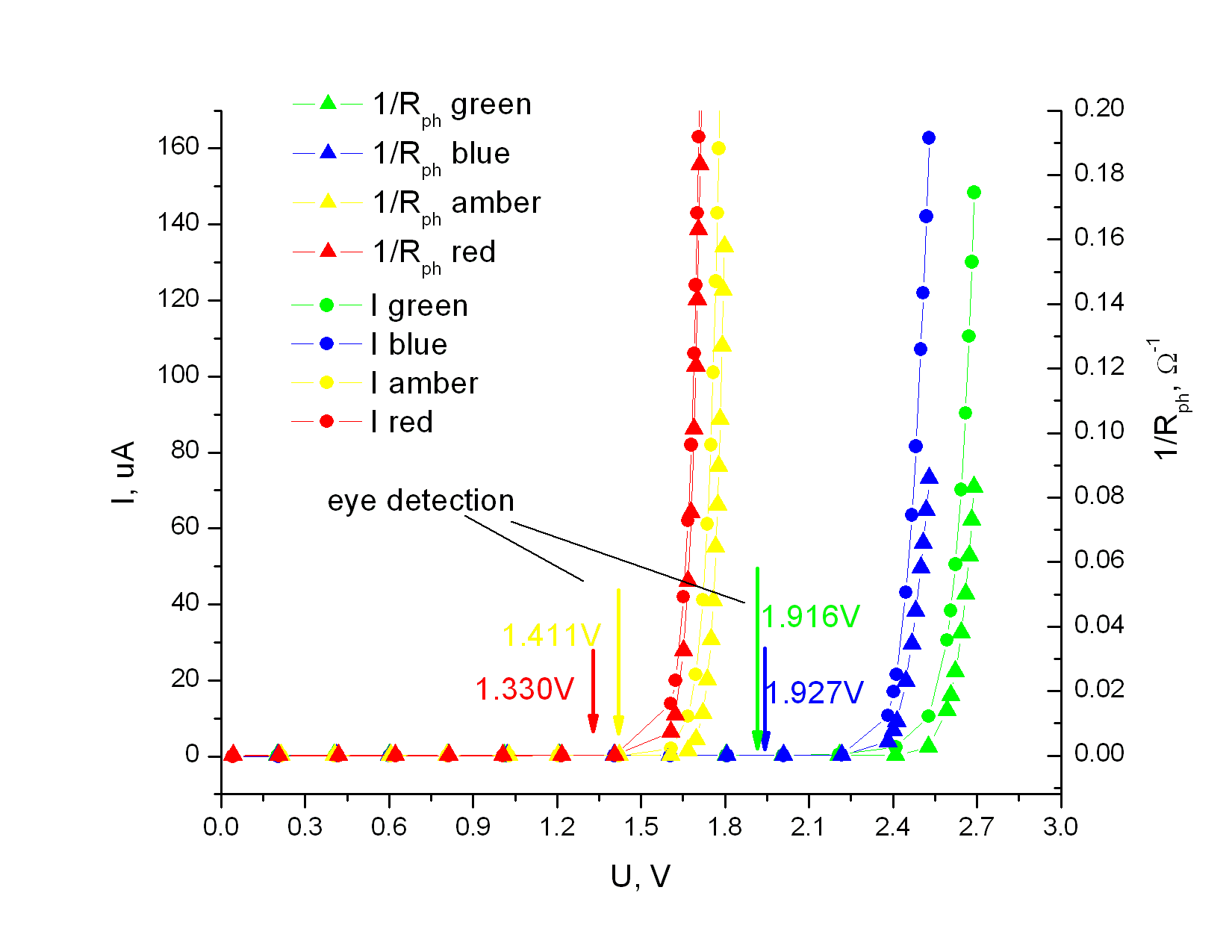}
\caption{
Зависимост на тока $I,$ кръгчета и скалата в ляво,
както и на проводимостта $R_\varphi,$ триъгълници и скалата в дясно,
като функции от напрежението $U.$ 
Стрелките показват праговите напрежения, при които окото вижда слаба светлина $U_c.$ Данните за различните диоди са показани с различен цвят: червен (red), кехлибарено-жълт (amber), зелен (green)
и син (blue). По форма зависимостите $I(U)$ и $R_\varphi(U)$ са неразличими.
        }
\label{Fig:BG_amper_siemens_vs_volt} 
\end{figure}

Данните от таблицата са представени на няколко фигури.
На Фиг.~\ref{Fig:BG_amper_siemens_vs_volt} тока $I$ през светодиода и проводимостта на фоторезистора $R_\varphi^{-1}$ са представени като функция от подаденото напрежение. 
Всички фигури са сходни и са само отместени по напрежение за различните цветове. 
Когато диода не свети, токът през него е пренебрежим 
и проводимостта на фоторезистора също е пренебрежима. 
Човешкото око е много чувствително и вижда слаба светлина при напрежения,  
при които не се забелязват особености на изследваните величини. 
Чупката, при която ВАХ започва да се издига стръмно е значително по-вдясно от критичното напрежение $U_c$, при което окото вижда слаба светлина.

\begin{figure}[ht]
\includegraphics[scale=0.6]{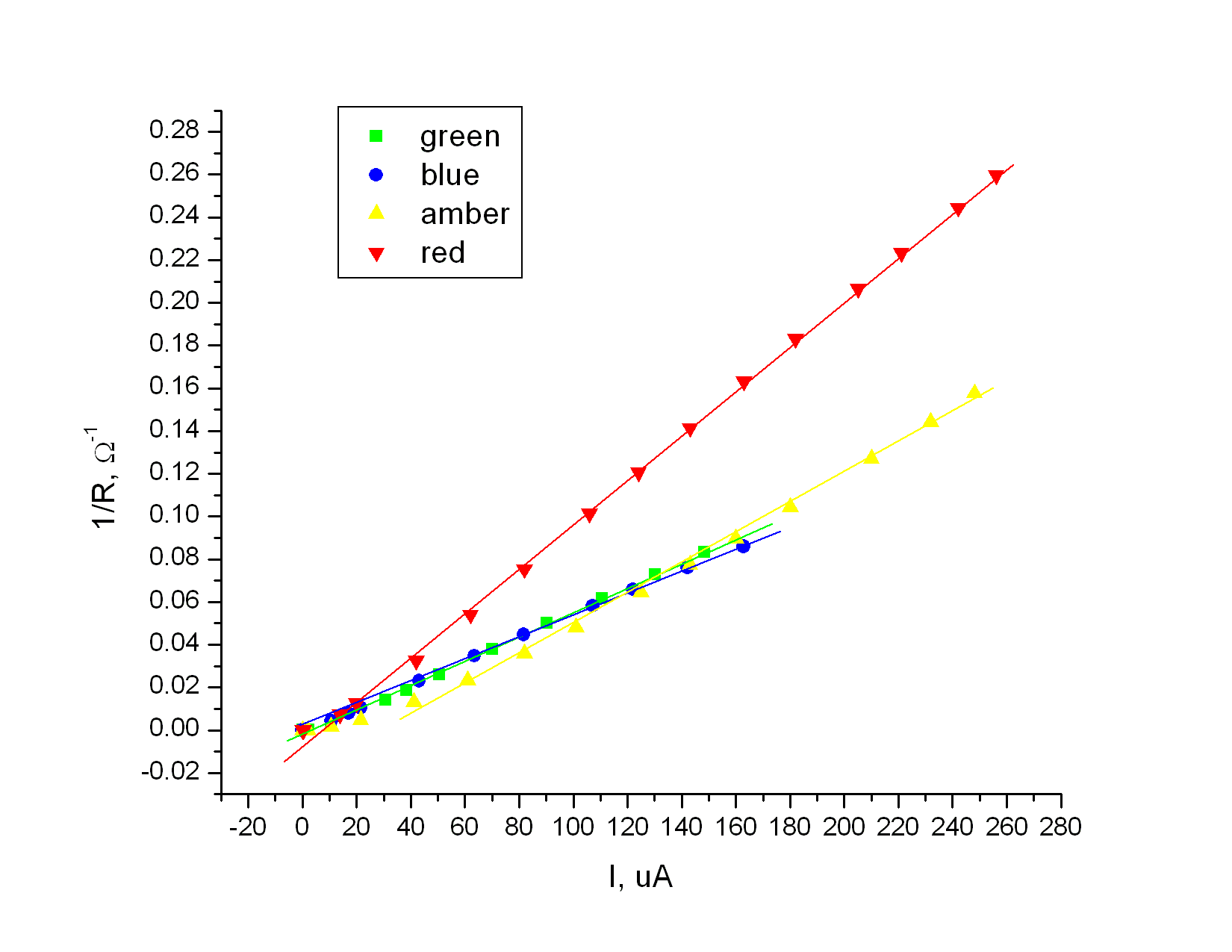}
\caption{Проводимост на фоторезистор $R_\varphi^{-1}$ като функция от тока на поставения до него светодиод $I.$ Инженерите, които не обичат гръцки букви използват съкращението $\mathrm{u A}$ за $\mathrm{\mu A}.$ 
За всички цветове: 
червен (red), кехлибарено-жълт (amber), 
зелен (green) и син (blue),
с висока точност се наблюдава линейна зависимост.
Около прага, токът през светодиода е пропорционален на интензивността на излъчената светлина, 
който на свой ред създава зарядови носители и електрична проводимост на фоторезистора.
            } 
\label{Fig:BG_Ampere_Siemens} 
\end{figure}

Подобието на ВАХ и $R_\varphi(U)$ показани на Фиг.~\ref{Fig:BG_amper_siemens_vs_volt} заслужава да бъде изследвано.
Нека обсъдим накратко природата на фоторезистора.
На тъмно той е добър изолатор, при който светлината създава носители на електричен ток, електрони и дупки. 
Когато подаденото напрежение е причина за възникналия ток е по-удобно закона на Ом да се представи чрез продимостта дефинирана като реципрочно съпротивление 
\begin{equation}
I_\varphi=\sigma U_\varphi,\qquad \sigma\equiv\frac{1}{R_\varphi}.
\end{equation}
За фоторезисторите, проводимостта е пропорционална на интензивността на светлината.
И ако фоторезистора е под постоянно напрежение 
токът през него е пропорционален на интензивността на светлината.
На Фиг.~\ref{Fig:BG_Ampere_Siemens} е показана зависимостта на проводимостта на фоторезистора $\sigma_\varphi(I_\mathrm{LED})$
като функция от тока през светодиода. Линейната зависимост показва,
че рекомбиниралите електрон-дупчести двойки в светодиода създават фотони,  които долитат до фоторезистора и там създават електрон-дупчести двойки и електрична проводимост. 
Единицата за проводимост се нарича още Сименс Sm=$\Omega^{-1}$ 
и техниците, които смесват физичните величини с единиците биха казали, 
че имаме  линейна сименс-амперна характеристика.

Както волт-амперните характеристики, така и сименс-амперните характеристики на светодиодите могат след известна обработка на експерименталните данни да дадат оценка на критичното напрежение
$\tilde{U}_c$ и енергията на квантовия преход $q_e\tilde{U}_c.$
Така намерените критични напрежения са представени в таблица~\ref{Table:BG_light_voltage_treshold}.

\begin{figure}[ht]
\includegraphics[scale=0.6]{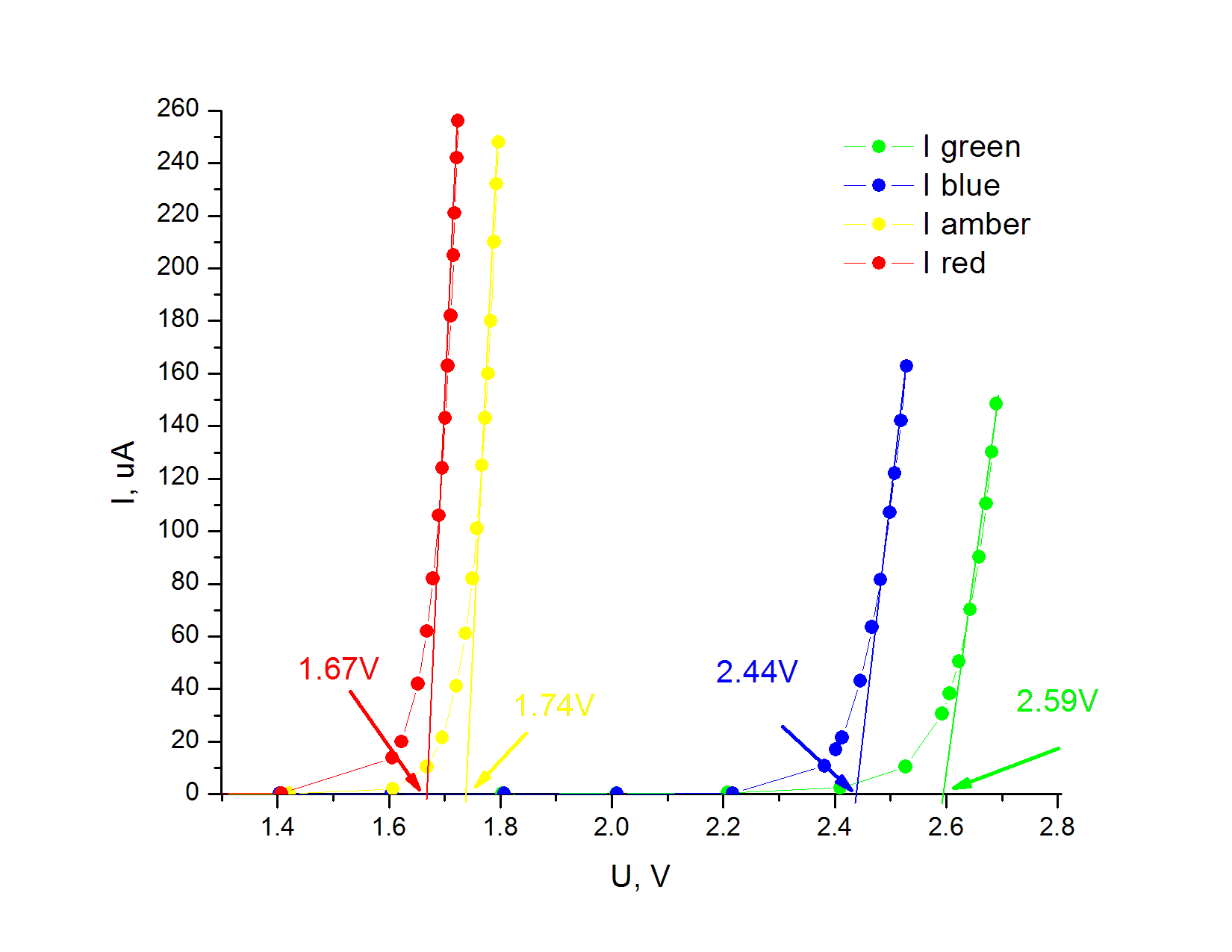}
\caption{ВАХ и тяхната линейна апроксимация за светодиодите с различни цветове: 
червен (red), кехлибарено-жълт (amber), 
зелен (green) и син (blue).
Линейната еекстраполация пресича абсцисата и дава едно критично напрежение $\tilde{U}_c$ получено чрез анализ на ВАХ, 
а не чрез субективното усещане кога диода светва.
$u$A $\equiv$ $\mu$A.
            }
\label{Fig:BG_v_threshold} 
\end{figure}

\begin{table}
\caption{Експериментални данни за праговото (критично) напрежение 
за светване (угасване) на светодиод от съответен цвят} 
\begin{tabular}{| l | c |}
\tableline 
Цвят & $U_\mathrm{c}$ [V] \\
\tableline \tableline
 червен & 1.330 \\
 жълт & 1.411 \\
 зелен & 1.916 \\
 син & 1.927 \\
\tableline 
\end{tabular}
\label{Table:BG_light_voltage_treshold}
\end{table}

\begin{table}[ht]
\caption{Други експериментални данни за дифракция 
и прагово напрежение за фотодиоди с различни цветове (производител CREE); 
$D=21\,\mathrm{cm}$, 
$d=1.5\,\mu\mathrm{m}$}
\begin{tabular}{| c| l | c| c | c | c |c | c | c | c |}
\tableline \tableline 
\No & Цвят &  $L$ [cm] & $\tg(\theta)$ & $\theta$, deg& $\sin(\theta)$ & $\lambda\,[\mathrm{nm}]$
 & $\nu\mathrm{[Hz]}$ & $U_c\mathrm{[V]}$ & $q_eU_c\mathrm{[J]}$\\
\tableline 
1&виолет&6.0& 0.294&15.5& 0.274& 411& $7.30\times 10^{14}$ & 1.46 & $3.70\times 10^{-19}$\\
2&син     &6.7& 0.319&17.7& 0.304& 456& $6.58\times 10^{14}$ & 1.55 & $2.10\times 10^{-19}$\\
3&зелен  &7.5& 0.357&19.7& 0.337& 505& $5.94\times 10^{14}$ & 1.99 & $1.94\times 10^{-19}$\\
4&жълт   &8.3& 0.395&21.6& 0.368& 552& $5.43\times 10^{14}$ & 2.10 & $1,55\times 10^{-19}$\\
5&червен&9.0& 0.429&23.2& 0.394& 591& $5.08\times 10^{14}$ & 2.23 & $1.40\times 10^{-19}$\\
\tableline 
\end{tabular}
\label{Table:BG_Valandovo}
\end{table}

\begin{table}[ht]
\caption{Критичните напрежения определени чрез око $U_c\equiv U^{\mathrm{eye}}_c$ 
или ВАХ $\tilde{U}_c\equiv U^{\mathrm{VAC}}_c$ са дадени в съседни колони.
Експериментални данни за дифракция и прагово напрежение за светодиоди от различни цветове серия 272 (широкоъгълни)~\cite{LED}, $D=20\,\mathrm{cm}$, $d=1.5\,\mu\mathrm{m}$}. 
\begin{tabular}{| c| l | c| c | c | c |c | c | c | c | c | c |}
\tableline 
\No & Цвят &  $L$ [cm] & $\tg(\theta)$ & $\theta$ [deg]& $\sin(\theta)$ & $\lambda\,[\mathrm{nm}]$
& $\nu\mathrm{[Hz]}$ & $U^{\mathrm{eye}}_c\mathrm{[V]}$ & $q_eU^{\mathrm{eye}}_c\mathrm{[J]}$
& $U^{\mathrm{VAC}}_c\mathrm{[V]}$ & $q_eU^{\mathrm{VAC}}_c\mathrm{[J]}$\\
\tableline \tableline
2&син     &6.5& 0.325&18.0& 0.309& 464& $6.47\times 10^{14}$ & 1.927 & $3.087\times 10^{-19}$ & 2.44 & $3.91\times 10^{-19}$\\
3&зелен  &7.6& 0.380&20.8& 0.355& 532& $5.63\times 10^{14}$ & 1.916 & $3.069\times 10^{-19}$ & 2.59 & $4.15\times 10^{-19}$\\
4&жълт   &8.9& 0.445&24.0& 0.407& 609& $4.92\times 10^{14}$ & 1.411 & $2.260\times 10^{-19}$ & 1.74 & $2.79\times 10^{-19}$\\
5&червен&9.5& 0.475&25.4& 0.429& 644& $4.66\times 10^{14}$ & 1.1.33 & $2.131\times 10^{-19}$& 1.67 & $2.67\times 10^{-19}$\\
\tableline 
\end{tabular}
\label{Table:BG_Valandovo}
\end{table}

\begin{figure}[ht]
\includegraphics[scale=0.6]{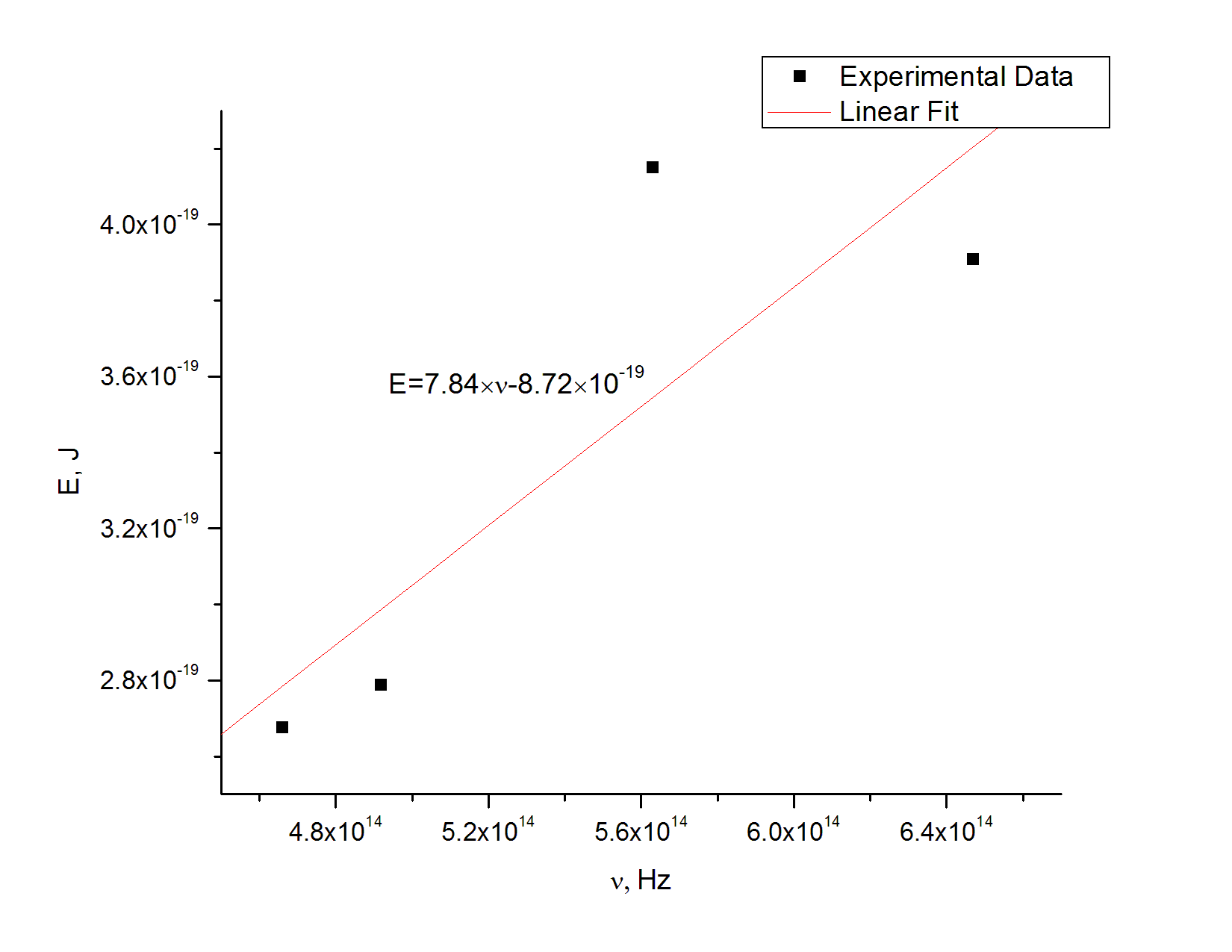}
\caption{Енергия на прехода $E$ определена чрез изследване на ВАХ 
и съответната честота на светлинната вълна $\nu$.}
Плътната права линия $E=(7.8\,\times\, 10^{-34}\,J)\nu+\mathrm{const}$ е прекарана така, че
сумата от квадратите на разстоянията по вертикала между експерименталните точки 
и правата линия да бъде минимална. Измерването дава 18\% 
по-голяма стойност от точното значение; 
като за първо измерване на константата на Планк това е добро начало.
Пунктирната линия е $E=(\dots\,\times\, 10^{-34}\,J)\nu+q_e(\dots\, \, \mathrm{V})$ 
е построена като в обработката на данните е пропусната зелената светлина.
\label{Fig:BG_Planck_VA} 
\end{figure}

\begin{figure}[ht]
\includegraphics[scale=0.6]{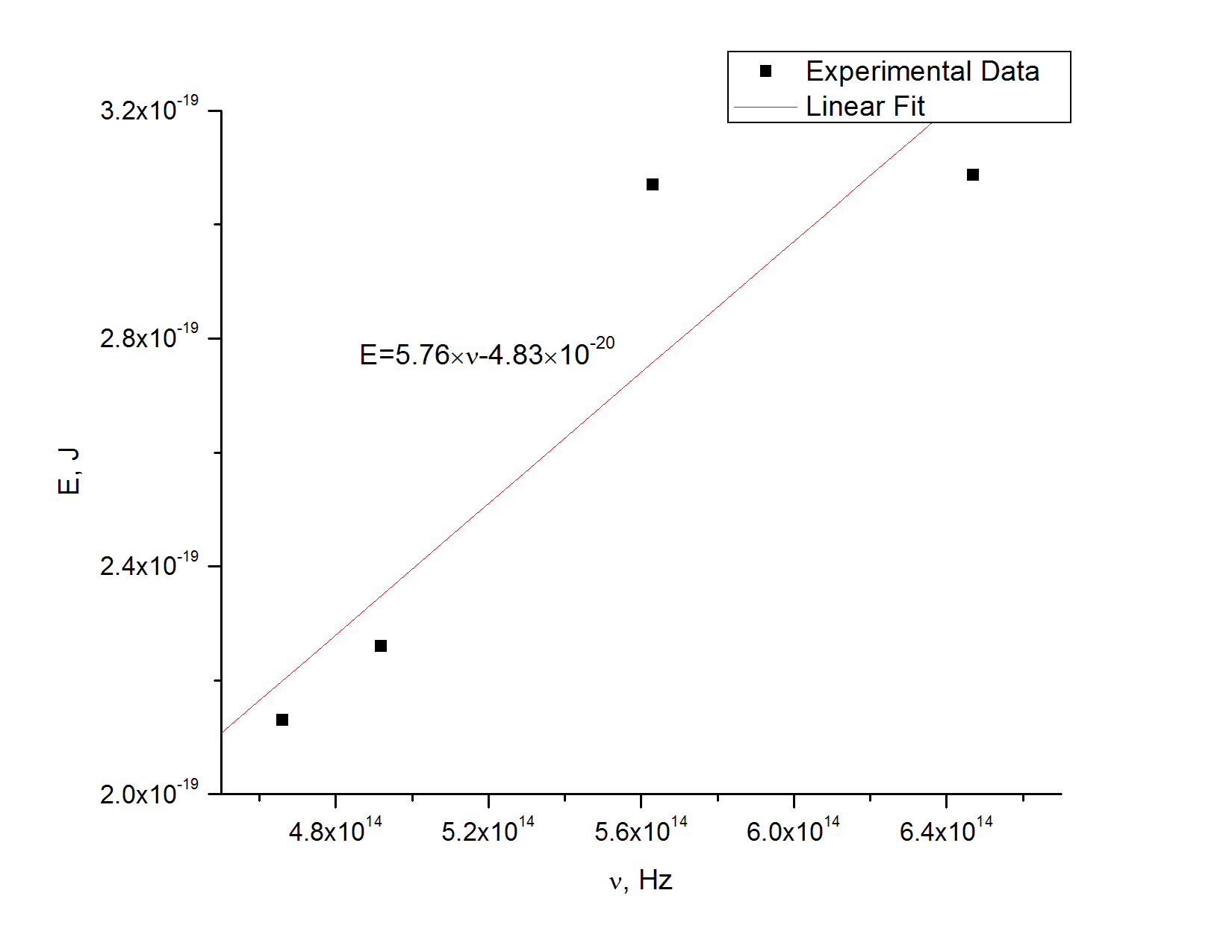}
\caption{Енергия на прехода определена по напрежението на светване и изгасване срещу честотата. За разлика от другата Фиг.~\ref{Fig:BG_Planck_VA} сега фитиращата (уйдурдисаната) права $E=(5.7\,\times\, 10^{-34}\,J)\nu+\mathrm{const}$} дава 15\% по-малка стойност. Това е съвсем прилична точност за първо измерване на константата на Планк. За нас е важно, че се докоснахме до измерване на една от важна фундаменталн константа която е вързана с квантовите явления.
\label{Fig:BG_Planck_eye} 
\end{figure}

\label{item:RGB} Нека за сенките показани на Фиг.~\ref{Fig:BG_RGB_shadow} съобразим кои цветове дават принос:
1) пурпур (магента) = червено + синьо,
2) червено
3) жълто = червено + зелено,
4) зелено
5) синьо-зелено (циан) = синьо + зелено
6) бяло = червено + синьо + зелено 
(само при добре балансирани с потенциометри 2 цвята).

\ref{BG_PlanckError}. 
Константата, която участва в уравнението за фотоефекта~\eqref{BG_photoeffect} е леко различна за полупроводниците на светодиодите, които светят с различен цвят. 
Особено силно това различие се проявява при зелените светодиоди. 
Освен това светването и изгасването на фотодиодите не е наблюдавано на тъмно. 
При подробно изследване на волтамперната характеристика критичното напрежение може да се определи по-точно,
но трябва да се отчита множителя на неидеалност (\textit{non ideality factor}) 
на светодиодите $n,$ което вече ще ни отведе дълбоко в електрониката.
Най-точно би било да се изследва ВАХ на вакуумен диод (това е радиолампа) осветяван от различни светодиоди.
Опита за наблюдаване на фотоефект с вакуумен диод е описан в учебниците.

\ref{BG_alpha}. Преди дифракцията, фотона има успоредна на дифракционната решетка проекция импулса $p\equiv(h/\lambda)\sin(\alpha)$. 
От дифракционната решетка с период $d$ фотона получава импулс $p_0\equiv h/d$ или кратен на него; 
за простота в това подусловие разглеждаме само първия дифракционен максимум. 
След дифракцията $y$-компонентата на импулса е 
$p^\prime\equiv(h/\lambda)\sin(\theta)$, 
а закона за запазване на импулса $p^\prime=p^\prime+p_0$ дава 
\begin{equation}
 \frac{h}{\lambda}\sin(\theta)
 =\frac{h}{\lambda}\sin(\alpha)+\frac{h}{d}.
\end{equation}
Разликата в оптичните пътища за съседни процепи на дифракционната решетка тогава е
\begin{equation}
 \lambda=\left[\sin(\theta)-\sin(\alpha)\right]d.
\end{equation}
Вълновото решение изисква добър чертеж, а механичното решение - познаване на законите за запазване. 

\ref{BG_RefractionIndeks}.
Когато светлината се движи в среда с показател на пречупване $n$
дължината на вълната става $n$ пъти по-малка. Във формулата за ъгъла на дифракция трябва да заместим дължината на вълната в среда
$\lambda_n\equiv\lambda/n=d \sin(\theta).$ За вода, синуса от ъгъла на дифракция намалява с около 25\% и това лесно може да се провери експериментално. 
Ако дифракционната решетка е на границата на две среди с показатели на пречупване $n_1$ и $n_2$, законът за запазване на импулса приема вида
\begin{equation}
 \frac{h}{\lambda_2}\sin(\theta)
 =\frac{h}{\lambda_1}\sin(\alpha)+i_m\frac{h}{d},
\qquad \lambda_1\equiv\frac{\lambda}{n_1}, 
\quad \lambda_2\equiv\frac{\lambda}{n_2},
\end{equation}
където цялото число $i_m=0,\pm 1,\, \pm 2,\, \dots$ 
е порядъка на дифракционния максимум, a $\lambda$
е дължината на светлината във вакуум.
При $i_m=0$, например, получаваме закона на Снелиус 
$n_1\sin(\alpha)=n_2\sin(\theta),$ като следствие от закона за запазване на импулса. 

\ref{BG_f<<R}. Нека приближено разгледаме диода като точков източник на светлина поставен във фокуса на лещата. 
Зад лещата излиза почти успореден сноп светлина. 
Радиусът на снопа е сравним с радиуса на лещата. 
За да може да разглеждаме браздите на диска като почти успоредни трябва радиуса на диска $R_\mathrm{CD}$ 
да бъде много по-голям от фокусното разстояние $R_\mathrm{CD}\gg f$. 
Затова точността на измерването се подобрява, 
ако лещата е поставена близо до периферията на диска. 
Ако разполагате само със слаба леща с фокусно разстояние по-голямо от диаметъра на диска, тогава трябва на поставите бленда до дифракционната решетка. 
Най-просто би било, ако фолиото се махне само от $1\;\mathrm{cm}^2.$ 

\ref{BG_CombinationalFrequency} 
Ако решетката се движи със скорост $v$ периода $d$ се изминава за време $d/v$ и честотата на преминаване на браздите спрямо неподвижен наблюдател е $\nu_0=v/d.$ 
Тази малка честота се добавя към честотата на падащия фотон
$\nu=1/T$ с период на вълната $T$. 
В този случай, за честотата на дифрактиралата светлина получаваме $\nu^\prime=\nu+\nu_0.$ 
Ако умножим с константата на Планк $h$ получаваме за енергията на разсеяния фотон 
$E^\prime=E+E_0,$ къдет $E^\prime\equiv h\nu^\prime,$  $E\equiv h\nu$ и $E_0\equiv h\nu_0.$ 
С други думи, движещата се дифракционна решетка предава своята енергия на фотона. 
Такъв е например, случая, когато светлина се разсейва от звукова вълна. 
Звуковата вълна е като една движеща се дифракционна решетка.
Честотата на разсеяната вълна може да се пресметне от закона за запазване на енергията.
Кванта на звука (фонона) и кванта на погълтатата светлина (фотона) 
отдават своята енергия на разсеяния фотон.
При много физични процеси, когато две вълни създават трета, 
например, Комптоново или Раманово разсейване, имаме едновременно запазване на енергията и импулса
\begin{equation}
 E^\prime=E+E_0, \qquad \mathbf{p}^\prime=\mathbf{p}+\mathbf{p}_0.
\end{equation}

\ref{BG_Hamilton}. Във формулата за фазовата скорост на фотона 
изразяваме дължината на вълната чрез импулса и периода чрез енергията 
\begin{equation}
 c=\frac{\lambda\!=\!h/p}{T\!=\!h/E}=\frac{E}{p}
\end{equation}
и получаваме връзката между енергията и импулса на фотона $E=cp.$
Това е общо съотношение между импулс и енергия за плоска електромагнитна вълна $p=E/c.$

\ref{BG_Chain}. Заместването във веригата от формули дава
\begin{equation}
 X=\left(V=\frac{p=E/c}{M}\right)\left(t=\frac{l}{c}\right)
   =l\left(m\equiv\frac{E}{c^2}\right)/M.
\end{equation}

\ref{BG_Muenhausen}.  Горното уравнение може да се запише като $MX=mL.$ 
Това е оргиналното разсъждение на Айнщайн за еквивалентност на масата и енергията $E=mc^2$ 
изведена като следствие от постоянното положение на центъра на масата. 
При разумни енергии, отместването на вагона е пренебрежимо малко, 
масата на нагрятото тяло е нищожно по-голяма, 
но от друга страна малки различия в масите съответсват на огромна енергия.
Така импулсът на фотона е дал тласък в развитието на ядрената физика и енергетика.
А в нашия мислен експеримент първоначално енергията $E$ е била химична енергия на лазерната показалка, 
a после се е превърнала в топлинна енергия на стената погълнала светлинния импулс. 
Покрай решението на задачата направихме кратък преговор на материала от учебниците по физика.

\newpage

\section{Translation into Macedonian: Услов на задачата}

\section{Услов на задачата}
\begin{figure}[ht]
\includegraphics[scale=0.74]{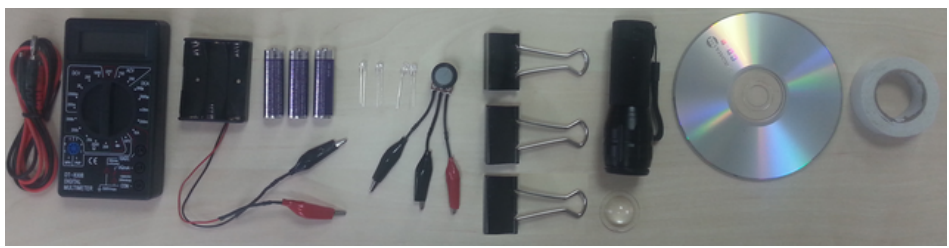}
\caption{Содржина на  комплетот за определување на константата на Планк $h$: 
Проводници со приклучок и штипка,
мултиметар,
држач за батерии и 3 батерии од 1.5 V,
4 лед диоди,
потенциометар со проводници и штипки,
3 држачи (штипки) за хартија,
фенерче со леќа,
компакт диск (CD)
и двострано леплива лента, 
фоторезистор, 
непрозрачна цевкичка,
пластелин,  
спајалица. 
}
\label{Fig:MK_Set_all} 
\end{figure}

\begin{figure}[ht]
\includegraphics[scale=0.74]{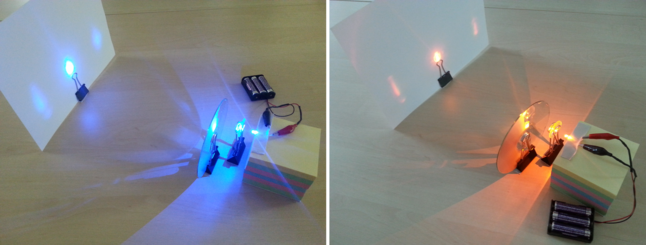}
\caption{Опис на апаратурата и методот  за определување на брановата должина на светлината од лед диода : 
Лед диода со помош на проводни жици се поврзува со електричен напон од изворот на струја-батериите.
Таа светнува а нејзината светлина се фокусира од леќата, 
поминува низ набраздената површина на проѕирен компакт диск
и паѓа на  екран, изгледајќи како пламенче од свеќа, при што може да се види  јасен лик на лед диодата.
Освен централниот максимум се гледаат и два бледи дифракциони максимуми.
Аглите на дифракцијата се различни за различни светлински бои, 
жолта и сина на снимките.
}
\label{Fig:MK_experimental_setup_blue_amber} 
\end{figure}

\begin{figure}[ht]
\includegraphics[scale=0.15]{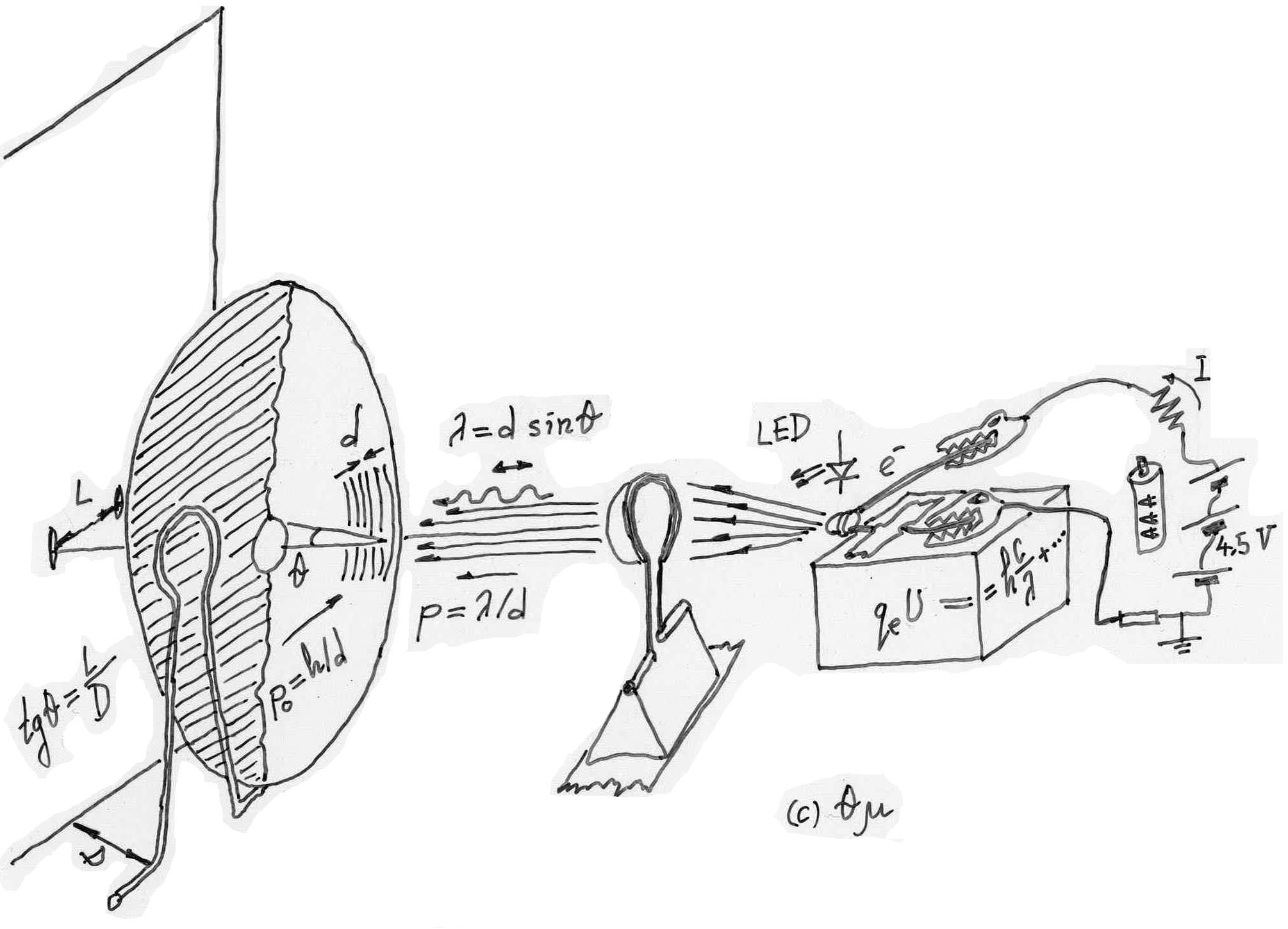}
\caption{Шематскиот приказ на апаратурата е даден на  Сл.~\ref{Fig:MK_experimental_setup_blue_amber}.
Лед диодата  (LED) е сврзана кон батериите со ``крокодил штипка''. 
Фотон от паралелниот светлински сноп  од леќата  има импулс $p=h/\lambda$, којшто е нормален на дискот.
При дифракција фотонот добива импулс $p_0=h/d$ во правец нормален на патеките набраздени на полиамидниот диск и се расејува под некој агол $\theta.$
$D$ e растојанието помеѓу дискот и екранот, 
а $L$ е растојанието помеѓу централниот максимум и првииот дифракционен максимум; $\tg\theta=L/D.$ 
Со помош на држачи-штипки се прицврстуваат леќата и ЦД-то како на шемата.
}
\label{Fig:MK_setup_scheme} 
\end{figure}

\begin{figure}[ht]
\includegraphics[scale=0.2]{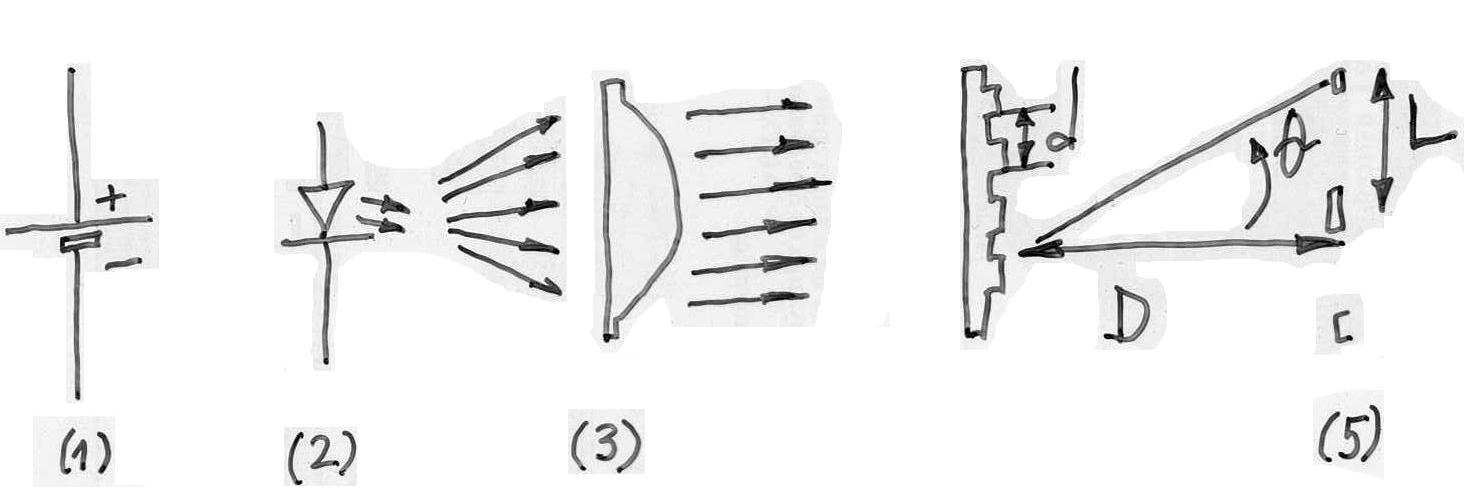}
\caption{Патот на енергијата: 1) хемсика во батериите 2) светлинска  3) фокусирање на светлината 4) дифракција 5) апсорпција и топлина.}
\label{Fig:MK_optical_scheme1} 
\end{figure}

\begin{figure}[ht]
\includegraphics[scale=0.95]{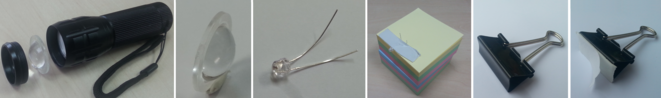}
\caption{Завртувачки прстен, леќа и фенерче.
лед диода.
Леќа на фенерчето.
Компакт диск со делумно  отстранета, одлепена  фолија.
Лед диода залепена со леплива лента врз ``постамент' ', пакет  хартиени ливчиња.}
\label{Fig:MK_set_diode_lens_disk} 
\end{figure}

Определете ја  константата на Планк $h$, 
искористувајќи го комплетот одуреди  
и апаратурата
покажани на 
Фиг.~\ref{Fig:MK_Set_all},  Фиг.~\ref{Fig:MK_experimental_setup_blue_amber},
Фиг.~\ref{Fig:MK_optical_scheme1},
Фиг.~\ref{Fig:MK_setup_scheme}  и
Фиг.~\ref{Fig:MK_set_diode_lens_disk}, 
имајќи ги во предвид
познатите вредности за брзината на светлината $c=299792458$~m/s 
и полнежот на електронот $q_e=1.602176565 \times$$10^{-19}$~C, 
како и разстојанието помеѓу патеките на компакт дискот $d=1.50\, \mu$m.


\section{Оценување на вашата работа  и организациони прашања}

Експерименталната апаратура ќе остане за Вас и нејзиното склопување нема да се оценува.
Ќе се оценуваат сите предадени листови  во кои се дадени: таблици со експерименталните резултати, графичките прикази, правите  кои минуваат најблиску до експерименталните точки,  анализата на експерименталните резултати и одговорите на теориските прашања поврзани со разбирањето на искористените формули за обработката на експерименталните податоци. 
 
Рангирањето ќе биде според возрасните групи, а распределбата  на условите по класови е следната:   
\begin{itemize}
\item 7--9 клас -- електричен дел: напон на гаснење на диодите, волтамперски карактеристики на лед диодите, ом-амперски карактеристики на осветлен фотоотпорник 
\item 10 клас -- оптички (мерење на брановата должина) и електричен дел (само напонот на гаснење); 
\textbf{определување  на Планковата константа}
\item 11--12 клас -- сите задачи
\end{itemize}

Теориските задачи се поврзани  со формулите искористени за обработката на експерименталните податоци и имаат за цел да провериме  дали експериментаторот го  разбира тоа што го мери. 
Тие  носат малку поени, 
но се во предност при  исти решенија на   експерименталните задачи. 
Теориските задачи носат повеќе поени при рангирањето на резултатите за најдобро решена домашна задача.

Ги советуваме учениците да ги решаваат задачите од соодветната возрасна група. 
Ако ученик од 7--9 клас ги реши сите задачи за неговата возрасна група, може да се обиде  да реши дел од оптичките експериментални или теорскии задачи, кои ќе му донесат дополнителни поени.
Исто така ученик от 10--12 клас може да реши задачи от електричниот дел, којшто се за 7--9 клас, при што ќе добие дополнителни поени.
Задачите за домашна работа може да започнете да ги решавате, ако  целосно сте ги решиле сите други задачи. 
Ќе има и дополнително рангирање по однос на решени  задачи за домашно.
Ако немате доволно  време, довршете ги  задачите за домашно, дома  и испратете ни го  одговорот денес на  е-маил адресата  epo@bgphysics.eu до 24:00. 

Ако имате некои дополнителни  прашања обратете се кон наставниците контролори.  
Времетраењето  за работа е  4 часа, при што во првите 2 часа не можете да ја  напуштите  просторијата. 
При решавањето на задачата можете да користете калкулатор.  
Ве замолуваме предадете ги мобилните телефони на наставниците контролори во просторијата.. 
Ако имате мобилен при вас ќе бидете дисквалификувани.

Освен мултиметрите,
другата апаратура 
 е подарок од организаторите за ``кабинетот по физика'' и учесниците се охрабруваат да го  демонстрираат решението на задачата пред своите соученици. Ако ја  дополните апаратурата за мерење  на $h$ со вакуумска  фотодиода ве замолуваме да се претставите на конкурсот за  уреди на 6 јуни 2015.

\begin{figure}[ht]
\includegraphics[scale=0.9]{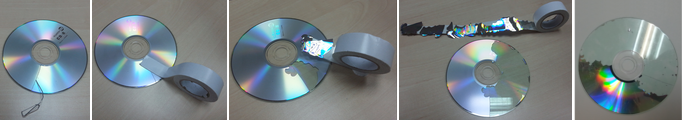}
\caption{``Технологија'' постапка  за отстранување  на фолијата од ЦД-то: 
1) Фолијата од ЦД-то се издраскува (сече, изгребува) радијално со помош на врвот од спојувалка, “спајалица“.
2) Врз гребнатината се залепува леплива лента “селотејп“.
3) Лентата се повлекува, дрпнува  и фолијата останува залепена за неа.
4)Постапката за лепење на лента и отстранување на фолијата од ЦД-то се повторува се додека половината од ЦД-то не се исчисти и стане проѕирно.}
\label{Fig:MK_set_disk_foil_detachment} 
\end{figure}

\section{Фотоефект. Вовед}

При фотоефектот, фотонот избива електрон от површината на метал 
и  според законот за запазување на енергијата имаме поврзаност на  енергијатата на фотонот 
$h\nu$, кинетичната енергија на електронот и излезната работа на електроните. 
Кај лед диодите  имаме обратен процес на внатрешниот фотоефект: 
електрон и празнина  со  мала енергија се рекомбинираат и се емитира фотон. 
Помеѓу граничниот напон  $U_c$, при кој диодата светнува и енергијата на фотонот важи следната релација
\begin{equation} 
 q_eU_c\approx h\nu+\mbox{const},
\label{MK_MK_photoeffect}
\end{equation}
каде $q_e$ е електричниот  полнежот на електронот. 
Зависноста на константата  од материјалот е мошне слаба  освен за зелените лед диоди. 
Кинетичните енергии на рекомбинирачките електрони се занемарливи,
т.е.се многу помали од енергијата на емитираниот фотон.
Задачата, која ја разгледуваме денес се заснива на фактот дека имаме   блиски вредности на константи за различни материјали,  
од кои се изработени лед диодите. 

\section{Електричен дел}

\subsection{Мерење на граничните напони $U_\mathrm{c}$ на светнување (“палење“) или гасење на лед диодите}
\begin{enumerate}
\item
\textit{Прецртајате ја шемата на Сл.~\ref{Fig:MK_Potenciometer}а без амперметар (заменете го со проводник). Составете ја таа шема со прикажаните елементи.}(7--12 клас)

Поставете ги батериите во кутијата со лежишта за батерии. 
Поврзете ја секоја диода според Сл.~\ref{Fig:MK_MaxLight}
Донесете и напон на диодата спојувајќи ги со крајот на проводниците (штипалки), 
така што струјата да поминува низ изолираните отпорници! 
Овие отпорници како составен дел на самиот проводник ја ограничуваат струјата на лед диодата за да не може истата да Ве заслепи.. 
Сепак не гледајте директно во лед диодите.

При овој услов не се поврзува амперметар, односно тој се заменува со проводник.
Поврзете сега некоја лед диода преку потенциометар, 
како што е прикажано на Сл.~\ref{Fig:MK_Potenciometer} или по-точно како на прецртаната од Вас шема без амперметар. 
Двата краја од  потенциометарот поврзете ги со штипките на проводниците од држачот на  батериите.
Можете да користете и  дополнителни проводници кои имаат штипалки. 
Поврзете го волтметарот со жиците што излегуваат од лед диодата. 
Вртете копчето на потенциометарот се додека диодата светне  и изгасне. 
Ако лед диодата не свети  или интензитетот на светлината не се променува при вртењето на потенциометарот можно е следното или диодата е сврзана во непропусна насока или пак неправилно ви е поврзан потенциометарот според шемата. Повторно проверете  дали  сите елемените се поврзани според нацртаната од Вас шема. 
[5~поени.]

\begin{figure}[ht]
\includegraphics[scale=0.61]{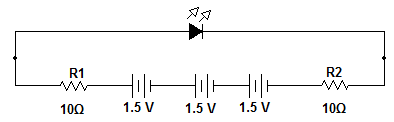}
\caption{
Напојување на лед диодите за експериментот со дифракција.
Оваа иста шема може да се искористи и за утвдување на поларитет на лед  диодите  во кој случај тие  светнуваат, односно пропуштат или не пропуштат струја. 
Електричната струја од 3-те сериско поврзани батерии поминува, тече низ лед диодата. 
Двата отпорника  $R_1$ и $R_2$ се изолирани и не се гледат.
Овие отпорници ја ограничуваат струјата за да  вклучената лед диода не прегрее и не ве заслепи.
Означените точки на вертикалните линии од шемата ги означуваат врските со штипалките (“крокодилчињата“).
}
\label{Fig:MK_MaxLight} 
\end{figure}

\begin{figure}[ht]
\includegraphics[scale=0.8]{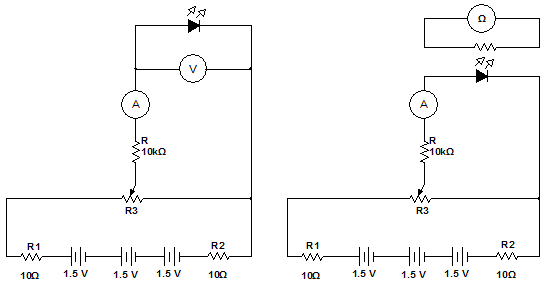}
\caption{
Потенциометриско сврзување на лед диодите за испитување  на зависимоста $I(U).$
Напон од краевите на   проводниците од кутијата во која се сместени батериите се донесува  на  потенциометарот. 
Диодата се поврзува  со  проводникот  кој се наоѓа на средината од потенциометарот и со некој  од крајните проводници  на потенциометарот. 
Отпорникот  $R$ е поврзан  за средниот приклучок на потенциометарот 
пред штипката (``крокодил'') и е скриен во црна изолаторска цевкичка. 
Овој голем отпор овозможува да се испита оној дел од волт-амперската карактеристика (ВАК) при кој  диодата светнува и изгаснува. 
Амперметарот е сериски поврзан со лед диодата , а волтметарот паралелно. 
Ако диодата не светне ни  при  една од крајните положби на потенциометарот,  променете го поларитетот на батериите.
Отпорниците  $R_1$ и $R_2$ се поставени во  црни изолаторски  цевкички пред “крокодилите“ на изводите од кутијата носач  на батерии.}
\label{Fig:MK_Potenciometer} 
\end{figure}

\item 
\label{Item:MK_Uc}
\textit{За 4-те светодиода измерете напрежението $U_c$, при което те изгасват или обратно започват да светят.} (7--12 клас)

Откако ќе ја составете шемата од потенциометриското сврзување на лед диодите  проверете дали при вртење на потенциометарското копче  диодата светнува и изгаснува. 
Решението на задачата бара  да набљудувате слаба светлина, поради што 
просториите се малку и затемнети. 
За секоја од лед диодите  направете по неколку мерења  и најдете средни вредности за секоја од нив. 
Пресметете и максимално отстапување од средната вредност,
односно дајте оценка  за точноста на мерењата. 
Резултатите представете  ги во табела , подредувајќи ги според измерениот напон, види табела~\ref{table:MK_U_c_table}
В табелата да има  : боја , средна вредност на напонот, 
максимално остапување од средната вредност, 
број на мерења за секоја лед диода. 
[10~т.]

\begin{table}[ht]
\caption{Експериментални резултати за граничните вредности на напоните на светнување или гасење на лед диодите} 
\begin{tabular}{| c| l | c |}
\tableline 
\No & Цвят & $U_c\mathrm{[V]}$  \\
\tableline \tableline
1& црвена н& \\
2& жолта   & \\
3& зелена  & \\
4& сина     & \\
\tableline 
\end{tabular}
\label{table:MK_U_c_table}
\end{table}

\subsection{Мерење на граничните напони  $\tilde{U}_\mathrm{c}$  преку испитување на волт- амперскиите карактеристики на лед диодите} 
\item 
\label{MK_CurrentVoltageCharacteristics} \textit{Снимете ја, направете волт-амперска карактеристика  за 4-те лед диоди } (7--9 клас)

Вклучете  го мултиметарот, кој го  носите со  себе , како  амперметар и поврзете го како на шемата од Сл.~\ref{Fig:MK_Potenciometer}а така, што да ја мери струјата која тече низ лед диодите. 

Испитајте ја  зависиноста помеѓу јачината  и напонот $I(U)$ (Волт-Амперска Карактеристика, ВАК) за дадените ви лед диоди.
За секоја од лед диодите треба да пополните посебна  табела со по 25 реда,
со реден број на мерење, јачина на струја  $I$ и напон $U$. 
Припремете си на друг лист слична табела како на онаа во пример табелата~\ref{Table:MK_I(U)}.

Првин, проверете дали во една от крајните положби на потенциометарот диодата светнува; 
може да има потреба да ја промените поврзаноста на изводите.
При максимално силно светната лед диода  запишете ги вредностите за напонот  и струјата
- тоа е првото мерење.
После вртете го  копчето на потенциометарот и намалувајте ја јачината на струја
за околу 20~$\mu$A; запишувајте ги резултатите во табелата,  се  додека се добие струја помала од 10~$\mu$A.
Наблудувајте го волтметарот може да се случи да има потреба од промена на неговото  мерното подрачје.
Ви препорачуваме како совет, да го вртете копчето на потенциометарот и да го запишувате напонот на лед диодата за околу   200~mV, 
 се додека не се стигне до нула напон т.е нула волти.
Пополнете ги колоните од табелата со резултати од мерењата  направени со  различни лед диоди. 
[20~поени. за првата лед диода и[20~поени . за трите останати лед диоди]

\begin{table}[ht]
\caption{Волтамперни карактеристики на лед диодите  околу напонот на запалување.
Последователните колони се за црвена, жолта, зелена и сина лед диода. За да се опишат добро волтамперните карактеристики ( ВАК) табелата  треба се извршатр најмалку од по 20 мерења за секоја лед диода. 
}
\begin{tabular}{| r | c | c | r | c | c | r | c | c | r | c |c |}
\tableline
\multicolumn{3}{|c|}{црвена}
& \multicolumn{3}{|c|}{жолта}
& \multicolumn{3}{|c|}{зелена}
& \multicolumn{3}{|c|}{сина}\\
\tableline 
\No & $I\;[\mu\mathrm{A}$] & $U$ [V] & 
\No & $I\;[\mu\mathrm{A}$] & $U$ [V] & 
\No & $I\;[\mu\mathrm{A}$] & $U$ [V] & 
\No & $I\;[\mu\mathrm{A}$] & $U$ [V] \\
\tableline \tableline
 1& & & 1& & & 1& & & 1& & \\
 2& & & 2& & & 2& & & 2& & \\
 ...& & & ...& & & ...& & & ...& & \\

\tableline 
\end{tabular}
\label{Table:MK_I(U)}
\end{table}

\item 
\label{MK_CurrentVoltageCharacteristics} \textit{ Претставете ја на график волтамперската карактеристика  на секоја од 4-рите лед диоди , искористувајќи ги резултатите од пополнетата табела~\ref{Table:MK_I(U)}.} (7--9 клас)

Представете ги на график   резултатите от табелите 
користејќи еден и ист координатен систем за различните лед диоди. 
Пред да почнете да го исцртувате графикот внимателно претставете го размерот на  величините на координатниот систем.  
Би било добро податоците да се претстават на цел лист од хартијата со квадратчиња или милиметарската хартија.  
Добро е да направете  претставените мерни точки за различните  лед диоди  се разликуваат. 
За таа цел искористете молив или пенкало со различна боја.
Податоците од жолтата диода означувајте ги со пенкало со црна боја.
На апцисата, хоризонталната оска од графикот нанесете ги податоците за напонот  $U[\mathrm{V}]$,
а на ординатата, вертикалната координатна оска  -јачината на струите $I[\mu\mathrm{A}]$.
[20~поени.]

\item 
\label{MK_CurrentVoltageCharacteristics_U_c_tilda} \textit{ Од графиците на ВАК определете го граничниот напон за 4-те  лед диоди $\tilde{U}_\mathrm{c}$.} (7--9 клас)

На ВАК (волт-амперната карактеристика), при јаки струи имаме праволиниски дел.
Нацртајте права линија  која ја допира ВАК до стрмниот наклон на ВАК при големи струи.
Каде  таа права линија ја сече апцисната оска  $I=0$ за различните лед диоди?
Најдете ги пресечните точки на правите за различните лед диоди и запишете ги соодветните  напони  $\tilde{U}_\mathrm{c}$ во  табела како што е покажано на  Табелата~\ref{table:MK_U_c__tilde_table}. 
Споредете ги тие гранични напони со напоните на светнување и гаснење  $U_c$, кои ги измеривте  во условот~\ref{Item:MK_Uc}, укажувајќи и на бојата на лед диодата. 
Колку изнесува максималната разлика од $\tilde{U}_\mathrm{c}-U_\mathrm{c}?$
[10~поени.]

\begin{table}[ht]
\caption{Експериментални резултати за критичните напони добиени преку проучување на ВАК на лед диодите.} 
\begin{tabular}{| c| l | c |}
\tableline 
\No & Боја & $\tilde{U}_c\mathrm{[V]}$  \\
\tableline \tableline
1& црвена& \\
2& жолта   & \\
3& зелена  & \\
4& сина     & \\
\tableline 
\end{tabular}
\label{table:MK_U_c__tilde_table}
\end{table}

\item 
\label{MK_CurrentVoltageCharacteristics_small_current} \textit{ Објаснете го, квалитативно, описно  наклонот на графикот на ВАК при мали струи.} (7--9 клас)

Сега насочете го вашето внимание на графикот од ВАК за мали струи.
Забележувате дека наклонот кај различните лед диоди е речиси еднаков.
Како можете да го објаснете и интерпретирате тој факт?
[10~поени .]

\subsection{Испитување  на интензитетот на светлината емитирана  од лед диода со помош на фотоотпорник. Алтернативен метод за мерење на граничните напони $\tilde{U}_\mathrm{c}$ на лед диодите}

\item
\label{MK_Siemens-Ampere}
\textit{ Измерете ја зависноста на електричниот отпор на фотоотпорник од јачината на струјакоја тече низ лед диода.} (7--9 клас)

За таа цел искористете ја шемата на Сл.~\ref{Fig:MK_Potenciometer}б.
Со амперметар се мери јачината на струја  која тече низ лед диодата. Волтметар не ви е потребен.
Приклучете го мултиметарот во улога на омметар,со кој ќе го мерите отпорот на фототпорникот  $R_\varphi.$
Доближете го  фоторотпорникот до диода која свети со максимален интензитет, 
за да видите как се менува неговиот електричен отпор.
Потребно е на некој начин да се елиминира влијанието на надворешната светлина .
Затоа, ставете ги лед диодата и  фотоотпорникот во цевкичка и затворете ја со пастелин, за да  не навлегува светлина. 
Исто како и во претходната задача  мерејќи ја ВАК,
запишувајте ја јачината на струјата која тече низ лед диодата  и отпорот на фотоотпорникот $R_\varphi.$
Во табелтата~\ref{Table:MK_R_varphi(I)}  додате уште една колона за реципрочната вредност на електричниот отпор  $1/R_\varphi,$
која се вика проводливост и се мери во единицата  $\Omega^{-1}.$
Вртете го потенциометарското копче за да ја намалувате јачината на струјата за околу 20$\;\mu$A и пак запишувајте ги податоците во табела.
Кога ќе стигнате до минимална вредност на струјата, нула , пробајте да го направете обратното и да испитувате со слаби струи со вредности  до 15$\;\mu$A.
Повторете ги тие мерења за сите лед диоди и запишете ги резултатите во различни табели.
По завршувањето на мерењата пресметајте ги и реципрочните вредности  $1/R_\varphi$  и внесете ги во 3-тата колона.
 [20~поени. за првата лед диода и 20~поени. за преостанатите трите лед диоди]
 
\begin{table}[ht]
\caption{Зависиност на отпорот на елктричниот отпор на фотоотпорник од јачината на струја низ лед диода. Табелата треба да има барем 25 мерења за секоја лед диода.
}
\begin{tabular}{| r | c | c  |c | r | c | c | c | r | c | c | c | r | c | c | c |}
\tableline
\multicolumn{4}{|c|}{црвена}
& \multicolumn{4}{|c|}{жолта}
& \multicolumn{4}{|c|}{зелена}
& \multicolumn{4}{|c|}{сина}\\
\tableline 
\No & $I\;[\mu\mathrm{A}$] & $R_\varphi\, [\Omega]$ &  $1/R_\varphi\, [\Omega^{-1}]$ &
\No & $I\;[\mu\mathrm{A}$] & $R_\varphi\, [\Omega]$ &  $1/R_\varphi\, [\Omega^{-1}]$ &
\No & $I\;[\mu\mathrm{A}$] & $R_\varphi\, [\Omega]$ &  $1/R_\varphi\, [\Omega^{-1}]$ &
\No & $I\;[\mu\mathrm{A}$] & $R_\varphi\, [\Omega]$ &  $/1R_\varphi\, [\Omega^{-1}]$  \\
\tableline \tableline
 1& & & &1& & & & 1& & & & 1& & & \\
 2& & & &2& & & & 2& & & & 2& & & \\
 ...& & & &...& & & & ...& & & & ...& & &\\

\tableline 
\end{tabular}
\label{Table:MK_R_varphi(I)}
\end{table}

\item 
\label{} \textit{Претставете ја на график зависноста на отпорот кај  фотоотпорникот од струјата која тече низ лед диодата искористувајќи ги резултатите  од пополнета табела~\ref{Table:MK_R_varphi(I)}.} (7--9 клас) 

Графичкиот приказ направете го така што на хоризонталната оска - на апцисата да биде струјата која тече низ лед диодата $I$, 
а на вертикалната оска, ординатата - реципрочната вредност на отпорот, проводливоста $1/R_\varphi$ в единици $\Omega^{-1}.$
На еден ист координатен систем претставете ги графиците за сите лед диоди.
Пожелно е точките на графикот  да ги означувате со различни бои.
Покрај точките исцртајте ги кривите на зависноста.
Анализирајте ги тие криви и обидете се да ги објаснете резултатите од овој експеримент. 
[20~поени. за една  лед диода  и 20~поени . за трите преостанати лед диоди]

\item 
\label{} \textit{Анализирајте како фотоотпорникот може да се искористи за определување на граничниот напон $\tilde{U}_\mathrm{c}$.}(7--9 клас) [10~поени.]

\section{Оптична част}

\subsection{Мерење на брановата должина на светлината емитирана од лед диодите. (10--12 клас)}

\item 
\textit{Отстранување на фолија од ЦД.} 

Ја искористувате спојувалката “спајалицата“ како нож, поточно со еден нејзин крај, врв, поминуватете т.е ја  пресечете, изгребете  фолијата од ЦД-то по нејзиниот радиус. 
Врз гребнатината залепете леплива  лента со чие повлекување 
 ќе ја  отсранете фолијата од ЦД-то. 
Фиг. ~\ref{Fig:MK_set_disk_foil_detachment}.
така, исчистете го ЦД-то половина.. 
Потоа фатете го ЦД-то со една стоечка штипка како е покажано на  
Сл.~\ref{Fig:MK_experimental_setup_blue_amber}. [0~поени.]

\item 
\textit{Составување и поставување на оптичката апаратура.}

Сега ви претстојува најсложениот дел од задачата-задржете го вашето трпение и ладнокрвност:
Врз една од штипките во улога на статив- држач залепете двојно леплива лента-таа лесно се откинува со рака. 
Бел картон, кој ќе игра улога на екран  ставете во штипката.  
Одлепете ја пластиката од двојно лепливата лента и фиксирајте, залепете  ја штипката со белиот картон во улога на екран, на работната маса пред вас. 
Екранот треба да е вертикално исправен.
На ист начин поставете паралелно пред екранот, вертикално исправено, претходно подготвеното, до половина исчистено од фолија ЦД,  
а оддалеченоста екран ЦД да биде приближно околу една педа.
Погрижете се  да бидете прецизни. 
Ако не сте задоволни од точната поставеност, 
 отсранете ги лентите и поставете  ги наново, повторно ЦД-то и екранот како што е пропишано.
Измерете го растојанието  $D$ помеѓу  екранот и ЦД-то со  точност до милиметар.
Пред дискот поставете оптичка  леќа сместена во штипката-статив, 
а пред леќата, отприлика на два прста оддалеченост, 
прицврстете со леплива лента една од лед диодите на “пиедесталот“, столб од ливчиња. 
Лед диодата треба да биде на иста висина со центарот на леќата.
Напојувањето на лед диодите во експериментот со дифракција е покажан на Сл.~\ref{Fig:MK_MaxLight}.

Вклучете лед диодата да свети. 
Придвижувајте го “постаментот“ со листови се додека на екранот не добиете јасно светло петно.
Тоа е реалниот лик на лед диодата низ леќата. 
Од постаментот извадувате ливчиња или  донесувате, регулирајќи ја  висината, се додека ликот на лед диодата не се проектира, 
 со точност 1-2 центиметри , 
 на истата висина како и лед диодата.

 Да го проследиме патот на светлината:
 таа е емитирана од лед диодата, 
која е речиси во фокусот на леќата, 
 како што е покажано на Сл. ~\ref{Fig:MK_experimental_setup_blue_amber}.
Паралелниот светлински сноп поминува низ проѕирниот дел од ЦД 
 и на екранот освен централниот максимум се набљудуваат и два дифракциони максимуми.

Придвижете го малку ЦД-то во штипката така , 
 да тесен светлински сноп поминува низ хоризонтален  радиус на дискот.. 
Тогаш дифракциониот максимум и неговиот централен максимум се добиваат на една хоиризонтала врз екранот. 
Тоа е важно за точното мерење на брановата должина, 
кој  е еден од критериумите за рангирање на олимпијадата.
Употребете пенкало во боја за да го одбележите централниот максимум  
и двата дифракционни максимума. 
При одбелеќување кај жолтата лед диода  искористете пенкало со црна боја.
Растојанието помеѓу двата дифракциони максимуми е  $2L,$ 
а растојанието помеѓу централниот максимум и секој од дифракционите маклсимуми е приближно $L.$
[0~поени.]

\item
\textit{Колку  е фокусното растојание  на  леќата  $f$ во метри и колку е  реципрочна вредност на фокусното растојание $1/f?$} [15~поени.]

\item 
\textit{Определување на дифракциони агли.}

Запишете го растојанието  $D$ помеѓу ЦД-то и екранот. Работете со точност до милиметар. 
Сукцесивно, т.е последователно измерете го за секоја од лед диодите растојанието $L$  од центарот на централниот максимум до центарот на првиот дифракционен максимум.  
Подобро е да ги обележите светлите петна на хартиениот екран а потоа определите според вас каде е центарот. Претставете ги резултатите во табела: боја, растојание $L,$ 
и откога ќе ги искористете граничните напони  $U_\mathrm{c}$
пополнете ја табелата~\ref{table:MK_Sample}.
[20~поени.]
 
\item 
\textit{Пресметување на синус од аголот на дифракција.}

За секоја од боите на светлината што дифрактира пресметајте синус од аголот на дифракција 
$\sin(\theta)=L/(L^2+D^2)^{1/2}=\sin(\arctg(L/D)).$ 
Можете да дојдете до решението и по графички пат. 
Нацртајте правоаголен триаголник со катети  $L$ и $D$ и за хипотенуза од  100~mm ја мерите  наспроти аголот $\theta$ катетата паралелна на  $L.$ 
Должината на таа катета во дециметри ја дава  $\sin(\theta).$ 
Во табелата ~\ref{table:MK_Sample} пополнете ги колоните со аголните величини и пресметајте ги соодветните бранови должини.
Колоните кои  содржат критичен напон $\tilde{U}_c$  добиен преку испитувањето на ВАК  остануваат непополнети. 
Оставете ја оваа задача  за домашна работа.
[15~поени.]

\item
\textit{Определување на  константата на Планк според експерименталните податоци.}
\label{MK_PlanckError} 

Претставете ги  графичнки  резултатите от последната 
табела~\ref{table:MK_Sample}.
На ординатата т.е на  вертикалната координатна оска нанесете ги $q_eU_c$, а на апцисата т.е хоризонталната оска  $\nu$. 
Нацртајте ја правата, која на најдобар можен начин, според ваше мислење, ја  опишува приближно равенката 
\begin{equation} 
 q_eU_c\approx h\nu+\mbox{const.}
\label{MK_MK_photoeffect}
\end{equation}
Од тангенсот на аголот на наклон 
$h=\Delta(q_eU_c)/\Delta(\nu)$ определете ја  константата на Планк. 
Овде $\Delta$ означува разлика,  отсечок од правата; 
разлика  помеѓу две вредности на апцисата и ралика помеѓу две вредности на ординатата. 
Колку измерената од Вас вредност за  $h=\dots\,\mathrm{J\,s}$ 
се согласува со познатата вредност 
на таа  фундаментална константа?Пополнете го празното место !
Колку проценти грешка сте направиле  $100\,(h_\mathrm{exp}-h)/h=\dots\%$?
Подвлечете го резултатот и ставете  го во рамка! 
[137~поени.]

\item
\textit{Определување на а $h$ преку волта-амперната карактеристика (ВАК).} (Само ако Ви остане време.)

Сега исполнете ги барањата под условот~\ref{MK_CurrentVoltageCharacteristics_U_c_tilda}
и извршете експериментална обработка на експерименталните податоци од волт-амперните карактеристики ВАК.
Повторете ја истата процедура за определување на константата на 
Планк искористувајќи ги не субјективно определените напони на светнување и гасењена лед диодите  $U_c,$ туку критичните напони добиени со анализа на ВАК $\tilde{U}_c$.
Кој од методите е подобар?
Посочете кои систематски грешки влијаат на резултатот  
и како би ја подобриле експерименталната апаратура,  
за да се зголеми  точноста на мерењето.

Константата, која учествува во равенката~(\ref{MK_MK_photoeffect}) 
за фотоефектот значително се разликува за зелената лед диода.
Затоа, само со помош на останатите 3 лед диоди определете ја  $h.$
[30~поени.]

\begin{table}[ht]
\caption{Експериментални податоци за дифракцијата и граничните напони за лед диодите кои емитираат светлина со различни бои ; 
$D=\dots\,\mathrm{cm}.$} 

\begin{tabular}{| c| l | c | c | c | c |c | c | c | c | c | c |}
\tableline 
\No & Боја &  $L$ [cm] & $\tg(\theta)$ & $\theta$ [rad]& $\sin(\theta)$ & $\lambda\,[\mathrm{nm}]$
& $\nu\mathrm{[Hz]}$ & $U_c\mathrm{[V]}$ 
& $\tilde{U}_c\mathrm{[V]}$ 
& $q_eU_c\mathrm{[J]}$ & $q_e\tilde{U}_c\mathrm{[J]}$ \\
\tableline \tableline
1& црвена& & & & & & $\qquad\times 10^{14}$ & & & $\qquad\times 10^{-19}$ & $\qquad\times 10^{-19}$ \\
2& жолта   & & & & & & $\qquad\times 10^{14}$ & & & $\qquad\times 10^{-19}$ & $\qquad\times 10^{-19}$ \\
3& зелена  & & & & & & $\qquad\times 10^{14}$ & & & $\qquad\times 10^{-19}$ & $\qquad\times 10^{-19}$ \\
4& сина     & & & & & & $\qquad\times 10^{14}$ & & & $\qquad\times 10^{-19}$ & $\qquad\times 10^{-19}$ \\
\tableline 
\end{tabular}
\label{table:MK_Sample}
\end{table}

\subsection{Оптички експеримент со мешање на бои.  
(7--12 клас)}
\item 
\label{item:MK_RGB_task}
\textit{Откријте ги боите на сенките}

На Сл.~\ref{Fig:MK_RGB_shadow} шематски се прикажани 3 извори  на светлина 
и различните сенки~\cite{Bib:RGB} врз екранот на даден осветлен објект. 
Одредете ги боите во различните области нумерирани од 1 до 6.
Можете да ги одредите боите во областите  со размислување  или практично соискористување на  постојните лед диоди. 
Зада се изведе  овој експеримент потребно е 3-те лед диоди да светат истовремено,
 како што е покажано на Сл.~\ref{Fig:MK_RGB_diodes}.
Црвената и жолтата лед диода даваат поинтензивна светлина
и за да може помешаната светлина да стане бела  тие треба да се  
поврзат со изворот на струја преку компензирачки отпори, види Сл.~\ref{Fig:MK_RGB_schema}.
Црвената и зелената диода треба да се наоѓаат на растојание  1.5~см од сината лед диода.
Искористете ја сенката на монета од 1 стотинка 
прицврстена како на Сл.~\ref{Fig:MK_RGB_holder} 
и испратете ни ја фотографијата до 24:00.
[20~поени.]

\begin{figure}[ht]
\includegraphics[scale=0.6]{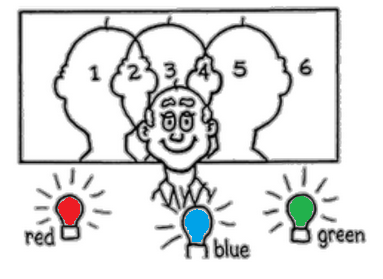}
\caption{
Сенки од 3 светлински извори: црвен (red), 
син (blue) и зелен (green) \cite{Bib:RGB}.
Означете ја бојата на сенките во различните области.
}
\label{Fig:MK_RGB_shadow}
\end{figure}

\begin{figure}[ht]
\includegraphics[scale=0.8]{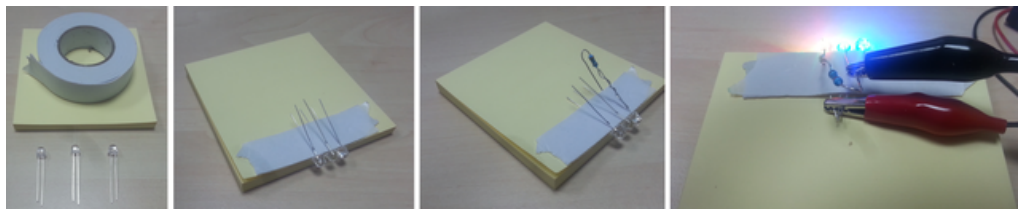}
\caption{Истовремено светење на црвената, зелената и сината лед диода:
1) Диоди, ливчиња и двојно леплива лента.
2) Диодите сеприцврстени со леплива лента врз ливчињата. 
Едната електрода на диодата е под лентата а другата е одозгора.
3) Црвената лед диода и синиот се поврзани преку отпорници 
за да може светлината од 3-те лед диоди да биде речиси бела.
4) “Крокодил“ штипките се паралелно споени со лед диодите.
Подолгите изводи на лед диодите се поврзуваат со негативниот пол т.е минус полот на батериите..
Ако  некоја од диодите не свети променете ги само врските со електродите.
}
\label{Fig:MK_RGB_diodes} 
\end{figure}

\begin{figure}[ht]
\includegraphics[scale=0.6]{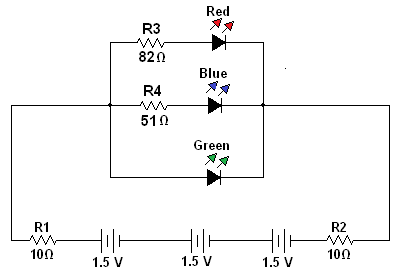}
\caption{Шематски приказ на на паралено поврзување на 3 лед диоди. 
Црвената лед диода е поврзана со  80~$\Omega$ отпорник,
а жолтата се поврзува соотпорник од  40~$\Omega.$
Отпорници од $10\;\Omega$, се поврзани со држачот на батериите,  
скриени односно поставени во црево. 
}
\label{Fig:MK_RGB_schema} 
\end{figure}

\begin{figure}[ht]
\includegraphics[scale=0.6]{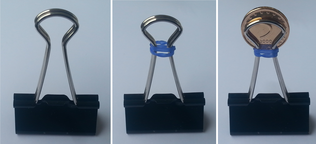}
\caption{Штипка, ластик и прицврстување на монетата.}
\label{Fig:MK_RGB_holder} 
\end{figure}

\section{Теориски дел. (Домашна задача за 11 и 12 клас)}

\begin{figure}[ht]
\includegraphics[scale=0.2]{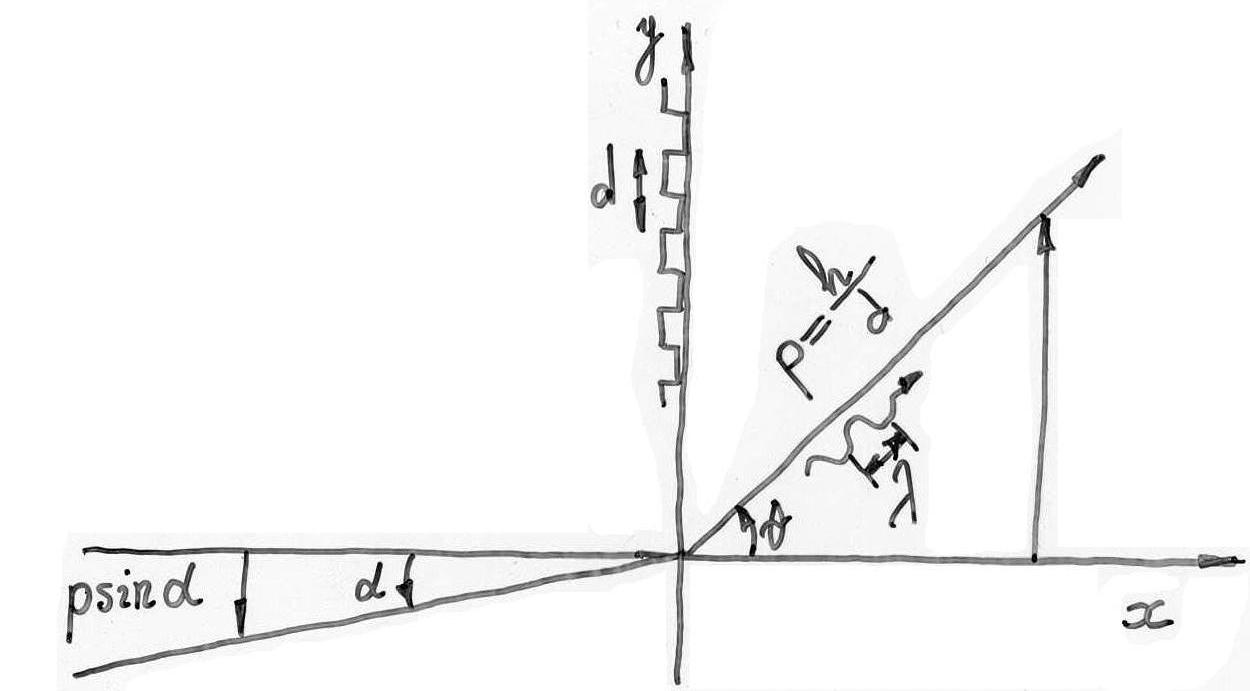}
\caption{Набраздувањето на дифракционата решетка е претставено на  $y$ оската.
Аглите ги читаме од  $x$ оската која е избрана да биде  по нормалата на дифракционата решетка.
Проекцијата на импулсот на фотонот паралелно на дифракционата решетка после расејувањето е $p\sin(\theta)$, 
а пред расејувањето  $y$-компонентата на импулса  била $p\sin(\alpha)$, 
каде според формулата на Де Броли  $p=h/\lambda$.
}
\label{Fig:MK_alpha}
\end{figure}

\item
\label{alpha}
\textit{Каков импулс може да има во една дифракциона решетка?}
Како ќе се промени формулата за агол на дифракција,  
ако светлината паѓа под агол $\alpha$ отчитан од нормалата на дифракционната решетка,како што е прикажано на Сл.~\ref{Fig:MK_alpha}? [5~поени.]

\item
\label{MK_RefractionIndeks}
\textit{Каков е импулсот на фотон во вода?}
Како ќе се промени аголот на дифракција при рефлектирачка светлина  за случај да имаме компактдиск потопен во вода? 
За индексот на прекршување на водата земете приближно $n=1.33.$ 
Заокружете ги точните одговори: [5~поени.]\\
	А. Формулата $\lambda=d \sin(\theta)$ не се променува.\\
	Б. Аголот на дифракција се зголемува.\\
	В. Аголот на дифракција се намалува  $n$ пати.\\
	Г. Синусот од аголот се намалува $n$ пати. 
Експериментално задачата можете  да ја решите дома,
при што потребен ви е сад со вода , а решението испратете го до 24:00 на е-маил адресата  
epo@bgphysics.eu 

\item
\label{MK_f<<R}
\textit{При каков сооднос помеѓу фокусното растојание на леќата $f$ 
и радиусот на компакт дискот $R_\mathrm{CD}$  поставениот експеримент е можен?} 
Може ли фокусното растојание да има должина колку еден лакт? [5~поени.]

\item
\label{MK_CombinationalFrequency}
\textit{Промена на фреквенцијата кај решетка во движење.}
Како се променува фреквенцијата на дифратираната светлина $\nu^\prime$, 
ако решетката се движи попречно на дифрационите процепи со брзина $v$?  [5~поени.]

\item
\label{MK_Hamilton} 
\textit{Импулс на рамен електромагмнетен бран.}
Искористете ги формулите за брзината на светлински бран $c=\lambda/T$, 
енергијата $E=h/T$ и импулсот на фотонот $p=h/\lambda$ 
и  докажете како импулсот на светлинскиот бран  $p(E)$ зависи од енергијата. [5~поени.]

\item
\label{MK_Chain} 
\textit{Да се најде поместувањето на вагон.}
За пресметување на аголот на дифракција најпрактично е да се искористи импулсот на фотонот.
Затоа, решете задача во која истовремено се разгледуваат енергијата и импулсот на фотонот.
Механиката во извесна смисла е и дел од оптиката.
Нека го изведеме следниот замислен експеримент шематски прикажан  
на Сл.~\ref{Fig:MK_GedankenExperiment1Stein}. 

\begin{figure}[ht]
\includegraphics[scale=0.6]{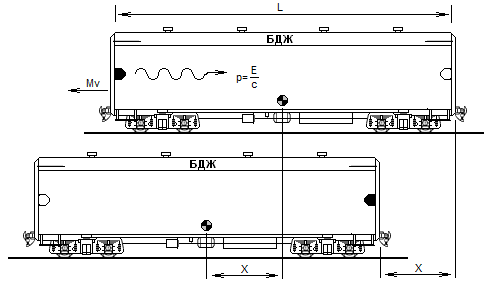}
\caption{Замислен експеримент: Вагон со маса $M$ и светлински импулс со енергија $E.$ 
Фотон се движи на десно со импулс $p=E/c,$
а вагонот се движи спротивно, на лево  со  импулс $MV.$
Механичната задача е аналогна за анализа како ``отскокнувањето '' на пушка или “трзај“ на некое оружје при пукање.
Додека светлината го прелетува вагонот со должина  $l,$  тој се поместува на лево на растојание $X.$
Претпоставуваме дека  $X\ll l.$
Точките го симболизираат патот на енергијата: првично енергијата во батеријата на ласерскиот покажувач, а потоа топлинско петно во кое светлината се апсорбирала.
}
\label{Fig:MK_GedankenExperiment1Stein}
\end{figure}

Вагон со маса $M$ и должина $l$ може да се движи по шини без триене . 
Во левиот крај на вагонот светнува ласерски покажувач со енергија на светлинскиот зрак  $E.$ 
Импулсот на светлината е $p(E)$. 
Формулата ја изведовте во претходниот услов. 
Во согласност со законот за запазување на импулсот вагонот се придвижува на лево со брзина  $V=p/M.$ 
Светлинскиот импулс  нека трае сосема кратко, 
прелетувајќи го вагонот за време $t=l/c.$ 
За тоа време вагонот се преместува на лево на растојание  $X=Vt.$ 
Па, откако светлината удира во десниот крај на вагонот, тој запира да се движи. 
Што се случува со светлината откако ќе удри во вагонот и се апсорбира? 
За колку се поместило тежиштето на вагонот? [5~поени.]

\item
\label{MK_Muenhausen}
\textit{Парадокс:Тежиштото е неподвижно. 
Кое од тврдењата е точно? Каде е грешката?}
Во претходното барање на задачата заклучивме дека 
вагонот се поместил на растојание $X.$
Тоа растојание можеби е мало, но сепак не е нула. 
Постои една строга теорема во механиката 
 дека едно неподвижно тело на кое не дејствуваат надворешни сили  
не може да му се промени тежиштето. 
Сликовито кажано: баронот Мјунхаузен не може да се подигне за косата. 
Баронот не може, но вагонот може - каде е грешката?
За потсетување да го разгледаме следниот едноставен пример:  
ако во левиот крај на вагонот имаме букова маса  со маса  $m$, 
тогаш при нејзиното преместување на десниот крај на вагонот на растојание  $l$, 
вагонот се поместува кон лево на кратко растојание  $X.$ 
Тие величини се поврзани со односот $M X=m l$ 
и тежиштето не се променува 
А зошто тогаш светлината може да го промени тежиштето на вагонот? 
[5~поени.] 
\item
\textit{Како си ги објаснувате боите на различните сенки опишани во точката~\ref{item:MK_RGB_task}?}
[5~поени.]

\item Домашна задача. 
Ако ја решите  задачата дадена во последниот подуслов информирајте нѐ по е-маил денес до  24:00 на следната адреса epo@bgphysics.eu,
а патем запишете ги името на консултантот или изворот на информација, 
како например Интернет страница или книги, кои сте ги користеле. 
Воопшто, испратете ни го домашното со сите подуслови што се ги решиле во попладневните часови. 
[премија 1-камен]

\section{Задача за конкурсот ``Уреди за кабинетот по физика, 6 јуни 2015''. (10, 11 и 12 клас)}

\item
\label{MK_ClassicalPhotoeffect}
\textit{Класичен експеримент за фотоефектот.}
Лед диодите на кои денес им ја меривте фреквенцијата и брановата должина на светлината што ја емитираат 
се добри извори на монохроматска светлина.
Испитајте ја волт-амперната карактеристика на вакуумска фотокелија ,
 како на пример  СЦВ-4 или RCA-930\cite{VaccumPhotoCell},  
 за случај кога катодата е осветлена со светлините емитирани од различните лед диоди.
Ако не најдете сличен елемент може да се искористи и фотомултипликаторската цевки, како на пример  ФЭУ-35\cite{PhotoMultiplier}, 
но во таков случај потребно е да ја искористите антимон - цезиумова емитерска  електрода (1) и првата забрзувачка електрода (2) без да додадете електрично напојување.
Најдете го закочниот напон  $U_c$ за различните бои;
овој експеримент е опишан во учебниците по физика.
Нанесете ги експерименталните податоци во рамнината енергија-фреквенција ($E$-$\nu$).
Повлечете права линија која максимално добро ги опишува експерименталните точки.
Од равенката $q_eU_c=h\nu+q_e A$,
пресметајте ја константата на Планк и излезната работа на металот со кој е обложена катодата.
Дадете оценка за точноста на мерењата.
 Овој метод треба да даде значително поголема точност
затоа што за разлика од различните лед диоди 
 кои имаат различни ширини на забранетите зони,
кај металот имаме  една и постојана излезна работа. 
Има повеќе современи прирачници, но сепак започнете со пребарување на интернет за класичната работа на Миликен \cite{Millikan:1916}.
Обидете се да испитувате волт-амперни карактеристики на полупроводнички фотодиоди.  
 осветлувани од различни лед диоди.
 Ве повикуваме да се пријавете на конкурсот \url{http://bgphysics.eu/} и демонстрирајте ја таа класична  апаратура. 
 Најголемиот успех на олимпијадата би бил доколку го направете  
и реанимирате овој експеримент во училиштата; обидете се да го направете денес за домашно. 
[137~поени.]

\end{enumerate}

\newpage
\section{Решение}

Знаењето за  фундаменталните константи на вселената е важен дел од развојот на човечката цивилизација.
Фундаменталните константи се мерат со доста скапи апаратрури 
и за нивно непрекинато уточнување работат врвни експериментатори. 
Планковата константа која изнесува  $6.62606957\times10^{-34}\;\mathrm{Js}$ е измерена со точност до милијардити делови.
Од друга страна  пак, овој фундаментален квант е дел од светот којшто нѐ опкружува. 
Лед диодите и компакт дисковите се наше секојдневие. 
Секој ученик ги видел, нешто знае за нив и е голем успех за нив, 
ако учениците успеат да ја определат таа фундаментална константа користејќи ги само школските знаења.
Доколку учениците разберат дека на Денот на фотонот ја измериле Планковата константа со точност до 20%
тогаш може да сметаме за голем успех  на нашиот образовен систем. 
Наученото во училиште успешно се трансформирало во вештина
и така младите ќе станат корисни за секоја квалификувана професија.

 По долу се опишани решенијата за секоја од барањата во условите на задачата. 
 Во коментарот често има и елементи на популарна статија, 
но тие се воглавном  наменети за наставниците.

\ref{MK_CurrentVoltageCharacteristics} и
\ref{MK_Siemens-Ampere}
Во табелата ~\ref{Table:MK_I(U)R_ph} е дадено целосното истражување на проблемот, 
при што  е малку повеќе од поставената задача на учениците,
но ја дава целосната претстава за физиката на проблемот.
За лед диодите кои емитираат различни бои се дадени податоци за напонот, $U,$ јачината на струјата $I$ 
и проводливоста на фотоотпорникот осветлен од лед диода $R^{-1}$

\begin{table}
\caption{Експериментални  податоци за напонот $U,$ јачината на струја $I$ и проводливоста на фотоотпорникот $R^{-1}$ за црвена, жолта, зелена и сина лед диода.За некои напони струите се многу мали или отпорот на фоторезисторот многу голем.}
\begin{tabular}{| r | c | c | c | c || c | c | c | c || c | c | c | c || c |c | c | c |}
\tableline  \tableline
\No & 
 $U$ [V] & $I[\mu\mathrm{A}]$ & $R_\varphi[\Omega]$ & $R^{-1}_\varphi[\Omega^{-1}]$ &
 $U$ [V] & $I[\mu\mathrm{A}]$ & $R_\varphi[\Omega]$ & $R^{-1}_\varphi[\Omega^{-1}]$ &
 $U$ [V] & $I[\mu\mathrm{A}]$ & $R_\varphi[\Omega]$ & $R^{-1}_\varphi[\Omega^{-1}]$ &
 $U$ [V] & $I[\mu\mathrm{A}]$ & $R_\varphi[\Omega]$ & $R^{-1}_\varphi[\Omega^{-1}]$ \\
\tableline
 1& 0.042 &   -   & 0    &   0    &  0.042 &   -   & 0.1 &   0    &    0.042 &   -   & 0    &   0    &      0.042 &   -   & 0 &   0     \\
 2& 0.207 &   -   & 0.1 &   0    &  0.215 &   -   & 0.1 &   0    &    0.2     &   -   & 0.1 &   0    &      0.204 &   -   & 0 &   0     \\
 3& 0.419 &   -   & 0.1 &   0    &  0.405 &   -   & 0.1 &   0    &    0.404 &   -   & 0.1 &   0    &    0.412 &   -   & 0.1 &   0     \\
 4& 0.623 &   -   & 0.1 &   0    &  0.605 &   -   & 0.1 &   0    &    0.601 &   -   & 0.1 &   0    &    0.602 &   -   & 0.1 &   0     \\
 5& 0.813 &   -   & 0.1 &   0    &  0.807 &   -   & 0.1 &   0    &    0.808 &   -   & 0.1 &   0    &    0.811 &   -   & 0.1 &   0     \\
 6& 1.006 &   -   & 0.1 &   0    &  1.028 &   -   & 0.1 &   0    &    1.011 &   -   & 0.1 &   0    &    1.022 &   -   & 0.1 &   0     \\
 7& 1.217 &   -   & 0.2 &   0    &  1.203 &   -   & 0.2 &   0    &    1.208 &   -   & 0.2 &   0    &    1.207 &   -   & 0.2 &   0     \\
 8& 1.407 &   -   & 0.3 &   0    &  1.421 &   -   & 0.2 &   0    &    1.404 &   -   & 0.2 &   0    &    1.404 &   -   & 0.2 &   0     \\
 9& 1.606 & 136.5 & 13.9 & 0.00733& 1.607 &   -   & 2.1 &   0    &    1.606 &   -   & 0.2 &   0    &    1.605 &   -   & 0.2 &   0     \\
10& 1.623 & 79.8 & 20 & 0.0125 &   1.668 & 617 & 10.5 & 0.00162 &    1.803 &   -   & 0.2 &   0    &    1.807 &   -   & 0.2 &   0     \\
11& 1.652 & 30.66 & 42 & 0.0326 &  1.696 & 202.5 & 21.5 & 0.00494 &  2.008 &   -   & 0.3 &   0    &    2.009 &   -   & 0.2 &   0     \\
12& 1.668 & 18.5 & 62 & 0.0540 &   1.721 & 76 & 41.2 & 0.0131 &     2.208 &   -   & 0.5 &   0    &    2.217 &   -   & 0.4 &   0     \\
13& 1.679 & 13.27 & 82 & 0.0753 &  1.737 & 42.6 & 61.1 & 0.0234 &   2.41 &   -   & 2.4 &   0    &     2.382 & 228 & 10.7 & 0.00439  \\
14& 1.690 & 9.87 & 106 & 0.101 &   1.75 & 27.75 & 82 & 0.0360 &     2.527 & 370.8 & 10.5 & 0.0027 &   2.402 & 127.7 & 17 & 0.00783  \\
15& 1.696 & 8.29 & 124 & 0.120 &  1.758 & 20.79 & 101 & 0.0481 &    2.593 & 71.5 & 30.6 & 0.0139 &   2.413 & 95.6 & 21.5 & 0.0104 \\
16& 1.701 & 7.07 & 143 & 0.141 &  1.767 & 15.46 & 125 & 0.0646 &   2.606 & 53.6 & 38.3 & 0.0186 &   2.446 & 43.3 & 43.1 & 0.0230  \\
17& 1.706 & 6.13 & 163 & 0.163 &  1.773 & 12.89 & 143 & 0.0775 &   2.623 & 38.43 & 50.5 & 0.0260 &  2.467 & 28.82 & 63.5 & 0.0347  \\
18& 1.711 & 5.46 & 182 & 0.183 &  1.778 & 11.15 & 160 & 0.0896 &   2.643 & 26.29 & 70.2 & 0.0380 &  2.482 & 22.35 & 81.6 & 0.0447  \\
19& 1.716 & 4.84 & 205 & 0.206 &  1.783 & 9.58 & 180 & 0.104 &    2.659 & 19.99 & 90.3 & 0.0500 &  2.499 & 17.17 & 107.1 & 0.0582  \\
20& 1.718 & 4.48 & 221 & 0.223 &  1.789 & 7.87 & 210 & 0.127 &    2.672 & 16.17 & 110.6 & 0.0618 & 2.508 & 15.16 & 122 & 0.0659  \\
22& 1.724 & 3.85 & 256 & 0.259 &  1.797 & 6.34 & 248 & 0.157 &    2.69 & 12.01 & 148.4 & 0.0832 &  2.529 & 11.64 & 162.8 & 0.0859  \\
21& 1.722 & 4.09 & 242 & 0.244 &   1.793 & 6.93 & 232 & 0.144 &     2.682 & 13.7 & 130.1 & 0.0729 &  2.519 & 13.15 & 142.1 & 0.0760  \\
\tableline  \tableline
\end{tabular}
\label{Table:MK_I(U)R_ph}
\end{table}

\begin{figure}[ht]
\includegraphics[scale=0.6]{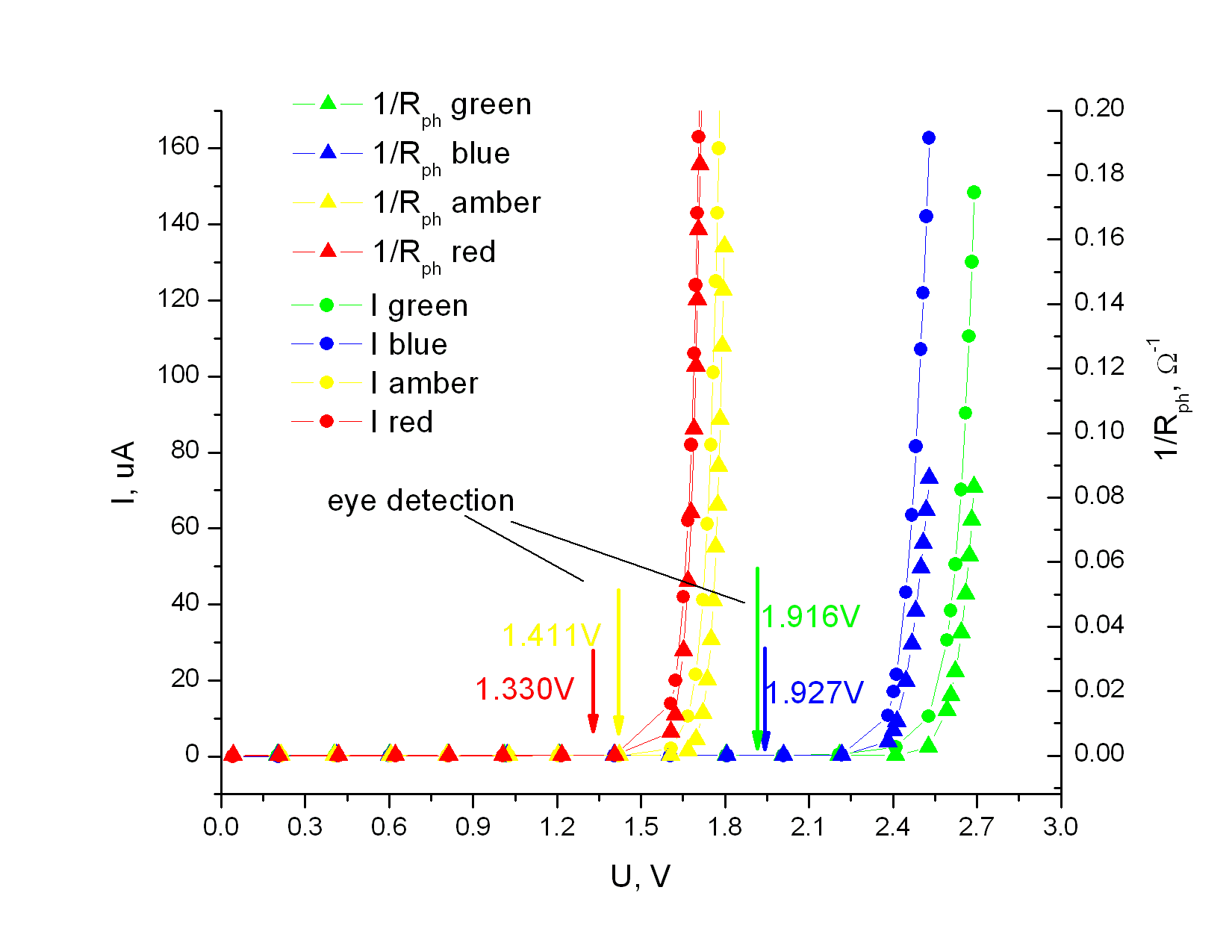}
\caption{
Зависноста на јачината на струја $I,$ кругче и скала во лево,
 како и на проводливоста  $R_\varphi,$ триаголници и скала на десно,
 како функции од напонот  $U.$ 
Стрелкитре ги покажуваат граничните напони, при кои окото може да види слаба светлина $U_c.$ Податоците за различните лед диоди се покажани со различна боја: црвена (red)  килибарно-жолта (amber), зелена (green) .
и сина (blue).Според формата зависностите  $I(U)$ и $R_\varphi(U)$ не се раликуваат.
        }
\label{Fig:MK_amper_siemens_vs_volt} 
\end{figure}

Податоците од табелите се претставени на неколку слики.
На Сл.~\ref{Fig:MK_amper_siemens_vs_volt} јачината на струјата $I$ што тече низ лед диодата и проводливоста на фотоотпорникот$R_\varphi^{-1}$  се претставени како функции од приложениот напон. 
Сите слики се слични и  само се разликуваат во однос на напонот за различните бои. 
За случаите кога лед диодата не свети  струјата што тече во неа е занемарливо мала 
и проводливоста на фотоотпорникот е занемарлива. 
Човечкото око е многу чувствително и забележува слаба светлина  и при  вредности на напони,  
 при кои не се забележуваат карактеристики на испитуваните величини. 
Подрачјето при кое волт-амперната карактеристика почнува стрмно да се издигнува е значајна по десно од критичкиот напон $U_c$, при коешто окото забележува слаба светлина.

\begin{figure}[ht]
\includegraphics[scale=0.6]{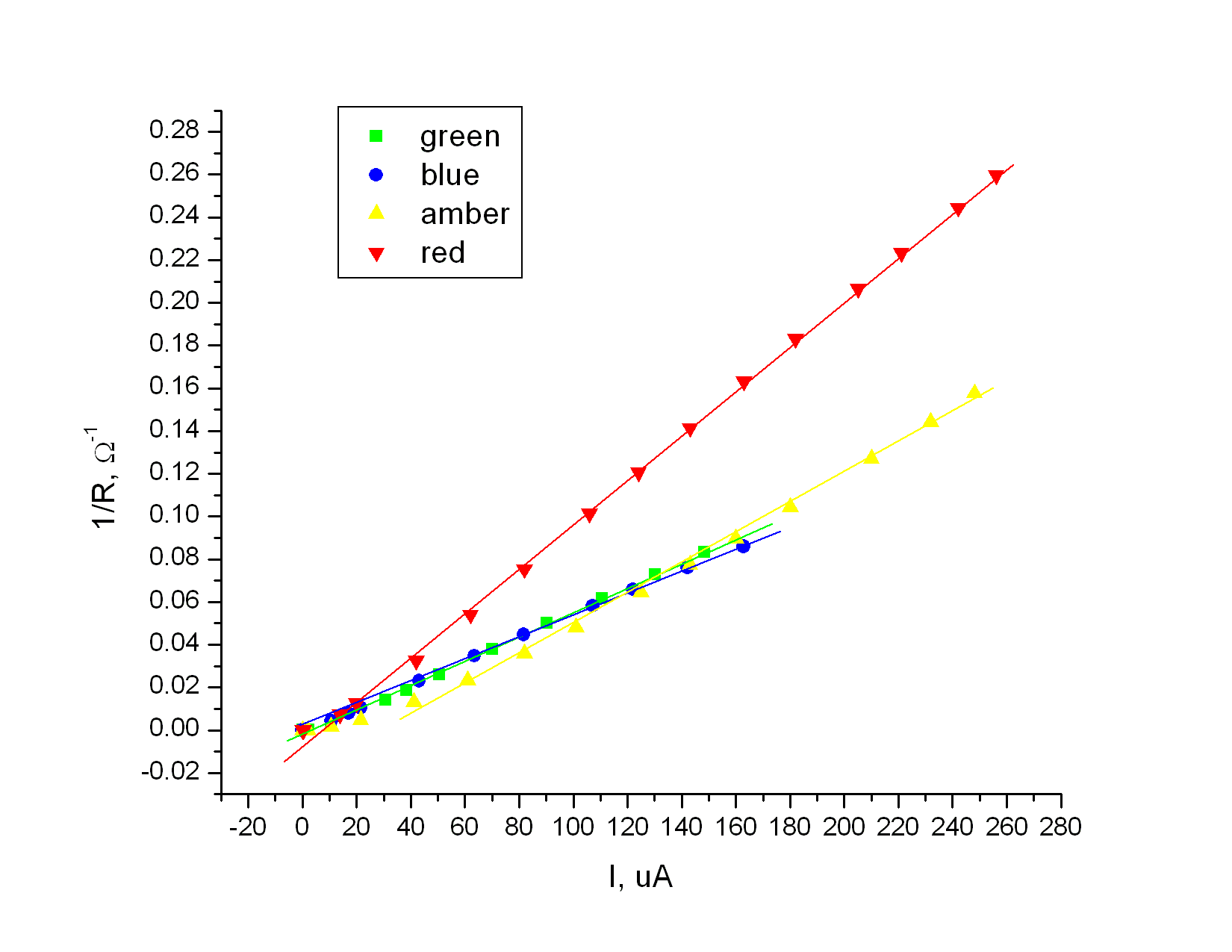}
\caption{Проводливоста на фотоотпорник $R_\varphi^{-1}$ како функција од јачината на струја од јачината на струја од до него поставената лед диода  $I.$  Инженерите кои не ги обожаваат грчките букви ја искористуваат кратенката  $\mathrm{u A}$ за $\mathrm{\mu A}.$ 
За сите бои: 
црвена (red), калибарно-жолта (amber), 
зелена (green) и сина (blue),
со висока точност се  наблудува линеарна зависност.
Околу прагот, јачината на струја низ лед диодата е пропорционален на интензитетот на емитираната светлина, 
што за возврат овозможува носители на електрична струја и електрична проводливост на фотоотпорникот.
            } 
\label{Fig:MK_Ampere_Siemens} 
\end{figure}

Сличностите на волта-амперните карактеристики  $R_\varphi(U)$ покажани на Сл.~\ref{Fig:MK_amper_siemens_vs_volt} заслужава да биде проучено.
Да ја објасниме накратко природата на фотоотпорникот.
Кога не е осветлен, тој е добар изолатор. Кога се осветлува пак, светлината создава носители на електрична струја: електрони и електронски празнини.  
Кога приложуваме напон, кој е причина за течење на струја тогаш е попрактично е да се претстави Омовиот закон преку проводливоста дефинирана како реципрочна вредност на отпорот 
\begin{equation}
I_\varphi=\sigma U_\varphi,\qquad \sigma\equiv\frac{1}{R_\varphi}.
\end{equation}
За фотоотпорниците, проводливоста  е пропорционална на интензитетот на светлината.
Ако фотоотпорникот е под постојан напон 
јачината на струјата што тече низ него е пропорционална на интензитетот на светлината.
На Сл.~\ref{Fig:MK_Ampere_Siemens} е прикажана зависноста на проводливоста на фотоотпорникот  $\sigma_\varphi(I_\mathrm{LED})$
во функција од јачината на струјата низ лед диодата.  Линеарната зависност покажува дека, 
рекомбинираните двојки електрон-празнина создаваат фотони кои пристигнуваат до фотоотпорникот и таму создаваат елктрон-празнина  кои овозможуваат електрична проводливост. 
Единицата за проводливост се нарекува Сименс Sm=$\Omega^{-1}$ 
и  техничарите кои ги помешуваат физичките величини со единиците често велат 
дека имаме линеарна сименс-амперна карактеристика.

Како што волт-амперните карактеристики, така и сименс-амперните карактеристики на лед диодите можат после извесна обработка на експерименталните податоци да дадат оценка на критичниот напон
$\tilde{U}_c$ и  енергијата на квантниот премин  $q_e\tilde{U}_c.$
Пресметаните критики напони се претставени во табелата~\ref{Table:MK_light_voltage_treshold}.

\begin{figure}[ht]
\includegraphics[scale=0.6]{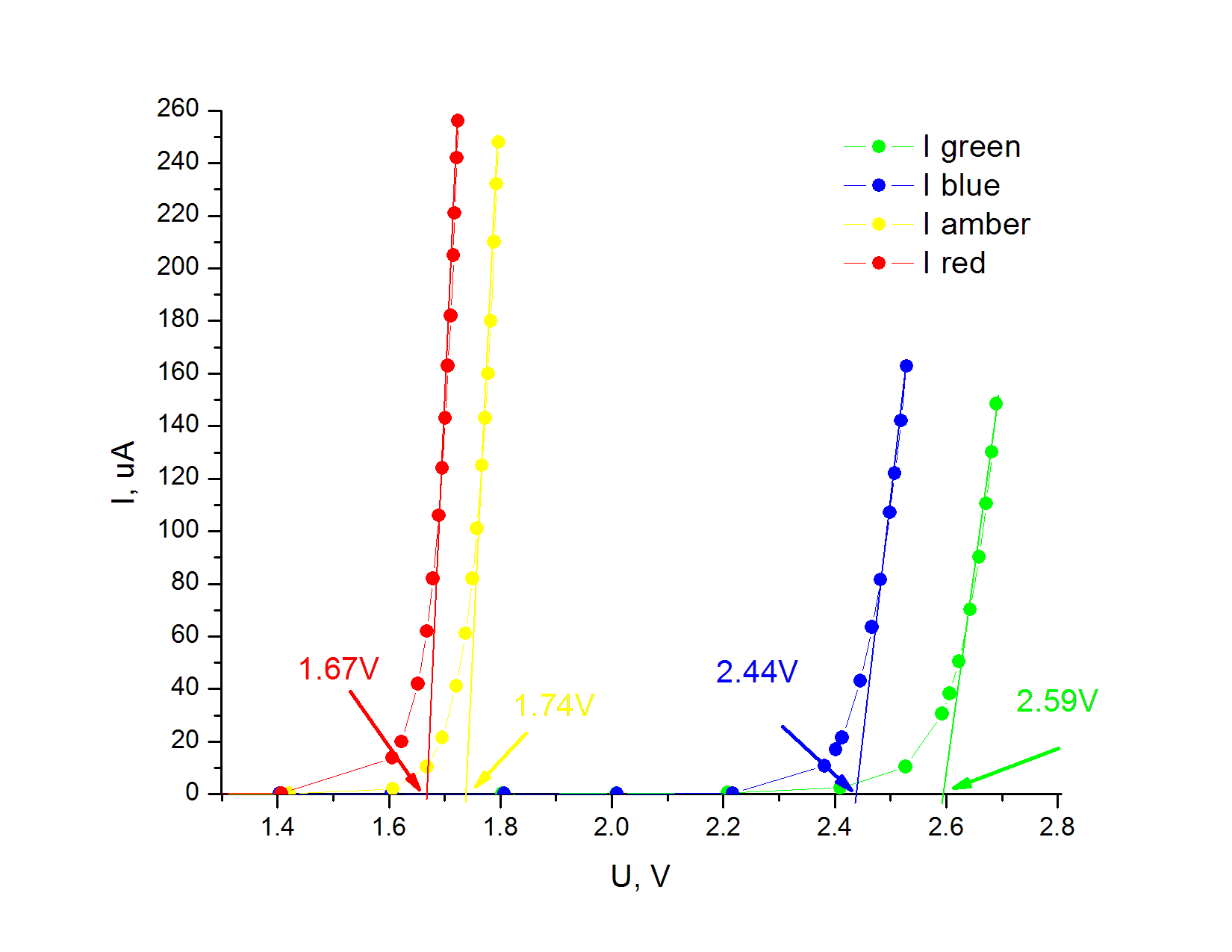}
\caption{Волт-амперни карактеристики и нивната линеарна апроксимација за различни по боја емитирачки лед диоди: 
црвена (red), калибарно-жолта (amber), 
зелена (green) и сина (blue).
Линеарната апроксимација ја пресекува апцисата и  го дава критичниот напон $\tilde{U}_c$  добиен преку анализа на волт-амперните карактеристики, 
а не  преку субјективниот осет кога диодата светнува.
$u$A $\equiv$ $\mu$A.
            }
\label{Fig:MK_v_threshold} 
\end{figure}

\begin{table}
\caption{Експериментални податоци за праговиот (критичкиот) напон 
на светнување (гасење) на лед диодите од соодветната боја} 
\begin{tabular}{| l | c |}
\tableline 
Цвят & $U_\mathrm{c}$ [V] \\
\tableline \tableline
 црвена & 1.330 \\
 жолта & 1.411 \\
 зелена & 1.916 \\
 сина & 1.927 \\
\tableline 
\end{tabular}
\label{Table:MK_light_voltage_treshold}
\end{table}

\begin{table}[ht]
\caption{Други експериментални податоци за дифракцијата 
и праговите, граничните напони за лед диоди со рзлично емитирачки  бои  (производител CREE); 
$D=21\,\mathrm{cm}$, 
$d=1.5\,\mu\mathrm{m}$}
\begin{tabular}{| c| l | c| c | c | c |c | c | c | c |}
\tableline \tableline 
\No & Цвят &  $L$ [cm] & $\tg(\theta)$ & $\theta$, deg& $\sin(\theta)$ & $\lambda\,[\mathrm{nm}]$
 & $\nu\mathrm{[Hz]}$ & $U_c\mathrm{[V]}$ & $q_eU_c\mathrm{[J]}$\\
\tableline 
1&виолетова&6.0& 0.294&15.5& 0.274& 411& $7.30\times 10^{14}$ & 1.46 & $3.70\times 10^{-19}$\\
2&сина     &6.7& 0.319&17.7& 0.304& 456& $6.58\times 10^{14}$ & 1.55 & $2.10\times 10^{-19}$\\
3&зелена  &7.5& 0.357&19.7& 0.337& 505& $5.94\times 10^{14}$ & 1.99 & $1.94\times 10^{-19}$\\
4&жолта   &8.3& 0.395&21.6& 0.368& 552& $5.43\times 10^{14}$ & 2.10 & $1,55\times 10^{-19}$\\
5&црвена&9.0& 0.429&23.2& 0.394& 591& $5.08\times 10^{14}$ & 2.23 & $1.40\times 10^{-19}$\\
\tableline 
\end{tabular}
\label{Table:MK_Valandovo}
\end{table}

\begin{table}[ht]
\caption{Критичните напони определени преку осетот на окото  $U_c\equiv U^{\mathrm{eye}}_c$ 
или ВАХ $\tilde{U}_c\equiv U^{\mathrm{VAC}}_c$  се дадени во соседните колони.
Експерименталните податоци за дифракцијата и критичните напони за лед диоди со различни бои од серијата  272 (широкоаголни)~\cite{LED}, $D=20\,\mathrm{cm}$, $d=1.5\,\mu\mathrm{m}$}. 
\begin{tabular}{| c| l | c| c | c | c |c | c | c | c | c | c |}
\tableline 
\No & Цвят &  $L$ [cm] & $\tg(\theta)$ & $\theta$ [deg]& $\sin(\theta)$ & $\lambda\,[\mathrm{nm}]$
& $\nu\mathrm{[Hz]}$ & $U^{\mathrm{eye}}_c\mathrm{[V]}$ & $q_eU^{\mathrm{eye}}_c\mathrm{[J]}$
& $U^{\mathrm{VAC}}_c\mathrm{[V]}$ & $q_eU^{\mathrm{VAC}}_c\mathrm{[J]}$\\
\tableline \tableline
2&син     &6.5& 0.325&18.0& 0.309& 464& $6.47\times 10^{14}$ & 1.927 & $3.087\times 10^{-19}$ & 2.44 & $3.91\times 10^{-19}$\\
3&зелен  &7.6& 0.380&20.8& 0.355& 532& $5.63\times 10^{14}$ & 1.916 & $3.069\times 10^{-19}$ & 2.59 & $4.15\times 10^{-19}$\\
4&жълт   &8.9& 0.445&24.0& 0.407& 609& $4.92\times 10^{14}$ & 1.411 & $2.260\times 10^{-19}$ & 1.74 & $2.79\times 10^{-19}$\\
5&червен&9.5& 0.475&25.4& 0.429& 644& $4.66\times 10^{14}$ & 1.1.33 & $2.131\times 10^{-19}$& 1.67 & $2.67\times 10^{-19}$\\
\tableline 
\end{tabular}
\label{Table:MK_Valandovo}
\end{table}

\begin{figure}[ht]
\includegraphics[scale=0.6]{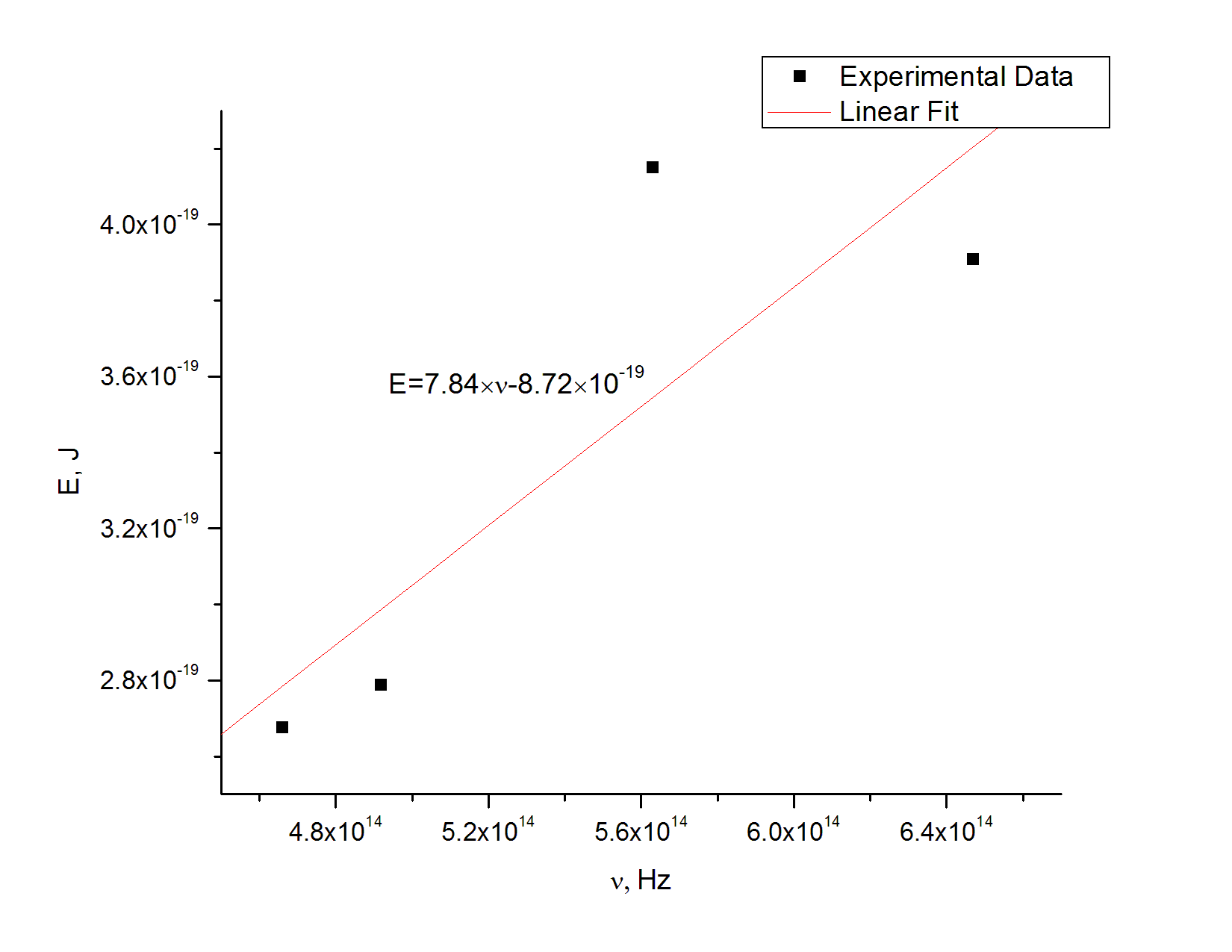}
\caption{Енергијата на премин  $E$ определена преку испитување на волт-амперните карактеристики 
и  соодветната фреквенција на светлинскиот бран  $\nu$.}
Полната права линија $E=(7.8\,\times\, 10^{-34}\,J)\nu+\mathrm{const}$ е  конструирана така што,
 сумата од квадратите на растојанијата по вертикала помеѓу експерименталните точки  
и правата линија да биде минимална. Мерењето ни покажува 18\% поголема 
 вредност споредена со точната вредност; 
како за прво мерење на Планковата константа претставува добар почеток.
Пунктирната линија е $E=(\dots\,\times\, 10^{-34}\,J)\nu+q_e(\dots\, \, \mathrm{V})$ 
е конструирана врз основа на  обработените податоци за зелената светлина.
\label{Fig:MK_Planck_VA} 
\end{figure}

\begin{figure}[ht]
\includegraphics[scale=0.6]{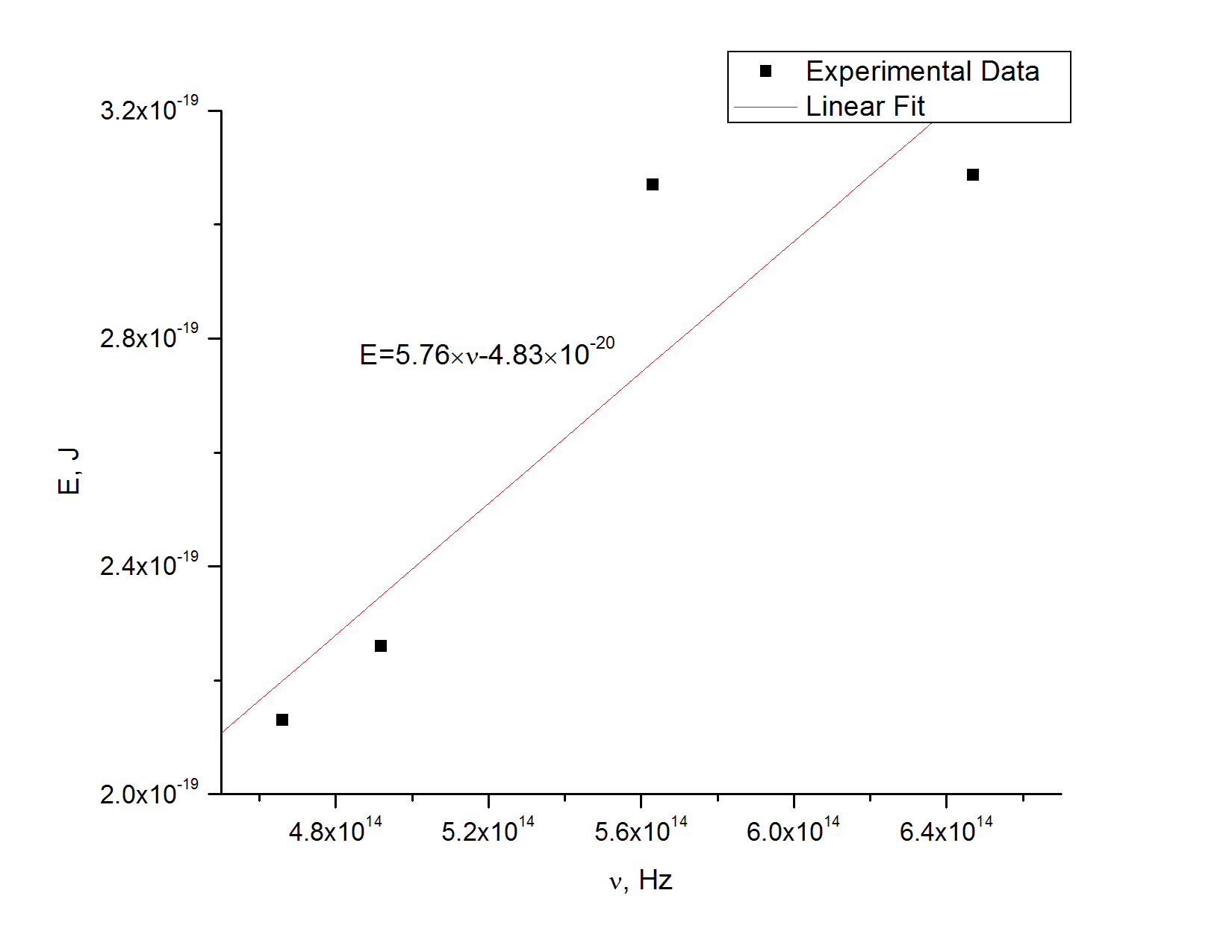}
\caption{Енергијата на премин определена според напонот на свeтење и гасење  во зависност од  фреквенцијата. За разлика от другата Сл.~\ref{Fig:MK_Planck_VA} сега фитуваната  права $E=(5.7\,\times\, 10^{-34}\,J)\nu+\mathrm{const}$} дава 15\% помала вредност. Тоа е  релативно добра точност при првото мерење на Планковата константа. За нас е важно, дека се соочивме со мерење на една од важните фундаментални константи поврзана со квантните појави.
\label{Fig:MK_Planck_eye} 
\end{figure}

\label{item:MK_RGB}На Сл.~\ref{Fig:MK_RGB_shadow} се покажани боите на сенките кои се добиваат:
1) пурпур (магента) =црвена + сина,
2) црвена
3)жолта= црвена + зелена,
4) зелена
5) сина-зелено (циан) = сина + зелена
6) бела= црвена+ сина + зелена 
(само при добро избалансирани со потенциометарот 2 бои).

\ref{MK_PlanckError}. 
Константата која учествува во равенката за фотоефектот~\eqref{MK_MK_photoeffect} за многу малку се разликува кај лед диодите кои светат со различна боја.  
Кај зелените лед  диоди таа константа значително се разликува во споредба со другите лед диоди. 
Осве тоа светењето и гаснењето на лед диодите не е набљудувано на темно. 
При подетално испитување на волта-амперната карактеристика  критичниот напон може да се определи со поголема точност,
но треба да се зема во предвид некој фактор, множител на неидеалност  (\textit{non ideality factor}) 
 на лед диодите  $n,$  нешто што би барало продлабочено изучување на електрониката.
Најточни мерења се добиваат со испитување на волт-амперни карактеристики на вакуумска диода осетлувана со различни лед диоди.
Експериментот за набљудување на појавата фотоефект е опишан во учебниците.

\ref{alpha}. Пред да настане дифракцијата фотонот има паралелно на дифракционата решетка проекција на импулсот  $p\equiv(h/\lambda)\sin(\alpha)$. 
Од дифракционата решетка со период  $d$ фотонот добива импулс  $p_0\equiv h/d$ или множител на овој импулс ; 
за поедноставност во тој услов на задачата го разгледуваме само првиот дифракционен максимум. 
По дифракцијата  $y$-компонентата на импулсот е 
$p^\prime\equiv(h/\lambda)\sin(\theta)$, 
а  според законот за запазување на импулсот  $p^\prime=p^\prime+p_0$ добиваме 
\begin{equation}
 \frac{h}{\lambda}\sin(\theta)
 =\frac{h}{\lambda}\sin(\alpha)+\frac{h}{d}.
\end{equation}
 Разликата  во оптичките патишта за соседните процепи на дифракционата решетка е
\begin{equation}
 \lambda=\left[\sin(\theta)-\sin(\alpha)\right]d.
\end{equation}
Брановото решение бара добра скица, а механичкото решение -знаење на законите за запазување. 

\ref{MK_RefractionIndeks}.
За случај кога светлината се движи во средина со индекс на прекршување $n$
 брановата должина станува $n$ пати помала. Во формулата за агол на дифракција треба да се замени брановата должина за во дадената средина 
$\lambda_n\equiv\lambda/n=d \sin(\theta).$ За вода, синусот од аголот на дифракцијата се намалува за околу 25\% кое лесно може и експериментално да се провери. 
Ако дифракционата решетка е на границата на две средини со индекси на прекршување  $n_1$ и $n_2$, законот за запазување на импулсот го добива видот 
\begin{equation}
 \frac{h}{\lambda_2}\sin(\theta)
 =\frac{h}{\lambda_1}\sin(\alpha)+i_m\frac{h}{d},
\qquad \lambda_1\equiv\frac{\lambda}{n_1}, 
\quad \lambda_2\equiv\frac{\lambda}{n_2},
\end{equation}
каде целиот број  $i_m=0,\pm 1,\, \pm 2,\, \dots$ 
го претставува  редот на дифракциониот максимум, a $\lambda$
е  брановата должина на светлината во вакуум.
На пример,при $i_m=0$, го добиваме Снелиусовиот закон 
$n_1\sin(\alpha)=n_2\sin(\theta),$  како логичка последица на законот за запазување на импулсот. 

\ref{MK_f<<R}. Нека диодата ја разгледуваме како точкаст светлински извор поставена во фокусот на леќата. 
Позади леќета се добива речиси паралелен сноп на светлина. 
Радиусот на светлинскиот сноп е еднаков со радиусот на леќата. 
За да може процепите (браздите, зарезите) на дискот да ги разгледуваме приближно  паралелени потребно е радиусот на дискот $R_\mathrm{CD}$ 
да биде многу поголем од фокусното растојание  $R_\mathrm{CD}\gg f$. 
За да се зголеми точноста на мерењата, 
потребно е леќата да биде поблиску до периферијата на дискот. 
Ако располагате само со леќа која  има поголемо фокусно растојание  од дијаметарот на дискот, потребно е да се постави бленда до дифракционата решетка. 
Најдобро би било ако фолијата од дискот се отсрани само од  $1\;\mathrm{cm}^2.$ 

\ref{MK_CombinationalFrequency} 
Ако решетката се движи  со брзина  $v$ периодот $d$ се поминува за време $d/v$ а фреквенцијата на преминување на процепите спрема неподвижен наблудувач е $\nu_0=v/d.$ 
Таа мала фреквенција се додава кон фреквенцијата на упадниот фотон
$\nu=1/T$ со период на бранот $T$. 
Во овој случај , за фреквенцијата на дифрактирана светлина добиваме $\nu^\prime=\nu+\nu_0.$ 
Ако ја помножиме со планковата константа  $h$  ја добиваме енергијата на расејаниот фотон 
$E^\prime=E+E_0,$ къдет $E^\prime\equiv h\nu^\prime,$  $E\equiv h\nu$ и $E_0\equiv h\nu_0.$ 
Со други зборови, дифракционата решетка во движење  ја предава својата енергија на фотонот. 
Таков случај имаме на пример кога светлината се рассејува од звучен бран. 
Звучниот бран претставува еден вид на движечка дифракциона решетка.
Фреквенцијата на рассејаниот бран може да се пресмета од законот за запазување на енергијата.
Квантот на звукот (фононот) и квантот на апсорбираната светлина  (фотонот) 
ја предаваат својата енергија на рассејаниот фотон.
При многу физички процеси кога два брана создаваат трет, 
на пример, кај  Комптоновото или Рамановоторассејување, имаме имаме истовремено запазување на енергијата и импулсот
\begin{equation}
 E^\prime=E+E_0, \qquad \mathbf{p}^\prime=\mathbf{p}+\mathbf{p}_0.
\end{equation}

\ref{MK_Hamilton}. Во формулата за фазна брзина на фотонот 
ја изразуваме брановата должина преку импулсот и преку периодот  преку енергијата 
\begin{equation}
 c=\frac{\lambda\!=\!h/p}{T\!=\!h/E}=\frac{E}{p}
\end{equation}
и  ја добиваме врската помеѓу  енергијата и импулсот на фотонот $E=cp.$
Тоа е релацјата на поврзаност помеѓу импулсот и енергијата за рамен електромагнетен бран $p=E/c.$

\ref{MK_Chain}. Со заменување од низата во формули се добива 
\begin{equation}
 X=\left(V=\frac{p=E/c}{M}\right)\left(t=\frac{l}{c}\right)
   =l\left(m\equiv\frac{E}{c^2}\right)/M.
\end{equation}

\ref{MK_Muenhausen}.Горната равенка може да се запише како $MX=mL.$ 
Тоа е оргиналното рассудување на Ајнштајн за еквивалентност на масата и енергијата  $E=mc^2$ 
 изведена како последица на постојаната положба на центарот на масата. 
При разумни енергии, поместувањето на вагонот е занемарливо мало, 
масата  на загреаното тело е незначително по-голема, 
но од друга страна мали разлики во масите соодветствуваат на огромна енергија.
Така импулсот на фотонот дал поттик во развојот на нуклеарната физика и енергетика.
А во нашиот замислен експеримент првичната енергија $E$  била хемиска енергија на ласерскиот покажувач,   
а потоа се претворува во топлинска енергија на ѕидот апсорбирана од светлинскиот импулс. 
Покрај решението на задачата направивме краток  преглед на материјалот  од учебниците по физика.


\begin{thebibliography}{9}
%

\bibitem{IYL2015}
  \textit{UNESCO, 2015 International Year of Light and Light-based Technologies}
 \url{http://www.light2015.org/Home.html}

\bibitem{EPO_IYL2015}
 \textit{Second Experimental Physics Olympiad: The Day of the Photon in the International Year of Light, Sofia, 25 April (2015)}\\
 \url{http://www.light2015.org/Home/Event-Programme/2015/Competition/Bulgaria-Second-Experimental-Physics-Olympiad--25-April-2015-in-Sofia.-The-Day-of-the-Photon-in-the-International-Year-of-the-Light.html}

\bibitem{LED}
 \textit{Светодиоди серия 272 (широкоъгълни)}
 \url{http://www.led-bg.com/bg/product/svetodiodi-seriya-272-shirokoagalni}

\bibitem{Bib:RGB}
 \url{http://dev.physicslab.org/Document.aspx?}
 \url{doctype=5&filename=Compilations_NextTime_Shadows2.xml}
PhysicsLAB,
HTML conversion,
Catharine H.~Colwell,
Mainland High School,
Daytona Beach,
FL 32114.

\bibitem{VaccumPhotoCell}
\url{https://www.pinterest.com/pin/285697170082704438/}
\url{http://www.vivatubes.com/nib-rca-usa-1p40-photocell-vacuum-tube-ip40-1/},\\
\url{http://www.istok2.com/data/2559/},\\
\url{http://ru.pc-history.com/wp-content/uploads/Fotoelement_SCB-4.jpg},\\
\url{http://ru.pc-history.com/fotoelement-scv-4.html},\\
\url{http://aukro.ua/fotoelement-scv-4-i5027393716.html},\\
\url{http://www.ebay.com/itm/2-VACUUM-TUBES-RCA-930-6E5/201322750445?_trksid=p2047675.c100005.m1851&_trkparms=aid%3D222007%26algo%3DSIC.MBE%26ao%3D1%26asc%3D29384%26meid%3Dbb80c151a256490fbe6f40df42bc9a3e%26pid%3D100005%26rk%3D3%26rkt%3D6%26sd%3D151256790743&rt=nc
},\\
\url{
http://www.ebay.com/itm/Vacuum-Tube-RCA-Type-6W6GT-Clear-Glass-Black-Rectangular-Plate-USED/321699509598?_trksid=p2047675.c100005.m1851&_trkparms=aid%3D222007%26algo%3DSIC.MBE%26ao%3D1%26asc%3D29384%26meid%3Dbb80c151a256490fbe6f40df42bc9a3e%26pid%3D100005%26rk%3D4%26rkt%3D6%26sd%3D151256790743&rt=nc
},\\
\url{
http://www.ebay.com/itm/RCA-930-PHOTOCELL-VACUUM-TUBE-BLACK-GLASS-/151256790743}.

\bibitem{PhotoMultiplier}
\url{http://en.wikipedia.org/wiki/Photomultiplier},\\
\url{http://www.eandc.ru/news/detail.php?ID=20954},\\
\url{http://www.eandc.ru/catalog/detail.php?ID=16654},\\
\url{http://www.pitersvet.ru/index.php?productID=501},\\
\url{http://www.istok2.com/data/2850/},\\
\url{http://pepina.org/pdf/tube/fey-35.pdf}.

\bibitem{Millikan:1916}
 R. Millikan, 
`A Direct Photoelectric Determination of Planck's ``h.'''
Physical Review \textbf{7}, Nr. 3, 355–388 (1 March 1916), \\
\url{http://journals.aps.org/pr/pdf/10.1103/PhysRev.7.355}

\bibitem{Patel:2014}
Neel V. Patel,
 IEEE Spectrum, (October 9, 2014)
``Nobel Shocker: RCA Had the First Blue LED in 1972''\\
\url{http://spectrum.ieee.org/tech-talk/geek-life/history/rcas-forgotten-work-on-the-blue-led},\\
\url{http://www.nobelprize.org/nobel_prizes/physics/laureates/2014/press.html},\\
\url{http://en.wikipedia.org/wiki/Light-emitting_diode#cite_note-134}.

\end{thebibliography}
\end{document}